\documentclass[aps,prx,superscriptaddress,onecolumn,notitlepage,longbibliography,nofootinbib,floatfix]{revtex4-2}
\usepackage{comment}
\usepackage{amsmath,amssymb,amsthm}
\usepackage{physics}
\usepackage{graphicx}
\usepackage{hyperref}
\usepackage{xcolor}
\usepackage{bm}
\usepackage{booktabs}
\usepackage{enumitem}
\usepackage{tikz}
\usepackage{orcidlink}
\hypersetup{colorlinks=true, linkcolor=green!60!black,
            citecolor=green!60!black, urlcolor=green!60!black}

\graphicspath{{./}{./figure/}{./tex/figure/}}

\usepackage{hyperref}
\hypersetup{
    colorlinks,
    linkcolor={green!60!black},
    citecolor={green!60!black},
    urlcolor={green!60!black}
}

\newcommand{\spinlabel}{\mathsf{s}}

\begin{document}

\title{Coherence dynamics in quantum many-body systems with conservation laws}

\author{Sreemayee Aditya\,\orcidlink{0000-0002-0412-7944}}
\email{asreemay@uni-koeln.de}
\affiliation{Institut für Theoretische Physik, Zülpicherstraße 77a, 50937 Köln, Germany}

\author{Emanuele Tirrito\,\orcidlink{0000-0001-7067-1203}}
\affiliation{Laboratory of Theoretical Physics of Nanosystems (LTPN), Institute of Physics}
\affiliation{Center for Quantum Science and Engineering, Ecole Polytechnique Fédérale de Lausanne (EPFL), CH-1015 Lausanne, Switzerland}

\author{Piotr Sierant\,\orcidlink{0000-0001-9219-7274}}
\affiliation{Barcelona Supercomputing Center, Plaça Eusebi Güell 1-3, 08034 Barcelona, Spain}

\author{Xhek Turkeshi\,\orcidlink{0000-0003-1093-3771}}
\affiliation{Institut für Theoretische Physik, Zülpicherstraße 77a, 50937 Köln, Germany}

\date{\today}

\begin{abstract}
We study how conservation laws shape the spreading of quantum coherence in many-body dynamics.
Focusing on $U(1)$-symmetric random circuits, charge-and-dipole conserving circuits, as well as ergodic Hamiltonian dynamics, we probe coherences both globally, via the participation entropy, and locally, via the relative entropy of coherence.
Combining exact vector evolution, matrix product state simulations, and replica tensor networks methods, we find that conservation laws replace the logarithmic saturation of unconstrained circuits with slow hydrodynamic relaxation of the global coherence measures.
Locally, symmetry-constrained circuits show a clean rise-peak-fall structure whose peak time grows algebraically with subsystem size. In contrast, ergodic Hamiltonians broaden the peak into an extended plateau at larger subsystems, highlighting a qualitatively distinct mechanism. Coherence thus emerges as a sensitive probe of symmetry-constrained thermalization, linking quantum resource dynamics to many-body transport.
\end{abstract}

\maketitle

\tableofcontents

\section{Introduction}
\label{sec:intro}

The dynamics of isolated many-body quantum systems is a central
theme at the interface of quantum information and condensed
matter~\cite{silva, fisher2023random, hangleiter2023computational, potter2022entanglement, dalessio2016quantum}.
Under generic unitary evolution, a simple  initial state
spreads over the exponentially large Hilbert space and eventually
mimics a typical (Haar-random) target~\cite{collins2006integration, hunterjones2019unitary, haferkamp2022, Haferkamp2022randomquantum, fava2025designs, grevink2025glueshortdepthdesignsunitary, schuster2025randomunitariesextremelylow, dalzell2024random}; how rapidly and in what manner this occurs depends crucially on the symmetry 
of the system.
In the absence of conservation laws, local random
circuits scramble efficiently. For example, entanglement grows
ballistically~\cite{nahum2017quantum, nahum2018operator, Keyserlingk2018operator, bertini2019exact, bertini2020scrambling, chan2018, ware2023sharp, zhou2019emergent, zhou2020entanglement, sierant2023ent, khemani2018operator, liu2022manybody, liu2024quantum,dowling2025freeindependenceunitarydesign},
while a global indicator of complexity, such as the participation
entropy, anticoncentrates on the much shorter
logarithmic scale $\log_2
L$~\cite{Hangleiter2018anticoncentration,christopoulos2024universal,fefferman2024anticoncentrationunitaryhaarmeasure, sauliere2025universalityanticoncentrationchaoticquantum, lami2025anticoncentration,   Anonymous_2026,magni2025anticoncentrationstatedesigndoped,p8dn-glcw,heinrich2026criticalbehaviorsmagicparticipation,heinrich2025anticoncentrationalmostneed,Magni2025quantumcomplexity,tirrito2024anticoncentrationmagicspreadingergodic, dalzell2022random, boixo2018characterizing, arute2019quantum, Bouland2018on, morvan2024phase, luitz2014universal, luitz2014participation, mace2019multifractal, sierant2022universal, backer2019multifractal, claeys2025fockspace, turkeshi2024hilbert, fefferman2024effect, mark2023benchmarking, mark2024maximum, ippoliti2023dynamical, Chan2022}.
Once a conservation law is imposed, however, the picture changes
qualitatively: hydrodynamic modes tied to the conserved charges
freeze the slowest sectors of the evolution and inject algebraic
tails that dominate the late-time relaxation of
entanglement~\cite{rakovszky2018diffusive, rakovszky2019sub, znidaric2020entanglement, Jonay24slow, turkeshi2025quantum,liu2024,tirrito2025universalspreadingnonstabilizernessquantum,iannotti2026nonstabilizernessu1symmetrychaotic}.

A particularly sharp and operationally motivated probe of this
symmetry-induced slowdown
is provided by the quantum coherence, the ability of a quantum state to exhibit superpositions
in a specified reference basis, and one of the most distinctive
nonclassical features of quantum
mechanics~\cite{chitambar2019quantum, baumgratz2014quantifying, streltsov2017colloquium}.
In an isolated many-body system, coherence is continuously generated
and reshuffled by the unitary evolution. Therefore, tracking its 
time evolution gives a direct readout of how rapidly the system loses memory of its initial product configuration and approaches the Haar-random limit set by the accessible symmetry sector.

In this work, we address the question of how symmetry constrains
the spreading of \emph{coherence} starting from an incoherent state under resourceful operations, both globally and locally.
We employ the resource theory~\cite{chitambar2019quantum} perspective on quantum coherence and probe its spreading with two complementary metrics: a global probe, the participation
entropy~\cite{luitz2014universal, luitz2014participation, mace2019multifractal, sierant2022universal, turkeshi2024hilbert},
which tracks how broadly the many-body wavefunction spreads in the
computational basis; and a local probe, the  relative
entropy of coherence~\cite{chitambar2019quantum, baumgratz2014quantifying, streltsov2017colloquium,saxena2020coherence}, which measures the
coherence retained by a subsystem of size $L_A$. We study three
paradigmatic settings: (i)~$U(1)$-symmetric random circuits for
spin-$\tfrac12$ and spin-$1$~\cite{Keyserlingk2018operator, nahum2018operator};
(ii)~random circuits that conserve both charge and
dipole~\cite{ feldmeier2020anomalous, Singh2021subdiffusion, iaconis2021multipole, moudgalya2022quantum, morningstar2020kinetically, nandkishore2019fractons, pretko2017subdimensional,pai2019localization,khemani2020localization},
where Hilbert-space fragmentation typically enforces subdiffusive transport in terms of the correlation function;
and (iii)~the ergodic mixed-field Ising chain 
representative of
local ergodic Hamiltonian
dynamics~\cite{bertini2020scrambling, deutsch1991quantum, srednicki1994chaos, rigol2008thermalization, dalessio2016quantum, kim2014testing, prosen2007chaos}.
We combine exact state-vector evolution with replica tensor-network
(RTN) simulations and matrix-product-state (MPS) methods~\cite{haegeman2011time, haegeman2016unifying, itensor, Yang20tdvp} to reach large scale
systems.

\emph{Summary of results.}
We find that conservation laws replace the logarithmic saturation of
unconstrained random circuits~\cite{tirrito2024anticoncentrationmagicspreadingergodic,aditya2025growthspreadingquantumresources} with a power-law approach to equilibrium.
Globally, the deviation of participation entropy from its long-time saturation value, quantified by $\Delta S_d(t)=S_d(\infty)-S_d(t)$ in symmetric circuits, exhibits
a two-stage decay: an intermediate power-law regime
$\Delta S_d\sim t^{-\beta_p}$ followed by a finite-size regime of exponential
decay with timescale $\tau_p\propto L^{\alpha_p}$. Consequently, the
time $t_\epsilon$ needed for $S_d$ to saturate to the Haar value (allowed by the symmetry sector) up to a fixed tolerance
$\epsilon$ grows as $t_\epsilon\propto L^{\alpha_\epsilon}$, in sharp
contrast with the $\log L$ scaling of unconstrained circuits~\cite{dalzell2022random, turkeshi2024hilbert, tirrito2024anticoncentrationmagicspreadingergodic}.
In local subsystems, the competition between the algebraic relaxation of $S_d$
and the ballistic growth of the Rényi-2 entanglement entropy $S_R$ produces a pronounced
rise--peak--fall profile of relative entropy of coherence, $C_d(t)$ in both $U(1)$- and
dipole-conserving circuits, with a peak time
$\tau_c^m\propto L_A^{\alpha_m}$ set by the exponent
$\alpha_m=1/(\beta_{S_d}+1)$, where the decay of local $S_d$ at intermediate times governed by $\Delta S_{d}\sim t^{-\beta_{S_d}}$. On the other hand, the local spreading in
Hamiltonian dynamics occur with a striking difference: the sharp
local peak observed at small $L_A$ broadens into an extended
plateau at larger subsystems, signalling a qualitatively distinct
local relaxation mechanism. We complement these numerics with
analytical results: Haar averages for both the diagonal and
subsystem purities, together with a rare-region analysis of
$S_R$ and $\Delta S_d$ based on the coupled symmetric simple exclusion processes (SSEP) effective
model~\cite{rakovszky2019sub, turkeshi2025quantum}, which explains
the exponent $\beta_{S_d}\approx 1$ as a product of two independent
diffusive modes. Taken together, these results show that coherence is
a sensitive probe of symmetry-constrained thermalization, linking
quantum resource dynamics directly to the underlying hydrodynamics
of conserved charges.

\emph{Outline.}
Section~\ref{sec:resource} reviews the resource-theoretic framework
and defines the global and local coherence measures.
Section~\ref{sec:models} introduces the three dynamical settings
and the numerical methods. Sections~\ref{sec:u1_results},
\ref{sec:dipole_results} and \ref{sec:mfim_results} collect our
results for $U(1)$ circuits, dipole-conserving circuits, and
Hamiltonian dynamics, respectively. Section~\ref{sec:summary}
concludes with a discussion of open directions. Technical derivations
are collected in Appendices~\ref{app:general_diag_renyi}--\ref{app:tn_coherence}.

\begin{figure}[h]
\includegraphics[width=0.7\columnwidth]{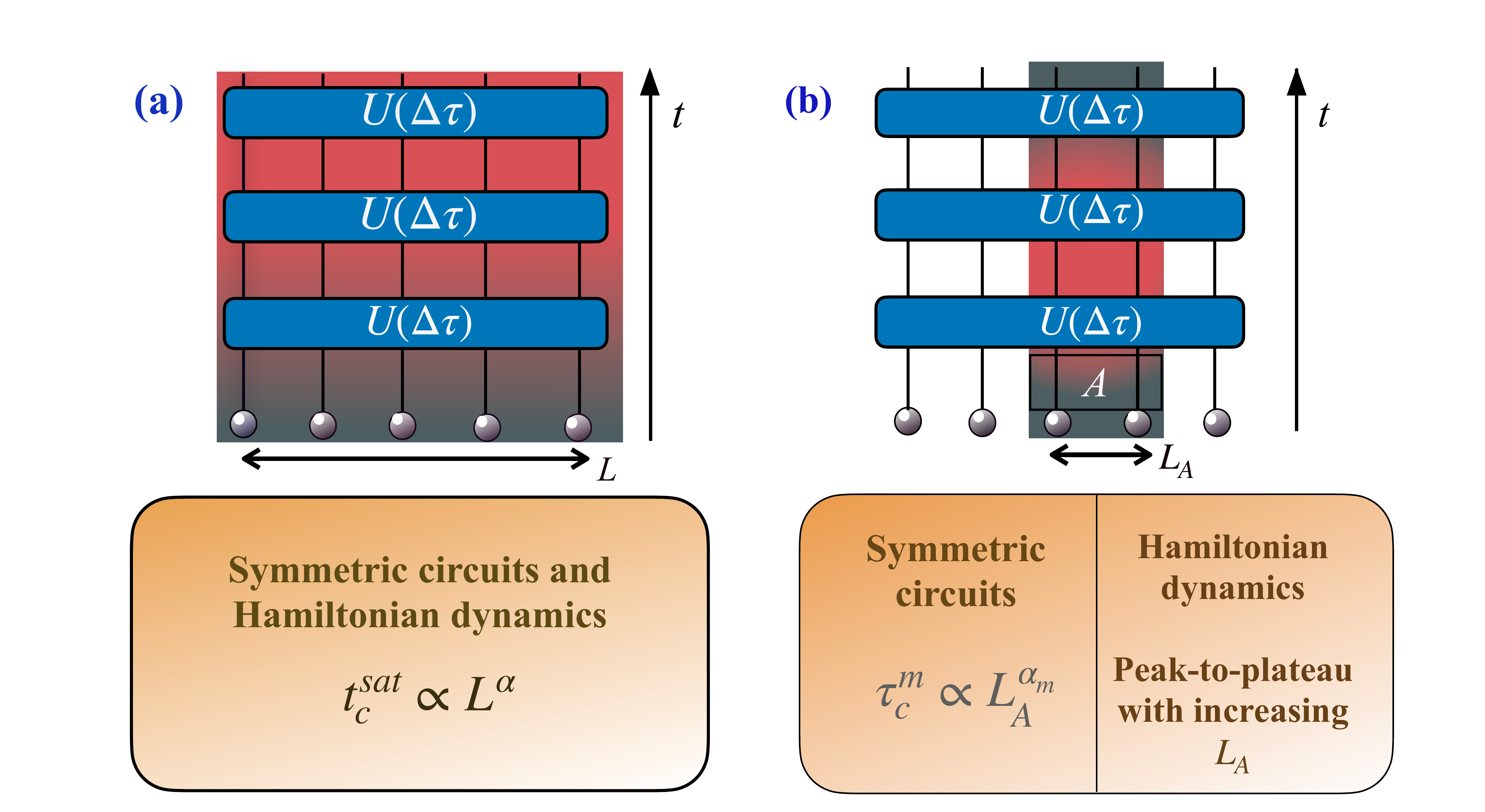}
\caption{\textbf{Global and local spreading of quantum coherence.} A product incoherent state on $L$ qubits is evolved by a brick-wall circuit of gates $U(\Delta\tau)$, realizing $U(1)$-symmetric or charge-and-dipole conserving random dynamics, or under the mixed-field Ising Hamiltonian. 
Red shading marks the region over which the global/ local resource is maximum, while the gray region indicates minimal coherence resource. 
\textbf{(a)} Global dynamics: $S_d$ saturates on the full chain at a time $t_c^{\,\mathrm{sat}}\propto L^{\alpha}$. \textbf{(b)} Local dynamics on a subsystem $A$ of size $L_A$: for symmetric circuits $C_d(A,t)$ peaks at $\tau_c^{\,m}\propto L_A^{\alpha_m}$ before relaxing to a free state, while there exists a peak-to-plateau crossover with $L_A$ in Hamiltonian evolution.}
\label{fig:schematic}
\end{figure}

\section{Resource theory of coherence and quantities of interest}
\label{sec:resource}

Resource theories provide a natural framework for quantifying nonclassical features of quantum states~\cite{chitambar2019quantum}. Intuitively, one separates states and operations that are \emph{free}, cheap or readily implementable under the physical constraints at hand, from those that are \emph{costly}, and hence genuinely useful, to produce.
Formally, a resource theory defines a set of \emph{free states and operations}, and regards as resourceful everything lying outside this ensemble. 
This framework naturally leads to the notion of a \emph{resource monotone} \(M(\rho)\), i.e.\ a function of a quantum state \(\rho\) that quantifies ``how much'' of a resource is needed to prepare 
the state. Resource monotones satisfy (i) \emph{faithfulness}, $M(\rho)=0$ if and only if $\rho$ is free, and (ii) \emph{monotonicity}, $M(\Lambda[\rho])\le M(\rho)$ for every free operation $\Lambda$. For coherence~\cite{chitambar2019quantum, baumgratz2014quantifying, streltsov2017colloquium}, the free states are the density matrices diagonal in a fixed reference (e.g., computational) basis, and the free operations are the so-called incoherent operations that do not create coherent superpositions when acting on basis states. 

In this work, we study the dynamics of coherence under symmetry-constrained evolutions from two complementary perspectives. The first is a \emph{global} viewpoint, which characterizes how broadly the full many-body wavefunction spreads in the computational basis. The second is a \emph{local} viewpoint, which quantifies the coherence retained in a subsystem. 

We consider a one-dimensional quantum spin-$\spinlabel$ chain of length $L$ with local Hilbert-space dimension $q=2\spinlabel+1$, so that its many-body Hilbert space has dimension $q^L$.
We define the spin-$\spinlabel$ operator $Z$ that is diagonal in the local computational basis $ \mathcal{B}_1:=\{ |(\spinlabel+1-n) (1+\mathrm{mod}(2\spinlabel,2))\rangle \; | \; n=1,\dots,q\}$, namely $Z|m\rangle=m |m\rangle$ for $|m\rangle\in\mathcal{B}_1$.
To quantify the global spreading of the wavefunction, we use the \emph{participation entropy}~\cite{luitz2014universal, luitz2014participation, mace2019multifractal, sierant2022universal, turkeshi2024hilbert}, which measures how much a pure state is distributed over a reference basis. 

For a many-body state $\lvert\Psi\rangle$ and computational basis $\mathcal B_L=\mathcal{B}_1^{\otimes L}$, the R\'enyi participation entropy is defined as
\begin{equation}
S_k^{\mathrm{PE}}(\lvert\Psi\rangle)
=
\frac{1}{1-k}\log_2\!\left(\sum_{x\in\mathcal B_L} p_x^k\right),
\qquad
p_x=\lvert\langle x\vert\Psi\rangle\rvert^2,
\end{equation}
and reduces to the Shannon entropy in the limit $k\to1$. It is non-negative and vanishes only when $\lvert\Psi\rangle$ is localized on a single basis configuration.
For generic many-body states, the participation entropy grows extensively with system size as $S_k^{\mathrm{PE}}=D_k L + c_k$, where $D_k$ is the multifractal dimension associated with R\'enyi index $k$ and $c_k$ is a subleading constant. The value $D_k=0$ characterizes a localized state, $D_k=1$ a fully ergodic and extended wavefunction, and $0<D_k<1$ a multifractal state provided that $D_k$ depends non-trivially on $k$~\cite{mace2019multifractal}. 

In this work, we will focus on the R\'enyi-$2$ case, which is directly related to the inverse participation ratio~\cite{BKramer1993} or, equivalently, to the collision probability $p_{\mathrm{col}}=\sum_x p_x^2$~\cite{dalzell2022random}, and thus probes the global Hilbert space delocalization, or anticoncentration, of the wavefunction.
To lighten the notation, we henceforth drop the label ``PE'' and denote the global R\'enyi-$2$ participation entropy simply by
\begin{equation}
S_d(\lvert\Psi\rangle)
:=S_2^\mathrm{PE}(|\Psi\rangle)= -\log_2 (p_{\mathrm{col}})=
-\log_2\!\left(\sum_{x\in\mathcal B_L} p_x^2\right),
\end{equation}
where the subscript ``$d$'' stands for \emph{diagonal} R\'enyi-$2$ entropy; the same symbol will be reused below for its subsystem version. 

While the participation entropy probes the global redistribution of wavefunction weight in the computational basis, it does not directly quantify how much coherence is retained by a given subsystem. To address this local aspect, we use the resource-theoretic notion of quantum coherence with respect to the same computational basis~\cite{chitambar2019quantum, baumgratz2014quantifying, streltsov2017colloquium,saxena2020coherence}. 
Consider a bipartition $A\cup B$ of the system with computational basis respectively $\mathcal{B}_{L_A}$ and $\mathcal{B}_{L_B}$. For a subsystem $A$ of $L_A$ spins with reduced density matrix $\rho_A:=\mathrm{tr}_B (|\Psi\rangle\langle \Psi|)=\sum_{y\in \mathcal{B}_{L_B}} \langle y|\Psi\rangle\langle \Psi|y\rangle$, let $\rho_{A,\mathrm{diag}}:=\sum_{x \in \mathcal{B}_{L_A}} \langle x|\rho_A|x\rangle |x\rangle\langle x|$ be the diagonal, or dephased, reduced density matrix, obtained by removing all off-diagonal elements in the local computational basis $\mathcal{B}_{L_A}$. 

Two natural second-moment quantities play a central role throughout this work.
We denote by
\begin{equation}
P_A:=\Tr[\rho_A^2],\qquad
P_{A,\mathrm{diag}}:=\Tr[(\rho_{A,\mathrm{diag}})^2]=\sum_{x\in\mathcal{B}_{L_A}}\langle x|\rho_A|x\rangle^{2},
\label{eq:purity_defs}
\end{equation}
respectively the \emph{subsystem purity} of $\rho_A$ and the \emph{diagonal} (or \emph{dephased}) \emph{purity}, i.e.\ the purity of the dephased reduced density matrix $\rho_{A,\mathrm{diag}}$. The diagonal purity coincides with the collision probability of the outcome distribution of a computational-basis measurement on $A$, and satisfies $0\le P_{A,\mathrm{diag}}\le P_A\le 1$. Their logarithms define the two R\'enyi-$2$ entropies of interest.
The Rényi-$2$ relative entropy of coherence then quantifies the local coherence content,
\begin{equation}
\label{eq:Cd_def}
C_d(\rho_A)=S_d(\rho_A)-S_R(\rho_A),
\end{equation}
where
\begin{equation}
S_d(\rho_A)=-\log_2 P_{A,\mathrm{diag}},
\qquad
S_R(\rho_A)=-\log_2 P_A,
\end{equation}
are the subsystem diagonal entropy and the subsystem entanglement entropy, respectively.
In other words, $C_d$ subtracts from the participation entropy of $\rho_A$ the contribution arising from entanglement with the complement. As a result, $C_d(\rho_A)$ quantifies the coherence stored locally in subsystem $A$~\cite{chitambar2019quantum}, complementing the global information contained in $S_d(\lvert\Psi\rangle)$.

The joint study of these two quantities provides a systematic way of characterizing how coherence spreads in many-body systems starting from an incoherent state under resourceful operation. The choice of R\'enyi-$2$ observables is motivated by three considerations. First, they are well suited for large-scale numerical simulations, since both the subsystem purity $\mathrm{Tr}\,\rho_A^2$ and its diagonal analogue $\mathrm{Tr}\,(\rho_{A,\mathrm{diag}})^2$ can be computed efficiently via replica techniques, tensor networks, or exact state-vector methods~\cite{zhou2019emergent, zhou2020entanglement, schollwock2011the, silvi2019the, orus2014a}. 
Second, this quantity is directly accessible in experiments, for instance, through Bell testing or shadow tomography~\cite{Huang2020predicting, Hangleiter2024bell}.
Finally,  the dynamics of R\'enyi-$2$ observables in ergodic systems with conservation laws is often considerably richer than that of their von-Neumann counterparts, displaying nontrivial hydrodynamic scaling and relaxation phenomena that are precisely the physics of interest here~\cite{Keyserlingk2018operator, rakovszky2019sub, huang2020dynamics,znidaric2020entanglement}.

\section{Methods and models}
\label{sec:models}

We characterize the dynamics of the global participation entropy $S_d$ and the local coherence measure $C_d$ in the three symmetry-constrained setups described above. Our focus is on the timescales governing the approach of $S_d$ to its stationary value, as well as on the temporal structure of $C_d(\rho_A)$ for subsystems of size $L_A$.
To quantify the relaxation of $S_d$ both globally and locally we introduce the deviation from saturation,
\begin{equation}
\Delta S_d(t)=S_d(\infty)-S_d(t),
\qquad
S_d(\infty)\equiv \lim_{t\to\infty} S_d(t),
\end{equation}
which measures the distance from complete Hilbert-space delocalization.
Analogously, the approach of the subsystem entanglement entropy to its stationary value is monitored via
\begin{equation}
\Delta S_{R}(t) = S_{R}(\infty)-S_{R}(t),
\qquad
S_{R}(\infty)=\lim_{t\to\infty} S_R(t),
\end{equation}
and quantifies how far subsystem $A$ is from its asymptotic entanglement structure.

As a reference point, we recall the behavior of these quantities in local one-dimensional random unitary circuits \emph{without} any conservation law~\cite{nahum2017quantum, Keyserlingk2018operator, fisher2023random, turkeshi2024hilbert, turkeshi2025magic, dalzell2022random}. For such circuits the deviation of $S_d$, both globally and locally, decays exponentially at large depth, $\Delta S_d(t)=A\,e^{-\alpha t}$, with $\alpha$ independent of system size and $A\propto L$. The global saturation value coincides with that of a Haar-random state $|\mathrm{Haar}\rangle$; for $k=2$,
\begin{equation}
S_d(|\mathrm{Haar}\rangle)\simeq \log_2(q^L+1)-1\simeq L \log_2(q)-1,
\end{equation}
where the approximation holds up to exponentially small correction in $L$ for \textit{any} Haar random state. 
The threshold condition $\Delta S_d(t)\le \epsilon$ is reached at times $t_{\mathrm{sat}}\propto \log_2 L$, signalling the rapid onset of anticoncentration~\cite{turkeshi2024hilbert, turkeshi2025magic}.
At the local level, these circuits exhibit a characteristic rise--peak--fall structure in $C_d$, with the coherence peak time scaling logarithmically with subsystem size, $\tau_c^m \propto \log_2 L_A$, followed by an exponential decay towards the free state~\cite{aditya2025growthspreadingquantumresources}. Random circuits without conservation laws thus provide a convenient benchmark against which the symmetry-constrained dynamics considered below should be compared.

\subsection{$U(1)$-symmetric random circuits}
\label{sec:u1_model}
To explore the impact of conserved quantities we first consider the coherence dynamics in a brick-wall $U(1)$-symmetric quantum circuit acting on $L$ qudits~\cite{Keyserlingk2018operator}. The time-evolution operator after depth $t$ is $U_t=\prod_{r=1}^{t}U^{(r)}$, where each layer consists of a brick-wall arrangement of two-site gates,
\begin{equation}
U^{(2m)} = \prod_{i=1}^{L/2-1} U_{2i,\,2i+1},
\qquad
U^{(2m+1)} = \prod_{i=1}^{L/2} U_{2i-1,\,2i}.
\end{equation}

Each two-site gate $U_{i,j}$ is independently drawn from the Haar measure over the subgroup of $\mathcal U(q^2)$ that commutes with the total magnetization,
\begin{equation}
\bigl[U_{i,j},\,Q_{i,j}\bigr]=0,
\qquad
Q_{i,j}=Z_i+Z_j,
\qquad
Q=\sum_{l=1}^{L}Z_l,
\end{equation}
where $Z_l$ is the local spin-$z$ operator on site $l$ and $Q$ denotes the globally conserved charge (total magnetization). 
In the following, we consider both $q=2$ and $q=3$, corresponding to spin-$\tfrac12$ and spin-$1$, respectively.  In these cases $Z|m\rangle=m|m\rangle$ with  $m=\pm 1$ for $\spinlabel=1/2$ and $m=\pm1,0$ for $\spinlabel=1$.
For the system of interest, we will also focus on open boundary conditions (OBC).

To probe the dynamics, we initialize the system in an incoherent Néel-modulated product state. For $q=2$ we take
\begin{equation}
|\Psi_0\rangle = |+1,-1,\cdots ,+1,-1\rangle,
\end{equation}
whereas for $q=3$ we consider
\begin{equation}
|\Psi_0\rangle = |-1,0,+1,\,-1,0,+1,\,\cdots,-1,0,+1\rangle.
\end{equation}
In both cases, the initial state lies entirely in the zero-magnetization sector $Q|\Psi_0\rangle=0$. In contrast to domain-wall configurations, which are known to be atypical, such alternating product states provide a natural, unbiased probe of the coherence and entanglement growth in typical, non-entangled, initial states. Our numerical analysis for $L\le 28$ ($L\le 15$) uses exact state-vector evolution for $q=2$ ($q=3$) averaged over at least $4000$ circuit realizations; beyond this regime, we employ the replica tensor-network (RTN) method~\cite{zhou2019emergent, zhou2020entanglement, turkeshi2025magic,turkeshi2025quantum}, which extends the accessible system sizes considerably. Details of the RTN approach are presented in Appendix~\ref{app:rtn}.

\subsection{Charge- and dipole-conserving random circuits}
\label{sec:dipole_model}

As a second setup, we consider random quantum circuits with simultaneous conservation of total charge and total dipole moment~\cite{sala2020ergodicity, khemani2020localization, pai2019localization,feldmeier2020anomalous}, corresponding to a $U(1)_Q\times U(1)_P$ symmetry. The local Hilbert space is that of a spin-$\tfrac12$ chain with computational-basis states $|\pm1\rangle$.
 Throughout, we will use open boundary conditions. The globally conserved quantities are
\begin{equation}
Q=\sum_{i=1}^{L}Z_i,
\qquad
P=\sum_{i=1}^{L} i\,Z_i,
\end{equation}
with local eigenvalues $Z_i\in\{\pm 1\}$. The dynamics is generated by a four-step Floquet brick-wall circuit,
\[
U_F = U^{(4)} U^{(3)} U^{(2)} U^{(1)},
\]
and the time-evolution operator after \(t\) Floquet periods is $U_t=(U_F)^t$.
Each layer consists of four-site unitaries acting on staggered blocks, namely
\begin{align}
U^{(1)} &= \prod_i U_{4i-3,\,4i-2,\,4i-1,\,4i},~~~
U^{(2)} = \prod_i U_{4i-2,\,4i-1,\,4i,\,4i+1},\nonumber\\
U^{(3)} &= \prod_i U_{4i-1,\,4i,\,4i+1,\,4i+2},~~
U^{(4)} = \prod_i U_{4i,\,4i+1,\,4i+2,\,4i+3},
\end{align}
with the allowed values of \(i\) determined by the system size and OBC. Each four-site gate is drawn independently from the Haar measure over the subgroup of \(\mathcal U(2^4)\) that preserves both total charge and total dipole moment on its support. 
These constraints already induce a nontrivial fragmentation of the local four-site Hilbert space. Although a four-site block has dimension \(2^4=16\), simultaneous conservation of $Q$ and \(P\) decomposes it into invariant subspaces that cannot be mixed by the gate. Most of these subspaces are one-dimensional, so the gate acts there only by phases. The only nontrivial resonant sector is the two-dimensional subspace spanned by $\lvert +1,-1,-1,+1\rangle$ and
$\lvert -1,+1,+1,-1\rangle$. Accordingly, the elementary allowed process is the dipole-preserving swap
\begin{equation}
|+1,-1,-1,+1\rangle \;\leftrightarrow\; |-1,+1,+1,-1\rangle.
\end{equation}

At the many-body level, this local constraint structure leads to \emph{strong Hilbert-space fragmentation}~\cite{sala2020ergodicity,moudgalya2022quantum,Aditya2024subspace,Ganguli2025East,Aditya2025East2}: even within a fixed global symmetry sector of charge $\mathfrak{q}$ and dipole $\mathfrak{p}$ (i.e., eigenvalues of the conserved $Q$ and $P$), the Hilbert space splits into exponentially many dynamically disconnected fragments, and the dimension of the largest fragment $D_F$ is exponentially smaller than that of the full $(\mathfrak{q},\mathfrak{p})$ sector, $D_{\mathfrak{q},\mathfrak{p}}$. To obtain analytical insight into the dynamics one must therefore specify not only the global symmetry sector but also the particular connected fragment selected by the initial state. Here we focus on the root configuration
\begin{equation}
|\Psi_0\rangle=|+1,-1,-1,+1,+1,-1,-1,+1,+1,-1,-1,+1,\cdots\rangle,
\end{equation}
which lies in the $(Q,P)=(0,0)$ sector and generates the largest connected fragment. Since this is among the largest fragments in the sector, it is a natural setting for probing the typical dynamical behavior compatible with the fractonic constraints. In this case, we compute the time-evolved state $|\Psi(t)\rangle$ using exact state-vector simulation restricted to the corresponding dynamical sector, which allows us to reach sufficiently large system sizes.

\subsection{Mixed-field Ising Hamiltonian dynamics}
\label{sec:mfim_model}
As a third setup, we consider a paradigmatic ergodic many-body system: the mixed-field Ising model (MFIM)~\cite{kim2013ballistic, bertini2020scrambling,tirrito2024anticoncentrationmagicspreadingergodic}, with Hamiltonian
\begin{equation}
H_{\mathrm{MFIM}} = b\sum_i X_i + \sum_i h_i Z_i + J\sum_i Z_i Z_{i+1} +\delta h_z\,(Z_1 - Z_L),
\end{equation}
where $X_i, Z_i$ are Pauli operators on site $i$, and $b$, $h_i$, $J$ and $\delta h_z$ are Hamiltonian parameters.
Throughout we fix $b=(5+\sqrt5)/8$, $h_i=(\sqrt5+1)/4$, $J=1$, and the symmetry-breaking boundary field $\delta h_z=0.25$, and employ open boundary conditions. With these parameters the model is non-integrable, quantum ergodic, and exhibits diffusive energy transport. The overall conclusions of our analysis are expected to be insensitive to the specific choice of $b$, $h_i$, and $J$ as long as the parameters are not fine-tuned to special points, such 
the classical limit $b=0$.

To probe the global and local spreading of coherence we use random $z$-basis product initial states $|\Psi_0\rangle$, further constrained to lie close to the centre of the many-body spectrum,
\begin{equation}
\frac{|\langle \Psi_{0}|H_{\mathrm{MFIM}}|\Psi_{0}\rangle - E_{\mathrm{mid}}|}{E_{\mathrm{max}}-E_{\mathrm{min}}}\le 0.05,
\qquad
E_{\mathrm{mid}}=\frac{E_{\mathrm{max}}+E_{\mathrm{min}}}{2},
\end{equation}
where $E_{\mathrm{max}}$ and $E_{\mathrm{min}}$ are the extremal eigenvalues of $H_{\mathrm{MFIM}}$. The dynamics is computed using Chebyshev time evolution~\cite{Ezer84, Sierant22chal} for $L\le 20$, and using the time-dependent variational principle (TDVP) combined with matrix-product-state techniques~\cite{schollwock2011the, silvi2019the, orus2014a, Yang20tdvp} for larger system sizes.

\section{Coherence spreading in $U(1)$-symmetric circuits}
\label{sec:u1_results}
\subsection{Global dynamics of $U(1)$-symmetric circuits}
\label{sec:u1_global}
We first investigate the global spreading of coherence under $U(1)$-symmetric dynamics. Since the time-evolved state $|\Psi(t)\rangle = U_t |\Psi_0\rangle$ remains confined to the zero-magnetization sector (we take $Q=0$ throughout), the symmetry strongly constrains the accessible Hilbert space~\cite{Keyserlingk2018operator}. To quantify its effect, we study the dynamics of $S_d(t)$ and, in particular, the timescale over which it approaches its long-time value.

The saturation value of $S_d$ is set by the Haar average within the accessible charge sector. Specializing the general formula derived in Appendix~\ref{app:general_diag_renyi} to a $U(1)$-symmetric chain at fixed magnetization $Q=\mathfrak{q}$, one obtains, up to corrections of order $O(1/D_\mathfrak{q})$,
\begin{equation}
    S_d\!\left(|\mathrm{Haar}_{U(1)}(\mathfrak{q})\rangle\right)\simeq \log_2\!\left(\frac{D_\mathfrak{q}+1}{2}\right),
\label{eq:haar_u1_saturation}
\end{equation}
where $D_\mathfrak{q}$ is the dimension of the magnetization-$\mathfrak{q}$ subspace. For the zero-magnetization sector considered throughout, $D_0=\binom{L}{L/2}$ for spin-$1/2$ ($q=2$) at half filling, and $D_0=\sum_{k=0}^{\lfloor L/2\rfloor} L!/[k!\,k!\,(L-2k)!]$ for spin-$1$ ($q=3$); in both cases $D_0$ grows exponentially in $L$, so Eq.~\eqref{eq:haar_u1_saturation} is already sharp for moderate chain lengths.

\begin{figure}[t!]
\centering
\includegraphics[width=0.7\textwidth]{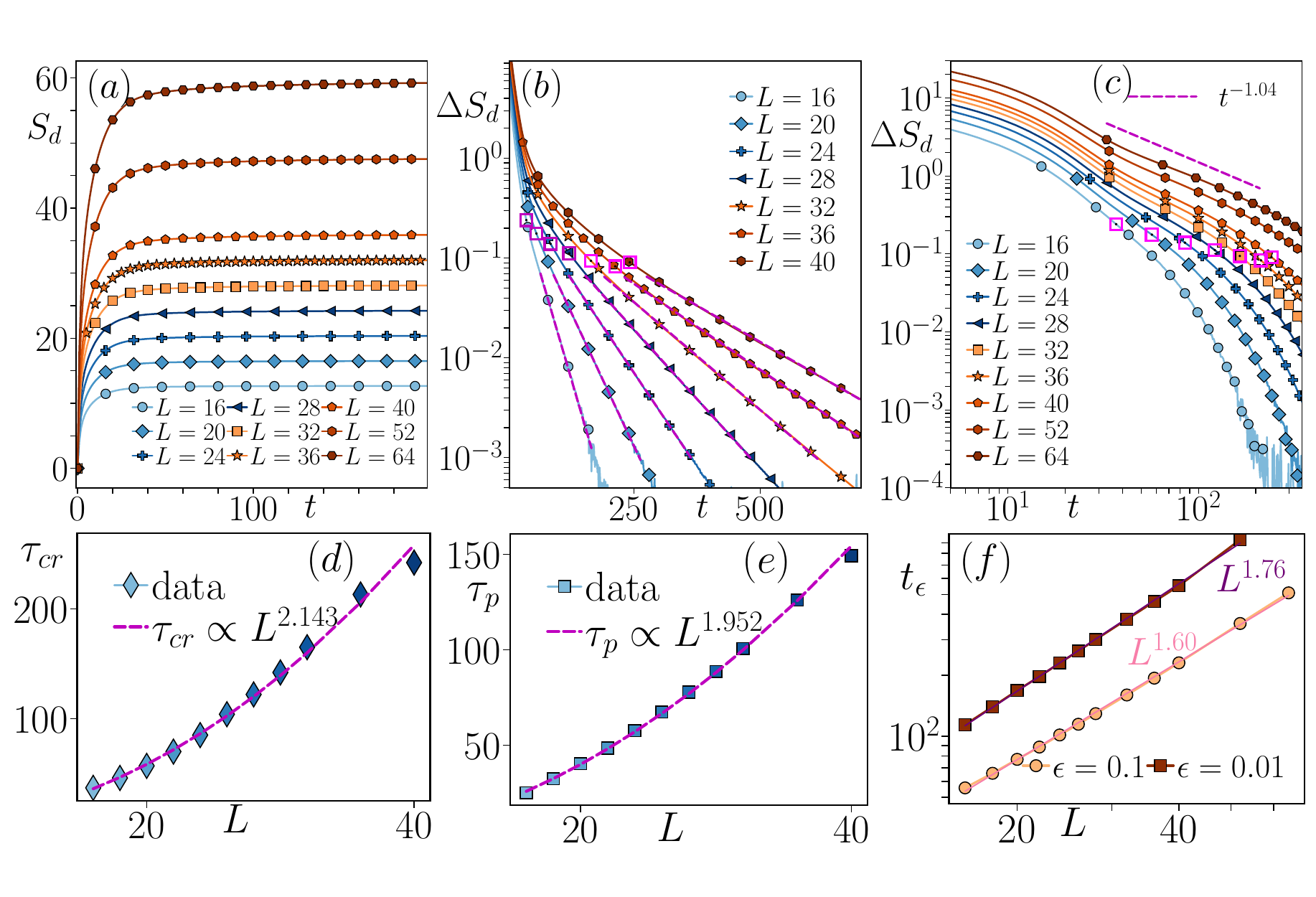}
\caption{{\bf Global $S_d$ dynamics of a spin-$\tfrac{1}{2}$, $U(1)$-conserving circuit with $L$ sites:}
(a) $S_d$ grows rapidly and saturates to the value set by the $U(1)$ symmetry.
(b) At late times, $\Delta S_d$ exhibits an exponential decay, $\Delta S_d = A_p e^{-t/\tau_p}$, with $\tau_p \propto L^{1.952}$ (see panel (e)). Magenta squares mark a dynamical crossover at intermediate times where the exponential fit deviates from the data.
(c) In the intermediate regime, $\Delta S_d \sim t^{-\beta_p}$ with $\beta_p \simeq 1.04$. The magenta squares (log--log scale) highlight the crossover from power-law to exponential decay.
(d) The crossover time scales diffusively, $\tau_{cr} \propto L^{\alpha_{cr}}$, with $\alpha_{cr} \simeq 2.143$ 
(f) The relaxation time to reach a fixed threshold, $\Delta S_d \le \epsilon \leq O(1)$, scales as $t_{\epsilon p} \propto L^{\alpha_\epsilon}$ in contrast $t_{\epsilon}\propto \log_2 L$ scaling in the case of random circuits without symmetry. For this analysis, we consider global Neel state and the results for $L=16-28$ are obtained using exact vector simulation average over at least $4000$ circuit realizations and $L\geq32$'s are obtained using RTN with $\chi=1024$. }
\label{fig:u1_global_sd}
\end{figure}

We now turn to the numerical results obtained from exact diagonalization and the replica tensor-network approach, focusing first on the case $q=2$. The participation entropy $S_d(t)$ grows rapidly at early times and approaches its symmetry-constrained saturation value, as shown in Fig.~\ref{fig:u1_global_sd}a. To characterize the approach to equilibrium more precisely, we analyze the deviation $\Delta S_d(t)$ which exhibits a clear two-stage relaxation dynamics. At intermediate times, $\Delta S_d(t)$ decays algebraically as $\Delta S_d(t) \sim t^{-\beta_p}$, as shown in Fig.~\ref{fig:u1_global_sd}c, with exponent $\beta_p \simeq 1.04$. At later times, once finite-size effects become relevant, the decay crosses over to an exponential form, $\Delta S_d(t) \sim e^{-t/\tau_p(L)}$, as shown in Fig.~\ref{fig:u1_global_sd}b. The corresponding relaxation timescale grows with system size approximately as $\tau_p \propto L^{\alpha_p}$ with $\alpha_p \simeq 1.952$, consistent with nearly diffusive scaling, as shown in Fig.~\ref{fig:u1_global_sd}e. In addition, the crossover time separating the algebraic and exponential regimes also exhibits a power-law dependence on system size, $\tau_{\mathrm{cr}} \propto L^{\alpha_{\mathrm{cr}}}$ with $\alpha_{\mathrm{cr}} \simeq 2.143$, as shown in Fig.~\ref{fig:u1_global_sd}d.  Here, $\tau_{\mathrm{cr}}$ is defined operationally as the latest time preceding the exponential-fit window at which the numerical data for $\Delta S_d(t)$ deviates from the fitted exponential by more than 10 percent in relative error; it therefore marks the onset of the exponential relaxation regime. Finally, as illustrated in Fig.~\ref{fig:u1_global_sd}f, the time required to reach a fixed threshold $\Delta S_d(t) \le \epsilon$ scales as a power law in system size, in sharp contrast to the logarithmic scaling observed in random circuits without conservation laws.

\begin{figure}[t]
\centering
\includegraphics[width=0.7\textwidth]{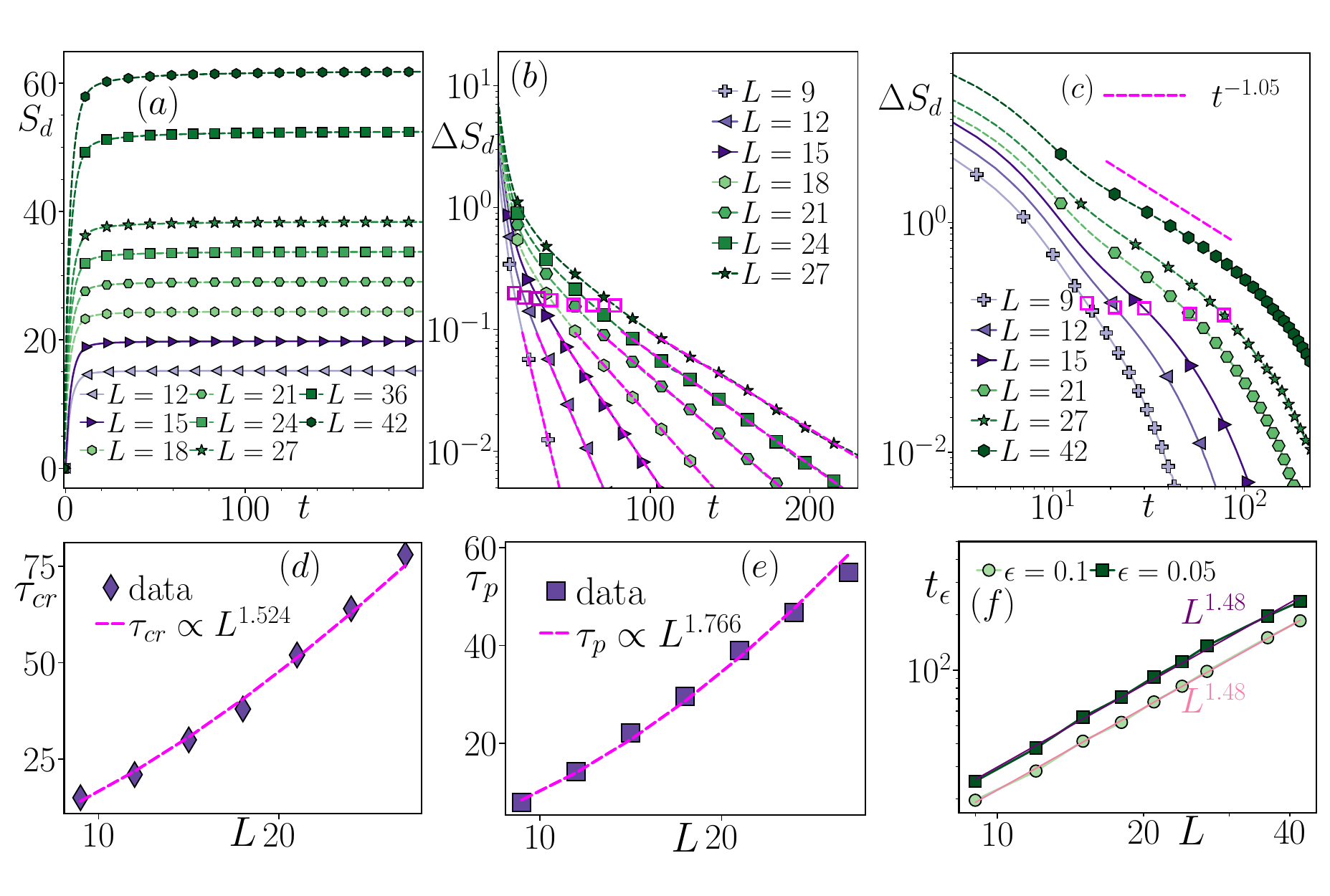}
\caption{{\bf Global $S_d$ dynamics of a spin-1, $U(1)$-conserving circuit with $L$ sites:}
(a) $S_d$ grows rapidly and saturates to the value set by the $U(1)$ symmetry.
(b) At late times, $\Delta S_d$ exhibits exponential decay, $\Delta S_d = A_p e^{-t/\tau_p}$, with $\tau_p \propto L^{1.766}$ (see panel (e)). Magenta squares mark a dynamical crossover at intermediate times where the exponential fit deviates from the data.
(c) In the intermediate regime, $\Delta S_d \sim t^{-\beta_p}$ with $\beta_p \simeq 1.05$. The magenta squares (log--log scale) highlight the crossover from power-law to exponential decay.
(d) The crossover time scales as a power law with system size, $\tau_{\mathrm{cr}} \propto L^{\alpha_{\mathrm{cr}}}$, with $\alpha_{\mathrm{cr}} \simeq 1.524$.
(f) The relaxation time to reach a fixed threshold, $\Delta S_d \leq \epsilon \leq O(1)$, scales as $t_{\epsilon} \propto L^{\alpha_\epsilon}$.
Results correspond to a period-3 spin-1 N\'eel initial state. Data for $L=12$--$15$ are obtained using exact state-vector simulations averaged over at least 4000 circuit realizations, while results for $L \geq 21$ are obtained using the RTN simulations with bond dimension $\chi=896$.}
\label{fig:u1_global_sd_spin1}
\end{figure}

This behavior admits a natural physical interpretation~\cite{rakovszky2018diffusive,rakovszky2019sub}. In a $U(1)$-symmetric circuit, the wavefunction spreads only within a fixed charge sector, and its approach to the Haar-random limit is constrained by the redistribution of the conserved charge~\cite{Haah2025short}. While local degrees of freedom scramble rapidly, the conserved charge remains the slowest mode in the problem and therefore acts as the bottleneck for equilibration. As a result, the relaxation of $S_d(t)$ is governed by diffusive charge transport. In particular, the intermediate-time power-law decay reflects the collective contribution of many slow hydrodynamic modes, rather than the dominance of a single microscopic relaxation rate. If the conserved density satisfies
\begin{equation}
\partial_t n(x,t)=D\,\partial_x^2 n(x,t),
\end{equation}
then each Fourier mode relaxes as $n_k(t)\sim e^{-Dk^2 t}$. In a finite system of size $L$, the spectrum of diffusive modes becomes 
discrete, with the smallest nonzero wavevector set by 
$k_{\min}\sim 2\pi/L$. At intermediate times, many of these modes 
contribute simultaneously and produce the observed 
algebraic decay. As time progresses, modes with larger $k$ relax first 
and drop out, until around $\tau_{\mathrm{cr}}\sim L^{2}$ the finiteness 
of the system halts further hydrodynamic relaxation: no modes with 
$k < k_{\min}$ are available. Beyond this point, the 
late-time decay is no longer governed by continuous hydrodynamics but 
by the slowest discrete diffusive mode alone, giving rise to the final 
exponential regime with 
$\tau_{p}(L)\sim 1/(D\,k_{\min}^{2})\sim L^{2}$. 
We now turn to the case of $q=3$, corresponding to spin-1 degrees of freedom. The overall dynamical features closely mirror those observed in the $q=2$ case. In particular, the participation entropy $S_d(t)$ again exhibits a rapid initial growth toward the symmetry-constrained saturation value, as shown in Fig.~\ref{fig:u1_global_sd_spin1}a. The deviation $\Delta S_d(t)$ displays a clear two-stage relaxation dynamics: at intermediate times, it decays algebraically as $\Delta S_d(t) \sim t^{-\beta_p}$ with $\beta_p \simeq 1.05$, as shown in Fig.~\ref{fig:u1_global_sd_spin1}c, while at later times it crosses over to an exponential decay $\Delta S_d(t) \sim e^{-t/\tau_p(L)}$, as shown in Fig.~\ref{fig:u1_global_sd_spin1}b. The associated relaxation timescale increases with system size as $\tau_p \propto L^{\alpha_p}$ with $\alpha_p \simeq 1.766$, as illustrated in Fig.~\ref{fig:u1_global_sd_spin1}e. 
While the qualitative structure of the dynamics remains unchanged, the scaling of the crossover time $\tau_{\mathrm{cr}}$ shows noticeable deviations from the nearly diffusive behavior observed in the $q=2$ case. In particular, we find $\tau_{\mathrm{cr}} \propto L^{\alpha_{\mathrm{cr}}}$ with $\alpha_{\mathrm{cr}} \simeq 1.524$, as shown in Fig.~\ref{fig:u1_global_sd_spin1}d. 
The late-time decay is still close to diffusive, as signalled by $\alpha_p\simeq 1.77$ not far from the hydrodynamic value $\alpha_p=2$. The deviation from the crossover exponent reflects the narrower window of system sizes ($L\lesssim 27$) accessible to exact evolution and RTN for $q=3$ case, which, combined with the 10\%-deviation criterion used to define $\tau_{\mathrm{cr}}$, makes this observable more sensitive to preasymptotic corrections.
The relaxation time required to reach a fixed threshold $\Delta S_d(t) \le \epsilon$ again exhibits a power-law dependence on system size, as shown in Fig.~\ref{fig:u1_global_sd_spin1}f. 

\subsection{Local dynamics of $U(1)$-symmetric circuits}
\label{sec:u1_local}
\begin{figure}[htb]
\centering
\includegraphics[width=0.7\textwidth]{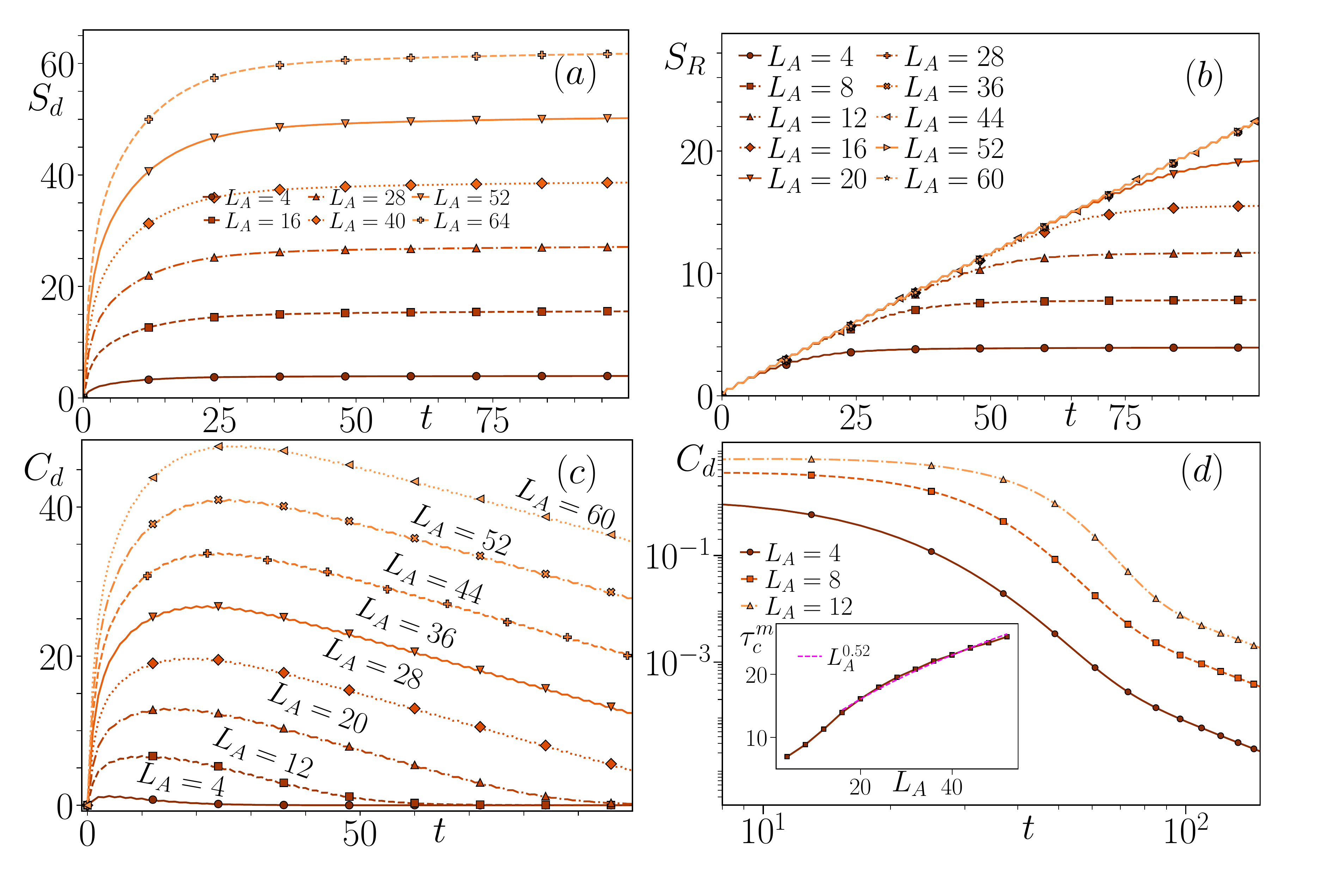}
\caption{{\bf Local coherence dynamics in a $U(1)$-conserving spin-$\tfrac{1}{2}$ circuit with $L$ sites (RTN):}
(a--c) Dynamics of local $S_d$, $S_R$, and $C_d$ for different subsystem sizes $L_A$ at $L=128$.
(d)The intermediate-time decay of $C_d$ shows power-law tails. The time scale at which $C_d$ attains its maximum exhibits power-law scaling with subsystem size, $\tau_c^m \propto L_A^{\alpha_m}$, with $\alpha_m \simeq 0.52$. (see inset).
All results correspond to a global N\'eel initial state; RTN simulations are performed with bond dimension $\chi=784$.}
\label{fig: cdspinhalf}
\end{figure}

We next turn to the local dynamics of coherence in $U(1)$-symmetric circuits, focusing first on the spin-$1/2$ case. To this end, we monitor $S_d$, $S_R$, and the corresponding relative entropy of coherence $C_d=S_d-S_R$ as functions of time~\cite{10.21468/SciPostPhys.15.6.250}. As shown in Fig.~\ref{fig: cdspinhalf}(a--c), the diagonal entropy $S_d$ grows rapidly at early times and approaches its saturation value, in close analogy with the global dynamics discussed previously. By contrast, $S_R$ exhibits a clear ballistic growth regime at early times, reflecting the fast spreading of entanglement under local unitary dynamics~\footnote{We note that diffusive entanglement growth~\cite{rakovszky2019sub, znidaric2020entanglement}, occurs \textit{only} for initial states supported on multiple charge sectors~\cite{Jonay24slow}, e.g., states of the type $\ket{\pm}^{\otimes L}$, where $\ket{\pm} = (\ket{+1} \pm \ket{-1})/\sqrt{2}$ . }.

The interplay between these two quantities gives rise to a distinctly non-monotonic behavior of the coherence measure $C_d$, which displays a characteristic rise--peak--fall structure, as shown in Fig. \ref{fig: cdspinhalf} (c). The initial increase of $C_d$ reflects the generation of coherence, while its subsequent decay is driven by the buildup of entanglement between the subsystem and its complement. This competition leads to a well-defined maximum at a characteristic time $\tau_c^m$, which depends strongly on the subsystem size $L_A$. As shown in Fig.~\ref{fig: cdspinhalf}d (inset), the peak time scales algebraically as $\tau_c^m \propto L_A^{\alpha_m}$ with $\alpha_m \simeq 0.52$, in sharp contrast to the logarithmic scaling observed in random circuits without conservation laws~\cite{aditya2025growthspreadingquantumresources}. At intermediate times, the decay of $C_d$ exhibits power-law behavior, indicating the presence of slow hydrodynamic modes associated with the conserved charge, as shown in Fig. \ref{fig: cdspinhalf}d.

\begin{figure}[t]
\centering
\includegraphics[width=0.7\textwidth]{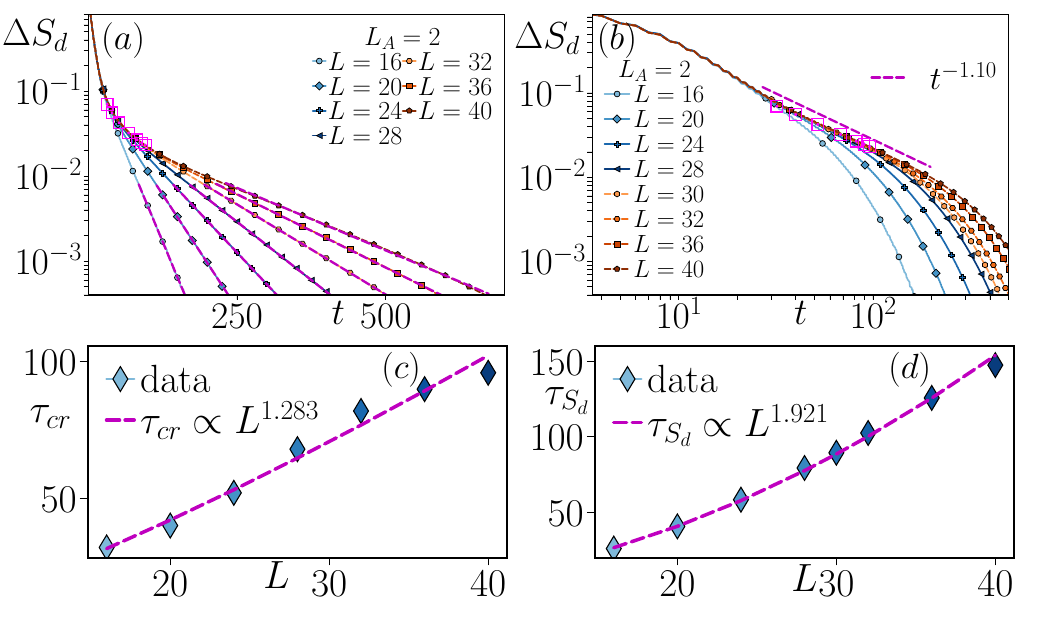}
\caption{{\bf Local $\Delta S_d$ decay in a $U(1)$-conserving spin-$\tfrac{1}{2}$ circuit with $L$ sites for a subsystem of size $L_A$:}
(a-b) The local $\Delta S_d$ like the global case exhibits a dynamical crossover from intermediate-time power-law decay, $\sim t^{-\beta_{S_d}}$ with $\beta_{S_d} \simeq 1.10$, to late-time exponential decay, $\sim A_{S_d} e^{-t/\tau_{S_d}}$. The crossover regime is highlighted by magenta squares.
(c) The crossover time scales as $\tau_{\mathrm{cr}} \propto L^{1.283}$.
(d) The late-time exponential decay-time scale grows diffusively, $\tau_{S_d} \propto L^{1.921}$.
Results are obtained for a N\'eel initial state. Data for $L=16$--$28$ are computed using exact state-vector simulations averaged over at least $4000$ circuit realizations, while results for $L \ge 32$ are obtained using RTN simulations with bond dimension $\chi=1024$.}
\label{fig:u1_local_decay}
\end{figure}

To further elucidate the underlying mechanism, we analyze the relaxation of the diagonal entropy through the deviation $\Delta S_d(t)$ for fixed subsystem size $L_A$ and varying system size $L$, where $S_{d}(\infty)$ value can be analytically computed (see Appendix-\ref{app:u1_q0} for detailed derivation). As shown in Fig.~\ref{fig:u1_local_decay} (a-b), the dynamics exhibits a clear two-stage relaxation. At intermediate times, $\Delta S_d(t)$ decays algebraically as $\Delta S_d(t) \sim t^{-\beta_{S_d}}$ with $\beta_{S_d} \simeq 1.10$, reflecting the contribution of slow hydrodynamic modes associated with charge conservation. At later times, the decay crosses over to an exponential form $\sim e^{-t/\tau_{S_d}(L)}$, as finite-size effects become relevant. 
Importantly, the timescale at which this dynamical crossover occurs grows with system size, indicating that the separation between the two regimes is itself controlled by hydrodynamic transport. As shown in Fig.~\ref{fig:u1_local_decay}c, the crossover time scales as $\tau_{\mathrm{cr}} \propto L^{\alpha_{\mathrm{cr}}}$ with $\alpha_{\mathrm{cr}} \simeq 1.283$, 
milder than the fully diffusive $\alpha_{\mathrm{cr}}\simeq 2.14$ obtained for the global observable in Fig.~\ref{fig:u1_global_sd}d. We interpret this as follows: while the late-time tail is still controlled by the slowest moving hydrodynamics mode so that the exponential timescale remains close to diffusive ($\alpha_{S_d}\simeq 1.92$, see below in Fig. \ref{fig:u1_local_decay}d), the crossover between the two regimes becomes more susceptible to preasymptotic scales arising due to interplay between $L_A$ and $L$, thus possibly exhibiting deviations from the typical diffusive scaling.

Next, we turn to the growth of the Haar-averaged Rényi-2 entanglement entropy $S_R(t)$, which exhibits linear growth for N\'eel initial states (or, more generally, for any $z$-basis product state), as can be seen in Fig.~\ref{fig: cdspinhalf}b. This behavior contrasts with the diffusive scaling observed for charge-inhomogeneous initial states evolving under $U(1)$-symmetric dynamics. The emergence of ballistic scaling for a single $z$-basis product state, such as the N\'eel state, can be understood from the failure of the rare-region mechanism~\cite{rakovszky2019sub, znidaric2020entanglement, huang2020dynamics} during the early-to-intermediate growth regime. Unlike a superposition state (e.g., $|+\rangle^{\otimes L}$), a single product configuration in the $z$ basis lacks the ensemble of spatial charge arrangements required to statistically guarantee a large inactive interval at the partition cut. 
For a typical configuration at finite density, the probability that the cut lies within an empty or fully occupied interval of length $\ell$ is exponentially suppressed, $P(\ell) \sim e^{-c\ell}$. Consequently, the distance from the cut to the nearest mobile charge is $O(1)$ with high probability, allowing the cut to become ``activated'' on microscopic time scales $t_{\mathrm{act}} \sim O(1)$. In the absence of a large inactive region that could delay entanglement growth (see Appendix~\ref{app:z_basis_average} for a detailed explanation), the system undergoes standard local scrambling, leading to the linear growth $S_{\alpha>1}(t) \propto t$ characteristic of ballistic spreading.
Nevertheless, even within this ballistic regime, the late-time approach to saturation remains sensitive to the underlying $U(1)$ symmetry. In particular, the global relaxation of the conserved charge eventually produces an algebraic tail, followed by exponential decay toward saturation, with the decay time showing a nontrivial dependence on the total size $L$ for a fixed $L_A$ due to the finite-size hydrodynamics mode, similar to $S_d$ (not shown here). 

The emergence of a well-defined peak in $C_d(t)$ can be understood from the competition between the diagonal entropy $S_d(t)$ and the entanglement entropy $S_R(t)$. Since $C_d(t) = S_d(t) - S_R(t)$, the time at which coherence reaches its maximum is determined by the condition $\partial_t C_d(t) = 0$, or equivalently when the growth rate of $S_R(t)$ matches the relaxation rate of $S_d(t)$. As seen in Fig.~\ref{fig: cdspinhalf}a--c, $S_d(t)$ grows rapidly and approaches its saturation value at early times, in close analogy with the global dynamics, while $S_R(t)$ exhibits an extended ballistic growth regime~\cite{nahum2017quantum, nahum2018operator}. More quantitatively, the relaxation of $S_d(t)$ is governed by the algebraic decay of $\Delta S_d(t)$ (Fig.~\ref{fig:u1_local_decay}b), with $\Delta S_d(t) \sim t^{-\beta_{S_d}}$, implying a decay rate $\partial_t S_d(t) \sim t^{-(\beta_{S_d}+1)}$. In contrast, the growth rate of $S_R(t)$ remains approximately constant in the ballistic regime. Equating the two rates at the peak leads to a characteristic timescale $\tau_c^m\sim L_A^{\alpha_m}$, where the exponent 
$\alpha_m=1/(\beta_{S_d}+1)$. This scaling is in excellent agreement with the numerical results shown in Fig.~\ref{fig: cdspinhalf}d, where the peak time exhibits a clear power-law dependence on subsystem size.

At late times, the repeated action of local $U(1)$-symmetric gates strongly entangles subsystem $A$ with its complement $\bar{A}$. Since the total charge is fixed to zero, entanglement can only be generated between states whose subsystem charges satisfy $q_A + q_{\bar{A}} = 0$. Consequently, the reduced density matrix $\rho_A$ remains block diagonal in the subsystem charge sectors. Within each sector, however, the dynamics scrambles efficiently, so that each block approaches a maximally mixed state. Accordingly, at long times one expects
\[
\rho_A \approx \bigoplus_{q_A} p(q_A)\,\frac{I_{A,q_A}}{D_{A,q_A}},
\]
where $D_{A,q_A}$ is the dimension of the charge-$q_A$ sector of subsystem $A$, and $p(q_A)$ is the probability of observing charge $q_A$, consistent with the global constraint.

In this limit, the reduced density matrix becomes effectively diagonal in the computational basis, implying that all off-diagonal coherences are suppressed. As a result, the diagonal entropy $S_d(t)$ and the entanglement entropy $S_R(t)$ approach the same asymptotic value, and their difference, namely the coherence $C_d(t)$, vanishes at long times, $C_d(t) \to 0$ as $t \to \infty$. This reflects the emergence of an incoherent state on the subsystem $A$, corresponding to a free state in the resource-theoretic sense.

\begin{figure}[t]
\centering
\includegraphics[width=0.7\textwidth]{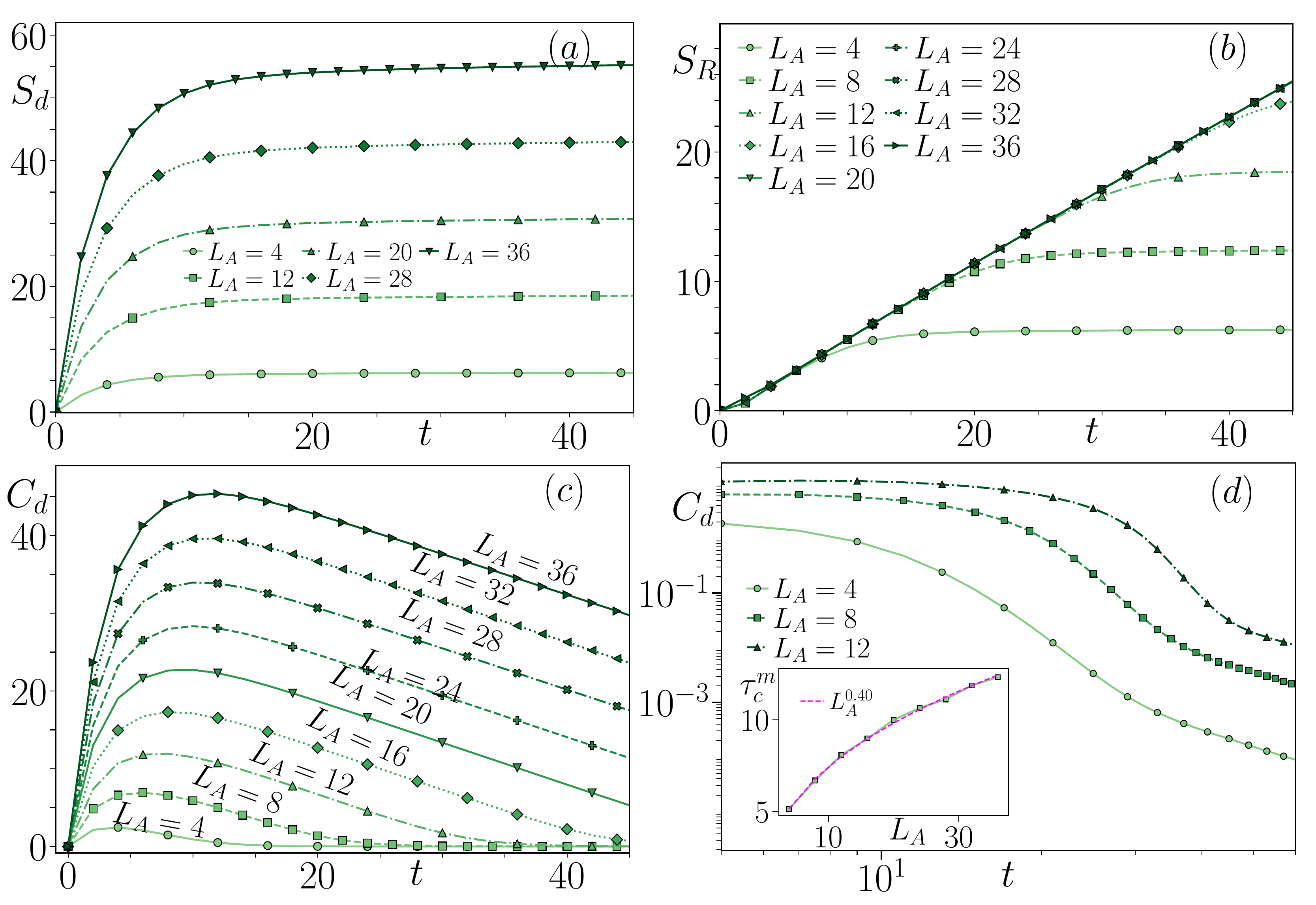}
\caption{{\bf Local coherence dynamics in a $U(1)$-conserving spin-1 circuit with $L$ sites (RTN):}
(a--c) Dynamics of local $S_d$, $S_R$, and $C_d$ for different subsystem sizes $L_A$ at $L=72$.
(d) The intermediate-time decay of $C_d$ exhibits power-law tails along with the time scale at which $C_d$ attains its maximum scales as a power law with subsystem size, $\tau_c^{m} \propto L_A^{\alpha_m}$, with $\alpha_m \simeq 0.40$ (see inset).
Results correspond to a period-3 spin-1 N\'eel initial state; RTN simulations are performed with bond dimension $\chi=640$.}
\label{fig:u1_local_sd_spin1}
\end{figure}

\begin{figure}[t]
\centering
\includegraphics[width=0.7\textwidth]{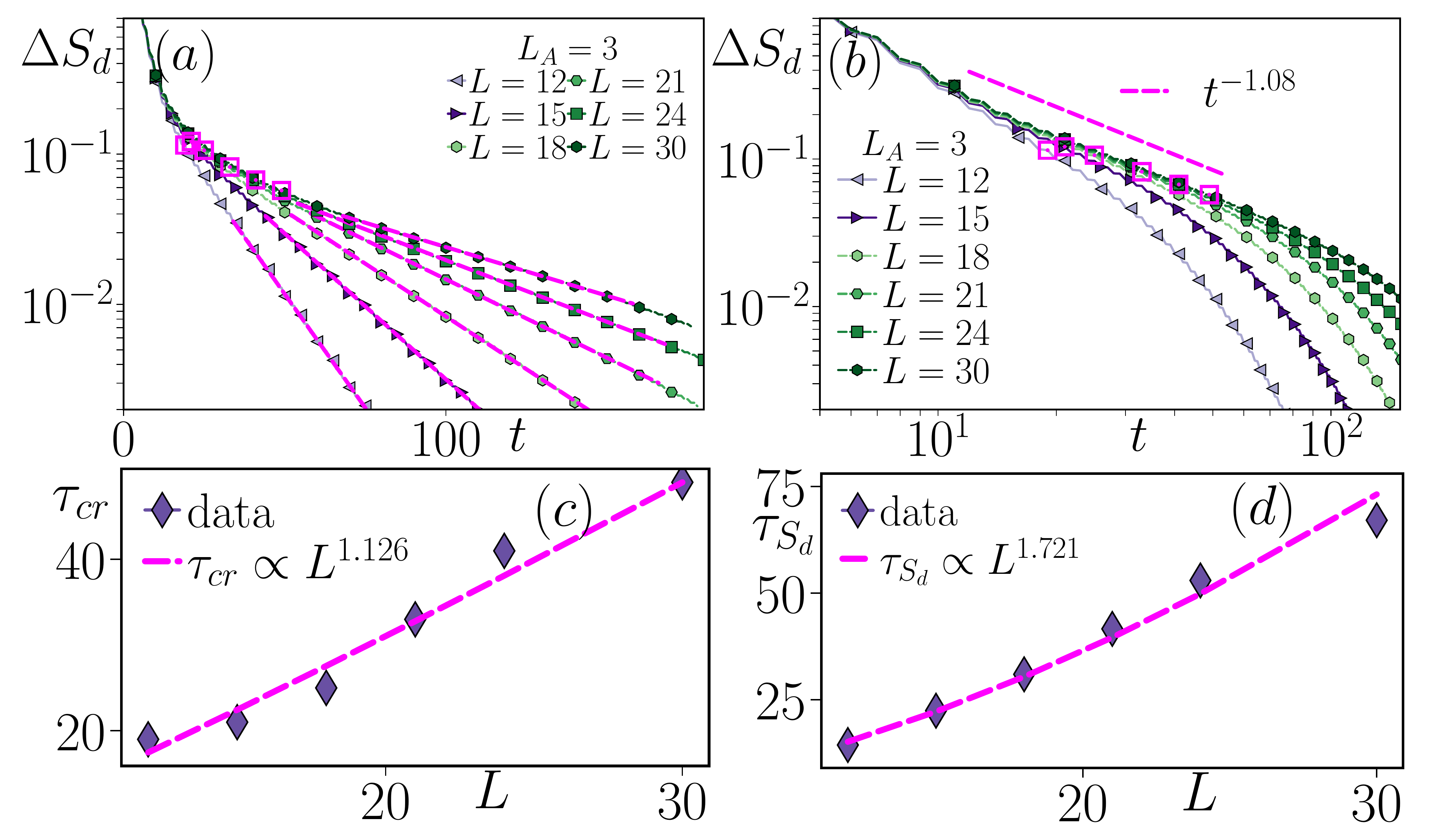}
\caption{{\bf Local $\Delta S_d$ decay in a $U(1)$-conserving spin-1 circuit with $L$ sites for a subsystem of size $L_A$:}
(a-b) The local $\Delta S_d$ similar to the spin-1/2 case exhibits a dynamical crossover from intermediate-time power-law decay, $\sim t^{-\beta_{S_d}}$ with $\beta_{S_d} \simeq 1.08$, to late-time exponential decay, $\sim A_{S_d} e^{-t/\tau_{S_d}}$. The crossover regime is highlighted by magenta squares.
(c) $\tau_{cr}$ exhibits power-law scaling with $L$ with exponent $1.126$.
(d) The late-time exponential decay-time scale grows, $\tau_{S_d} \propto L^{1.721}$.
Results are obtained for a period-3 N\'eel initial state. Data for $L=12$--$15$ are computed using exact state-vector simulations averaged over at least $4000$ circuit realizations, while results for $L \ge 21$ are obtained using RTN simulations with bond dimension $\chi=1152$.}
\label{fig:u1_local_decay_spin1}
\end{figure}

We now turn to the $q=3$ case. As shown in Fig.~\ref{fig:u1_local_sd_spin1}a--c, the qualitative structure of the local dynamics remains unchanged: the diagonal entropy $S_d(t)$ (panel a) grows rapidly and approaches its saturation value, while the entanglement entropy $S_R(t)$ (panel b) exhibits an early-time ballistic growth regime. Consequently, the coherence $C_d(t)$ (in Fig.~\ref{fig:u1_local_sd_spin1}c) displays the same characteristic rise--peak--fall profile, with the peak time scaling algebraically as $\tau_c^m \propto L_A^{\alpha_m}$ with $\alpha_m \simeq 0.40$, as shown in the inset of Fig.~\ref{fig:u1_local_sd_spin1}d with $C_d$ showing power-law decay at intermediate times.

This behavior is further supported by the relaxation of $\Delta S_d(t)$ ( Fig.~\ref{fig:u1_local_decay_spin1}a,b), which exhibits a two-stage decay similar to the $q=2$ case. At intermediate times, $\Delta S_d(t) \sim t^{-\beta_{S_d}}$ with $\beta_{S_d} \simeq 1.08$, while at later times it crosses over to exponential relaxation ($S_d(\infty)$ can again be analytically computed, shown in Appendix-\ref{app:u1_q0}). The corresponding decay timescale scales with system size as $\tau_{S_d} \propto L^{\alpha_{S_d}}$ with $\alpha_{S_d} \simeq 1.721$, as shown in Fig.~\ref{fig:u1_local_decay_spin1}d (close to slowest moving diffusive hydrodynamics mode). The dynamical crossover time scale, $\tau_{cr}$ again showcases a power-law scaling with $L$, as shown in Fig. ~\ref{fig:u1_local_decay_spin1}c.

\section{Coherence spreading in charge and dipole conserving circuits}
\label{sec:dipole_results}

We now turn to the dynamics of coherence spreading in random quantum circuits that conserve both the total charge $U(1)_Q$ and the dipole moment $U(1)_P$. Such circuits belong to the class of \emph{fractonic} quantum circuits, where particle motion is strongly constrained by the simultaneous presence of these conservation laws, typically leading to Hilbert-space fragmentation and subdiffusive hydrodynamics in terms of correlation functions (when averaged over many initial states from various fragments) ~\cite{feldmeier2020anomalous, Singh2021subdiffusion,Feldmeier2021criticalslow,Burchards2022coupledhydro}.

\subsection{Global dynamics of $S_d$}
\label{sec:dipole_global}
\begin{figure}[h]
\centering
\includegraphics[width=0.7\textwidth]{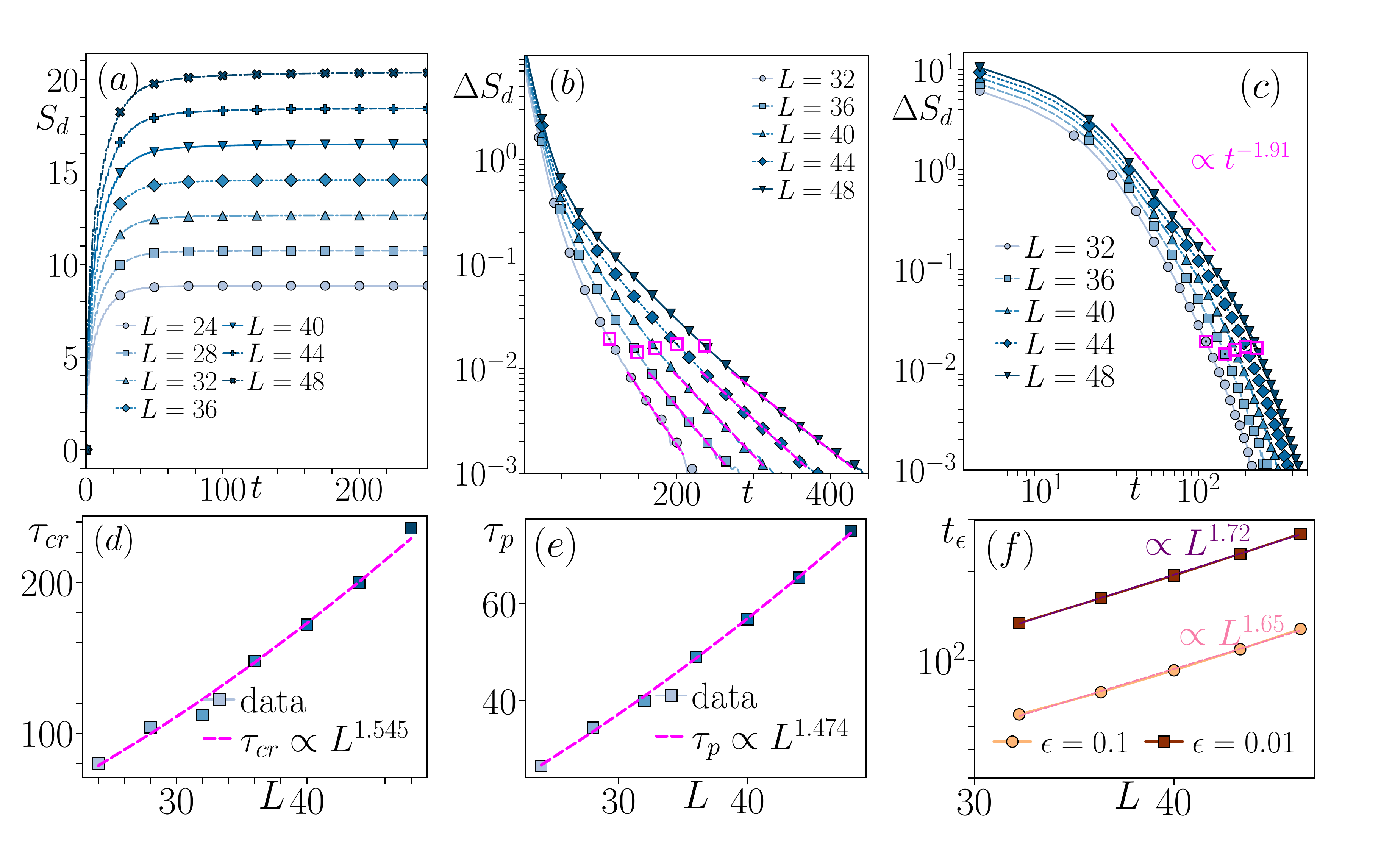}
\caption{{\bf Global $S_d$ dynamics in a spin-$\tfrac{1}{2}$ charge- and dipole-conserving circuit with $L$ sites:}
(a) Global $S_d$ grows rapidly and approaches the saturation value set by the fragmentation structure.
(b,c) $\Delta S_d$ exhibits a dynamical crossover from intermediate-time power-law behavior, $\Delta S_d \sim A_p t^{-\beta_p}$ with $\beta_p \simeq 1.91$, to late-time exponential decay, $\sim A_p e^{-t/\tau_p}$.
(c) The decay time scales with system size as $\tau_p \propto L^{\alpha_p}$, with $\alpha_p \simeq 1.474$ (Fig.~e). The magenta squares indicate a crossover between two dynamical regimes, as in the $U(1)$ case, where the crossover time $\tau_{\mathrm{cr}}$ scales in a power-law manner with $L$, $\tau_{\mathrm{cr}}\propto L^{1.545}$. (f) The relaxation time to reach a fixed threshold, $\Delta S_d \leq \epsilon \leq O(1)$, scales as $t_{\epsilon} \propto L^{\alpha_\epsilon}$. For all the analysis, we choose the initial state $|+1,-1,-1,+1,\cdots,+1,-1,-1,+1\rangle$, which is the root state generating the largest fragment within the $(Q,P)=(0,0)$ sector. All the results are obtained using the exact vector simulation and averaged over at least $10^4$ circuit realizations.}
\label{fig:dipole_global_sd}
\end{figure}

We now analyze the global dynamics of $S_d(t)$ in circuits that conserve both charge and dipole moment. In contrast to the $U(1)$ case, the dynamics is strongly constrained by Hilbert-space fragmentation~\cite{sala2020ergodicity,moudgalya2022quantum, Aditya2024subspace,Ganguli2025East,Aditya2025East2}, such that the evolution remains confined to a specific connected Krylov fragment determined by the initial state. We therefore focus on the largest typical fragment generated by the root state
$|\Psi\rangle = |+1,-1,-1,+1,\cdots,+1,-1,-1,+1\rangle$, which belongs to the $(Q,P)=(0,0)$ sector.
As shown in Fig.~\ref{fig:dipole_global_sd}a, the participation entropy $S_d(t)$ grows rapidly at early times and approaches a saturation value set by the dimension of the fragment. To analyze further, we map the state to a dimerized description, where one can argue that many features observed in the numerics are explained. In doing so, we introduce two elementary dimers, $A \equiv |+1,-1\rangle$ and $B \equiv |-1,+1\rangle$. Within the Krylov subspace rooted at $|\Psi\rangle$, the only allowed transition is $AB \leftrightarrow BA$, ensuring that the dynamics is restricted to sequences of $A$ and $B$ dimers. As a result, the original chain of length $L$ maps onto an effective chain of $N_d = L/2$ dimers. The conservation laws fix the imbalance between the two species, corresponding to an effective magnetization $K = L/4$. The dimension of the resulting Krylov fragment is therefore
\begin{equation}
D_F = \binom{N_d}{K},
\label{dim:fragment}
\end{equation}
which counts the number of configurations with $K$ $A$-dimers. Since $D_F \ll D_{0,0}$, the dynamics remains confined to a much smaller subspace, enabling large-scale exact simulations up to $L=48$. For $U(1)_Q \times U(1)_P$ conserving circuits restricted to this fragment, the global saturation value of $S_d$ is
\begin{equation}
S_d(\infty) = \log_2\left(D_F + 1\right) - 1,
\qquad \text{with} \qquad
D_F = \binom{L/2}{L/4}.
\end{equation}

To characterize the relaxation dynamics, we now analyze the deviation $\Delta S_d(t)$ using $S_d(\infty)$ computed analytically, which exhibits a clear two-stage behavior. As seen from the symbols in Fig.~\ref{fig:dipole_global_sd}(b,c), $\Delta S_d(t)$ decays algebraically at intermediate times, $\Delta S_d(t) \sim t^{-\beta_p}$, with $\beta_p \simeq 1.91$, indicating the presence of slow collective modes within the fragmented Hilbert space. At later times, the dynamics crosses over to an exponential decay, as highlighted by the linear behavior on the semi-log scale in Fig.~\ref{fig:dipole_global_sd}b. The crossover between these regimes is marked by the magenta squares in Fig.~\ref{fig:dipole_global_sd} (b-c), and the corresponding crossover timescale grows with system size as $\tau_{\mathrm{cr}} \propto L^{1.545}$, as shown in Fig.~\ref{fig:dipole_global_sd}d. The late-time relaxation is governed by a characteristic timescale $\tau_p$, which scales with system size as $\tau_p \propto L^{\alpha_p}$ with $\alpha_p \simeq 1.474$, as shown in Fig.~\ref{fig:dipole_global_sd}e. Furthermore, the time required to reach a fixed threshold $\Delta S_d(t) \le \epsilon$ also exhibits a power-law dependence on system size, as shown in Fig.~\ref{fig:dipole_global_sd}f. Altogether, these results demonstrate that while the global structure of the dynamics remains similar to the $U(1)$ case, the effective scaling is strongly modified by fragmentation and the constrained motion of dipolar degrees of freedom.

At the kinematic level, the fragment dynamics is indistinguishable from local $U(1)$-conserving dynamics on a half-filled chain of $N_d = L/2$ dimers: each configuration is a binary string with conserved particle number, and the only elementary move, $AB \leftrightarrow BA$, is a nearest-neighbor exchange. The analogy to $U(1)$-circuit dynamics therefore predicts diffusive scaling of the finite-size exponents. Numerically extracted exponents, however, deviate from this prediction;
the finite-size quantities $\alpha_p \simeq 1.47$ and $\alpha_{\mathrm{cr}} \simeq 1.55$, obtained from the crossover to the exponential regime, likely remain sensitive to preasymptotic effects due to the limited system sizes accessible numerically ($L \le 48$). We therefore interpret them as effective exponents characterizing the accessible range of system sizes and time scales, and expect the exponents to drift toward the asymptotic diffusive prediction at larger $L$.

\subsection{Local spreading of coherence with charge and dipole conservation}
\label{sec:dipole_local}

\begin{figure}[t!]
\centering
\includegraphics[width=0.7\textwidth]{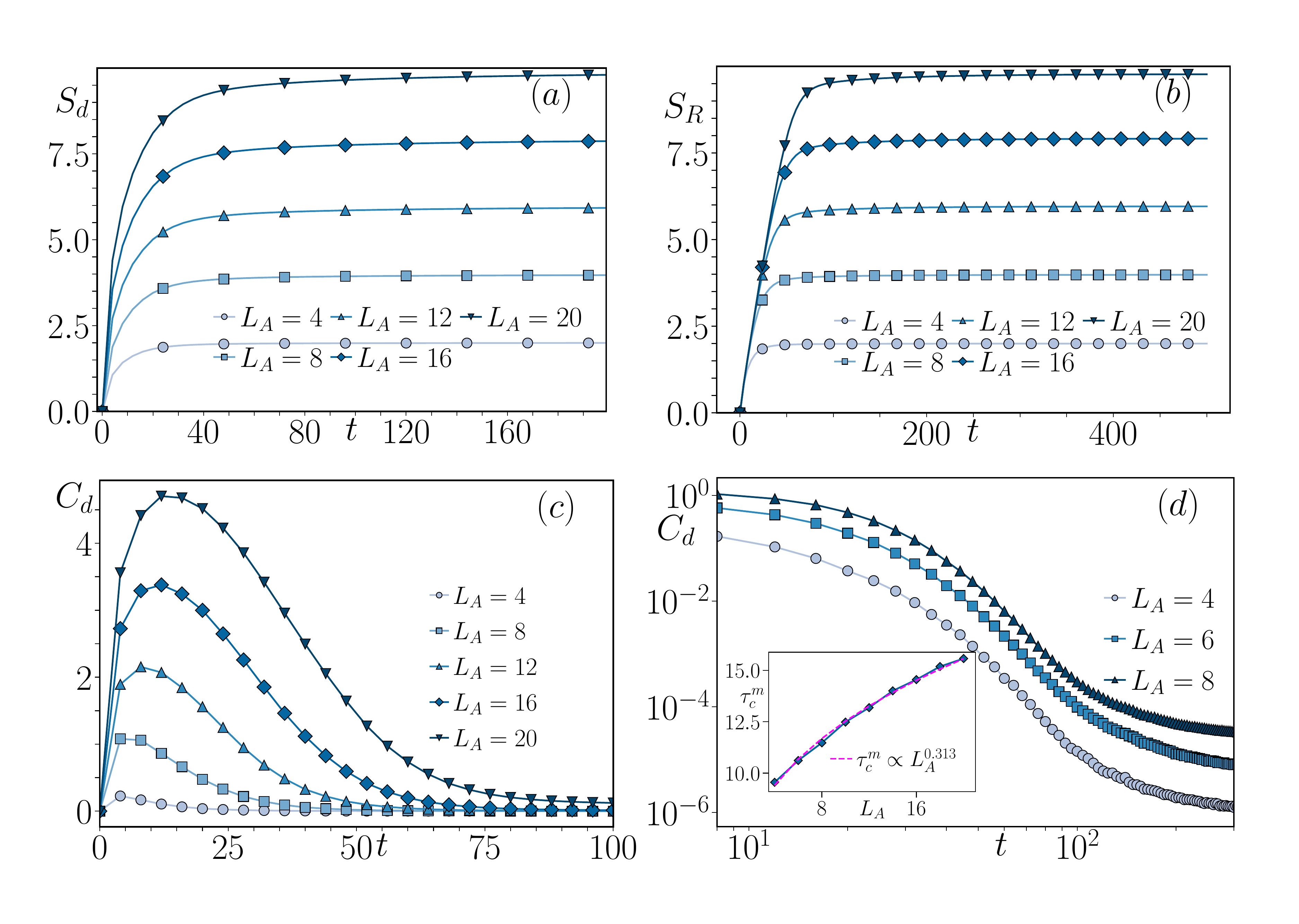}
\caption{\textbf{Local coherence dynamics in the spin-$1/2$ 
charge- and dipole-conserving circuit with $L$ sites.}
(a--c) Dynamics of the local $S_d$, $S_R$, and $C_d$ for different 
subsystem sizes $L_A$ at $L=48$.
(d) The peak time of $C_d$ scales as 
$\tau_c^{m} \propto L_A^{\alpha_m}$ with $\alpha_m \simeq 0.313$.
Results use the same initial state as in 
Fig.~\ref{fig:dipole_global_sd} and are obtained from exact 
state-vector simulations averaged over at least $10^{4}$ circuit 
realizations.}
\label{fig:dipole_local_sd}
\end{figure}

We now turn to the local resource content in the same fractonic 
circuit, starting from the same initial state in the $(Q,P)=(0,0)$ 
sector and focusing on subsystems of size $L_A$. As shown in 
Fig.~\ref{fig:dipole_local_sd}(a--c), the qualitative structure 
closely parallels the $U(1)$ case: the diagonal entropy $S_d(t)$
grows rapidly toward its saturation value, while the entanglement entropy $S_R(t)$ exhibits an extended regime of
ballistic growth. As a result, the coherence $C_d(t)$ displays the 
characteristic rise--peak--fall profile. The time at which $C_d(t)$
attains its maximum scales algebraically with subsystem size,
$\tau_c^m \propto L_A^{\alpha_m}$ with $\alpha_m \simeq 0.313$, as
shown in Fig.~\ref{fig:dipole_local_sd}(d). As in the $U(1)$ case, 
this scaling arises from the competition between the algebraic 
relaxation of $S_d(t)$ and the faster, approximately ballistic 
growth of $S_R(t)$.

To analyze this behavior quantitatively, we study the relaxation of
the local diagonal entropy toward its long-time value,
\begin{equation}
\Delta S_d(t) = S_d(\infty) - S_d(t),
\qquad
S_d(\infty) = -\log_2 \overline{P_{A,\mathrm{diag}}},
\label{eq:DeltaSd_frag}
\end{equation}
where $\overline{P_{A,\mathrm{diag}}}$ is the Haar-averaged
diagonal purity of the reduced density matrix within the fragment.
In the dimer mapping with $N_d=L/2$ and $K=L/4$, the Haar computation
in Appendix~\ref{app:fragment_resolved} yields the closed-form
saturation value
\begin{equation}
\overline{P_{A,\mathrm{diag}}}=
\frac{1}{D_F(D_F+1)}\!\left[D_F+\sum_{r}\binom{n}{r}\,W(r)\right],
\qquad n=\lfloor L_A/2\rfloor,
\label{eq:PAdiag_sat_main}
\end{equation}
with $D_F=\binom{N_d}{K}$ and
\begin{equation}
W(r)=
\begin{cases}
\displaystyle\binom{N_d-n}{K-r}^{\!2}, & L_A=2n\ \text{(even)},\\[6pt]
\displaystyle\binom{N_d-n-1}{K-r}^{\!2}+\binom{N_d-n-1}{K-r-1}^{\!2}, & L_A=2n+1\ \text{(odd)},
\end{cases}
\label{eq:Wr_main}
\end{equation}
which distinguishes bipartitions that fall between two dimers from those that cut through a single dimer. The detailed derivation, including the corresponding formula for $\overline{\Tr\rho_A^2}$, is given in Appendix~\ref{app:fragment_resolved}.

\begin{figure}[t]
\centering
\includegraphics[width=0.7\textwidth]{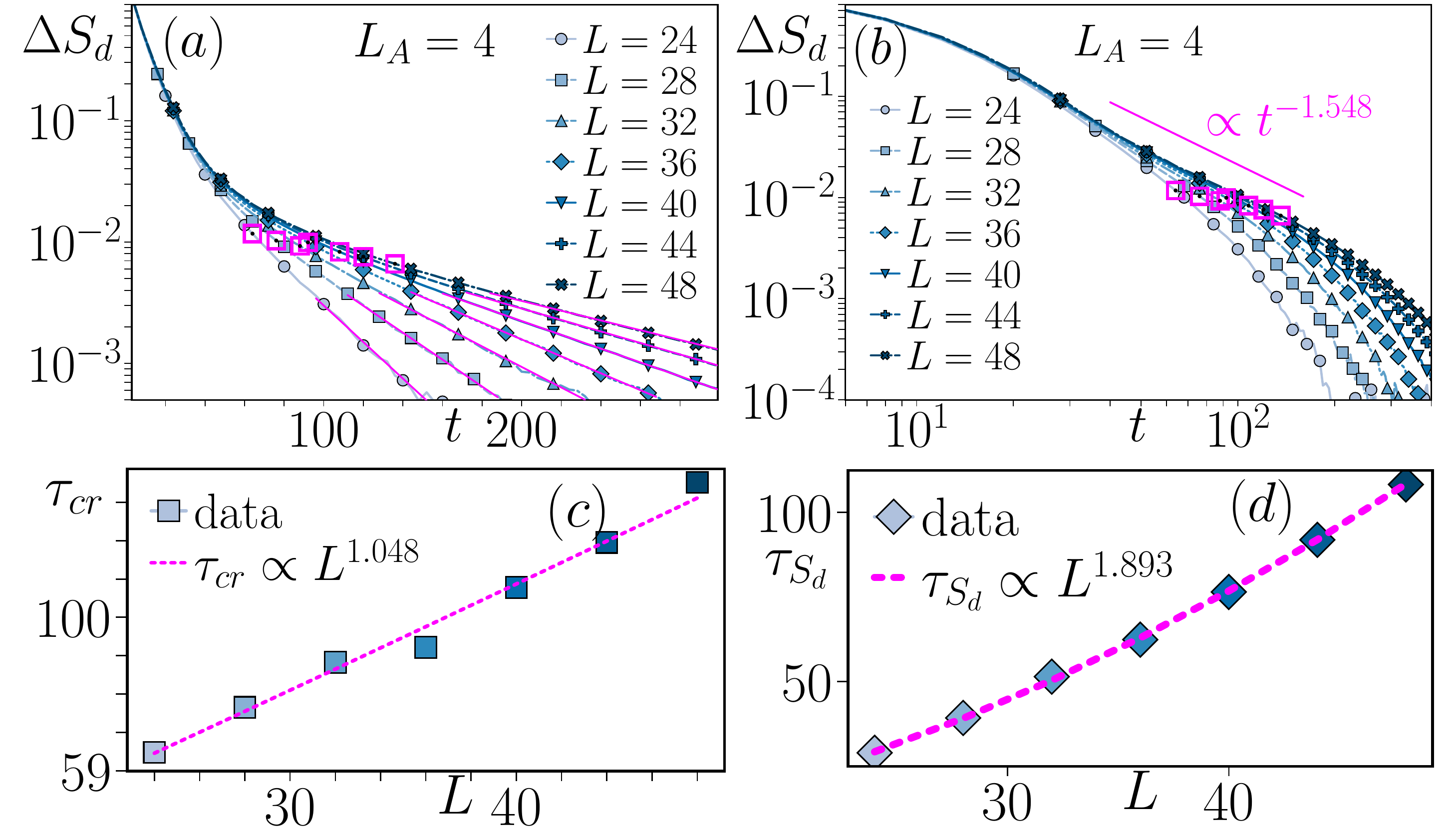}
\caption{\textbf{Local $\Delta S_d$ decay in charge- and 
dipole-conserving circuits with $L$ sites and subsystem size $L_A$.} 
(a,b) The relaxation exhibits an intermediate power-law regime 
$\Delta S_d(t) \sim t^{-\beta_{S_d}}$ with 
$\beta_{S_d}\simeq 1.548$, crossing over to a late-time exponential 
decay $A_{S_d}\, e^{-t/\tau_{S_d}}$ with 
$\tau_{S_d}\propto L^{1.893}$ (panel d). In panels (a,b), magenta 
squares mark the crossover between the two regimes; the crossover 
time scales as $\tau_{\mathrm{cr}} \propto L^{1.048}$ (panel c). 
Results use the same initial state as in 
Fig.~\ref{fig:dipole_global_sd} and are obtained from exact 
state-vector simulations averaged over at least $10^{4}$ circuit 
realizations for $L =24- 48$ and $L_A=4$.}
\label{fig:dipole_local_decay}
\end{figure}

Using the analytically determined saturation value, we now analyze 
the relaxation of local $\Delta S_d(t)$. As shown in 
Fig.~\ref{fig:dipole_local_decay} (a-b), the dynamics exhibits a 
clear two-stage behavior: at intermediate times $\Delta S_d(t)$ 
decays algebraically as $\Delta S_d(t) \sim t^{-\beta_{S_d}}$ with 
$\beta_{S_d} \simeq 1.548$, while at late times it crosses over to 
an exponential relaxation. The crossover between the two regimes 
is highlighted by the magenta markers in 
Fig.~\ref{fig:dipole_local_decay} (a-b); the corresponding crossover 
time grows in a power-law manner with $L$, as
$\tau_{\mathrm{cr}} \propto L^{1.048}$ 
(Fig.~\ref{fig:dipole_local_decay}(c)). Furthermore, the late-time decay is 
governed by a characteristic timescale 
$\tau_{S_d} \propto L^{\alpha_{S_d}}$ with 
$\alpha_{S_d} \simeq 1.893$, as shown in 
Fig.~\ref{fig:dipole_local_decay}(d).

The intermediate-time power-law decay can be understood 
quantitatively in terms of the constrained dimer-swap dynamics 
within the fragmented Hilbert space spanned by the dimer alphabet
$A \equiv |+1,-1\rangle$, $B \equiv |-1,+1\rangle$.
A useful phenomenological description of the slow sector is in 
terms of domain-wall pairs separating locally staggered regions 
of the form $ABAB\cdots$ and $BABA\cdots$, naturally characterized 
by two coordinates $(x_L, x_R)$ corresponding to the boundaries 
of a locally reversed segment. For a local observable such as the 
diagonal entropy of a subsystem, the absolute position of this 
pair is irrelevant: the reduced density matrix is modified only 
when the rearrangement occurs within the subsystem or in its 
immediate vicinity. The uniform translation mode of the 
domain-wall pair therefore does not contribute to the relaxation 
of $\Delta S_d(t)$.
Consequently, the leading contribution arises from spatially 
varying fluctuations of the slow mode. At the coarse-grained level, 
expanding the observable in long-wavelength modes and exploiting 
insensitivity to the $k=0$ component implies that the spectral 
weight at small momentum is suppressed as $k^2$. Assuming the 
long-wavelength dynamics of these rearrangements is approximately 
diffusive over the accessible timescales, each mode relaxes as 
$e^{-Dk^2 t}$, yielding
\begin{equation}
\Delta S_d(t) \;\sim\; \int\! dk\, k^{2}\, e^{-D k^{2} t}
\;\sim\; t^{-3/2},
\label{eq:tm32_prediction}
\end{equation}
where the second equality follows from the rescaling $k = u/\sqrt{t}$.
This prediction is in excellent agreement with the numerically 
extracted exponent $\beta_{S_d} \simeq 1.548$, indicating that the 
relaxation of local coherence is governed by spatially varying 
fluctuations of the constrained dipole-swap dynamics. At late times, 
finite-size effects discretize the spectrum of slow modes, driving 
the crossover to the exponential regime 
$\Delta S_d(t) \sim e^{-t/\tau_{S_d}(L)}$, with 
$\tau_{S_d}(L) \propto L^{1.893}$ consistent with an approximately 
diffusive finite-size cutoff over the accessible system sizes.

The early-time growth of the entanglement entropy shown
in Fig.~\ref{fig:dipole_local_sd}(b) is approximately ballistic
within this fragment, rather than subdiffusive. The underlying 
reason is that entanglement growth is controlled by local coherent 
resonances rather than by long-wavelength transport of conserved 
quantities. The elementary dynamical process in the fragment is 
the local dipole-preserving swap $AB \leftrightarrow BA$, which 
acts across neighboring sites and immediately generates 
entanglement whenever it overlaps the bipartition in close analogy to the $U(1)$-case. Once such a 
local resonance occurs near the cut, the resulting entanglement 
propagates outward through successive local rearrangements, giving 
rise to an effective growth as 
$S_R(t) \sim v_E t$. Subdiffusive growth would instead require strong 
bottlenecks (large inactive regions or rare configurations that
suppress local dynamics near the cut), which are absent with high
probability in the fragment generated from our root configuration, 
since flippable dipolar motifs occur at finite density.

The late-time decay of $S_R$ to its saturation value, quantified as $\Delta S_R$, is still
governed by slow hydrodynamic modes, although this regime sets in
only at parametrically later times. The saturation value
$S_R(\infty)=-\log_2\!\bigl(\overline{\mathrm{Tr}\,\rho_A^2}\bigr)$ is obtained
analytically from the Haar average within the fragment Hilbert space,
with closed-form expressions for $\overline{\mathrm{Tr}\,\rho_A^2}$ (separate
formulas for even and odd $L_A$) derived in Appendix~\ref{app:fragment_resolved}. Taken together, these
results show that $\tau_c^m$ follows an algebraic scaling with
$L_A$ even in the fragment-resolved case, mirroring the $U(1)$
scenario.

Finally, as the dynamics generated by charge- and dipole-preserving 
Haar-random gates remains confined to the connected Hilbert-space 
fragment selected by the initial state, the late-time state can be 
regarded as a Haar-random pure state within that fragment. Tracing 
out subsystem $B$ yields the Haar-averaged reduced density matrix
\begin{equation}
\overline{\rho_A(\infty)}
= \mathrm{Tr}_B\!\left(\frac{\Pi_F}{D_F}\right),
\end{equation}
where $\Pi_F$ is the projector onto the fragment. 

Writing the computational-basis states across the bipartition as $|a\rangle_A \otimes |b\rangle_B$, the fragment projector can be expressed as $\Pi_{F}
=\sum_{(a,b)\in F}
|a\rangle\langle a|\otimes |b\rangle\langle b|$,
which tracing out subsystem $B$ then yields
\begin{equation}
\overline{\rho_A(\infty)}
=
\sum_a \frac{N_{F}(a)}{D_{F}}\,|a\rangle\langle a|,
\end{equation}
where $N_{F}(a)$ denotes the number of fragment-compatible completions in subsystem $B$. Thus the late-time reduced density matrix is diagonal in the computational basis, and hence block diagonal in the subsystem symmetry sectors $(q_A,p_A)$ compatible with the global $(Q,P)=(0,0)$ constraint. However, the weights within those blocks are generally fragment-dependent and need not be uniform. The asymptotic reduced state is thus again a free state under resource theory, leading to $C_d\to 0$ at late times.

Before proceeding further, we would like to make a broader comment on 
fragment-dependence of these results. In strongly 
fragmented systems such as the charge- and dipole-conserving 
circuit considered here, the dynamical behavior of coherence 
spreading is inherently fragment-dependent: the Krylov sector 
selected by the initial state determines both the set of 
accessible configurations and the effective slow modes, so 
quantitative features such as the exponents 
$\alpha_m, \beta_{S_d}, \alpha_{S_d}$, and $\alpha_{\mathrm{cr}}$ 
are not expected to be universal across fragments. A direct imprint of this fragmentation is visible 
already at the level of the long-time saturation values themselves: they lie well below the value 
that would be predicted from charge and dipole conservation 
alone (analytical derivation given in Appendix \ref{app:qp_full_sector}), reflecting the fact that the dynamics explores only the 
connected Krylov sector rather than the full $(Q,P)=(0,0)$ 
symmetry subspace. 

Nevertheless, 
one robust qualitative conclusion emerges from our analysis: 
the presence of charge conservation, together with the diffusive 
or near-diffusive relaxation of the associated hydrodynamic tails, 
produces a phenomenology of coherence spreading that is 
qualitatively similar to the $U(1)$ case within the largest 
fragment, characterized by a rise--peak--fall profile of
$C_d(t)$, algebraic scaling of the peak time with subsystem size,
and a two-stage (power-law to exponential) relaxation of
$\Delta S_d(t)$, yet drastically different from the behavior
found in random circuits without any conservation law.
The fragment structure thus modulates the quantitative exponents, 
but the qualitative separation between symmetric and 
non-symmetric circuits remains a robust organizing principle.

\begin{figure}[t!]
 \centering
 \includegraphics[width=0.7\columnwidth]{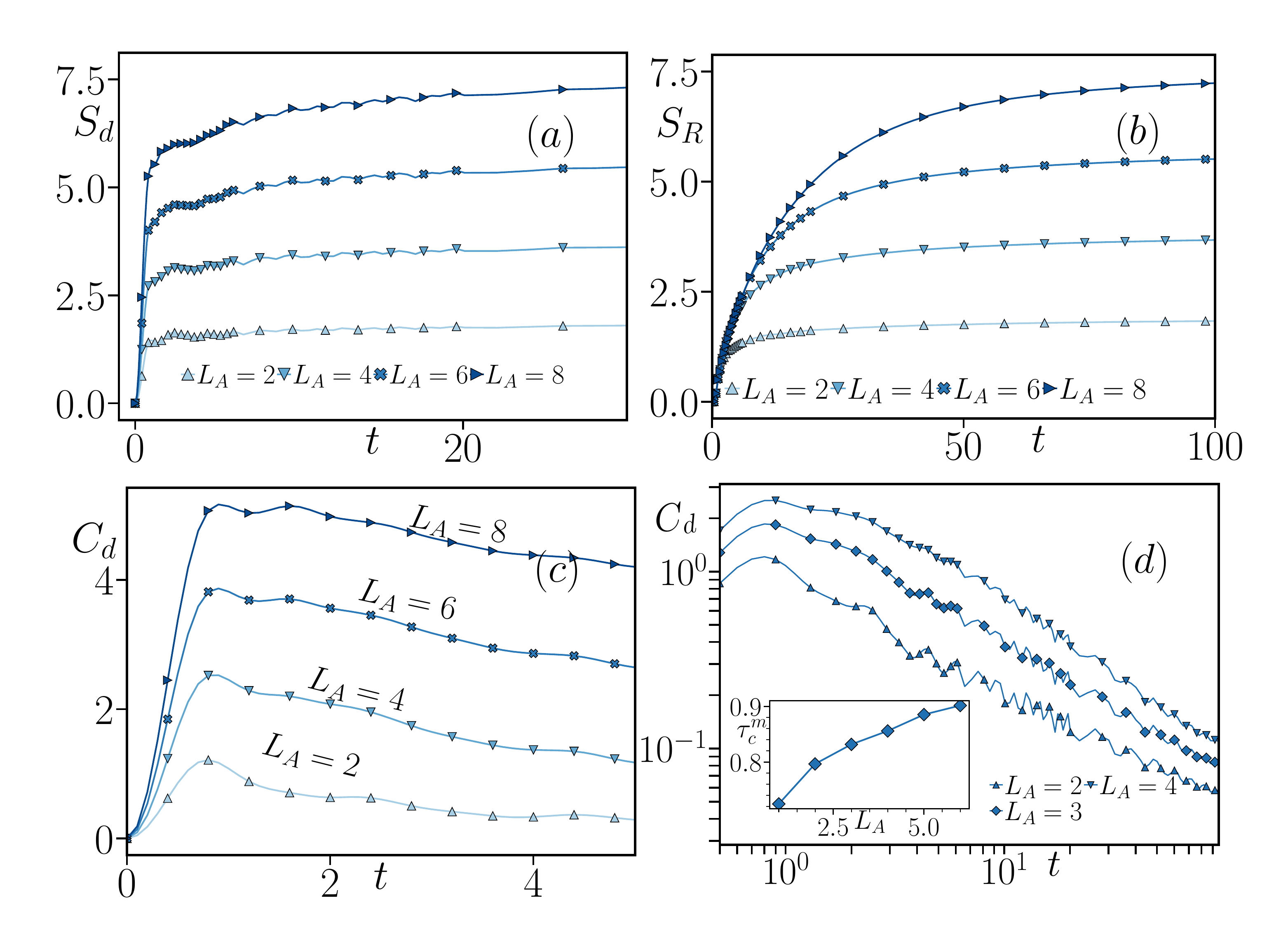}
\caption{\textbf{Local coherence dynamics for the MFIM.} 
(a--c) Local $S_d$, $S_R$, and $C_d$ dynamics for various 
subsystem sizes $L_A$ in the one-dimensional MFIM with $L=20$. 
(d) The intermediate-time relaxation of $C_d$ is governed by a 
power-law decay. Inset: the peak time as a function of $L_A$ for 
the first few subsystems for which a prominent peak exists. All 
results are obtained using the Chebyshev method and averaged over 
$200$ product initial states.}
\label{fig:Ham_ED}
\end{figure}

\begin{figure}[t!]
\centering
\includegraphics[width=0.7\textwidth]{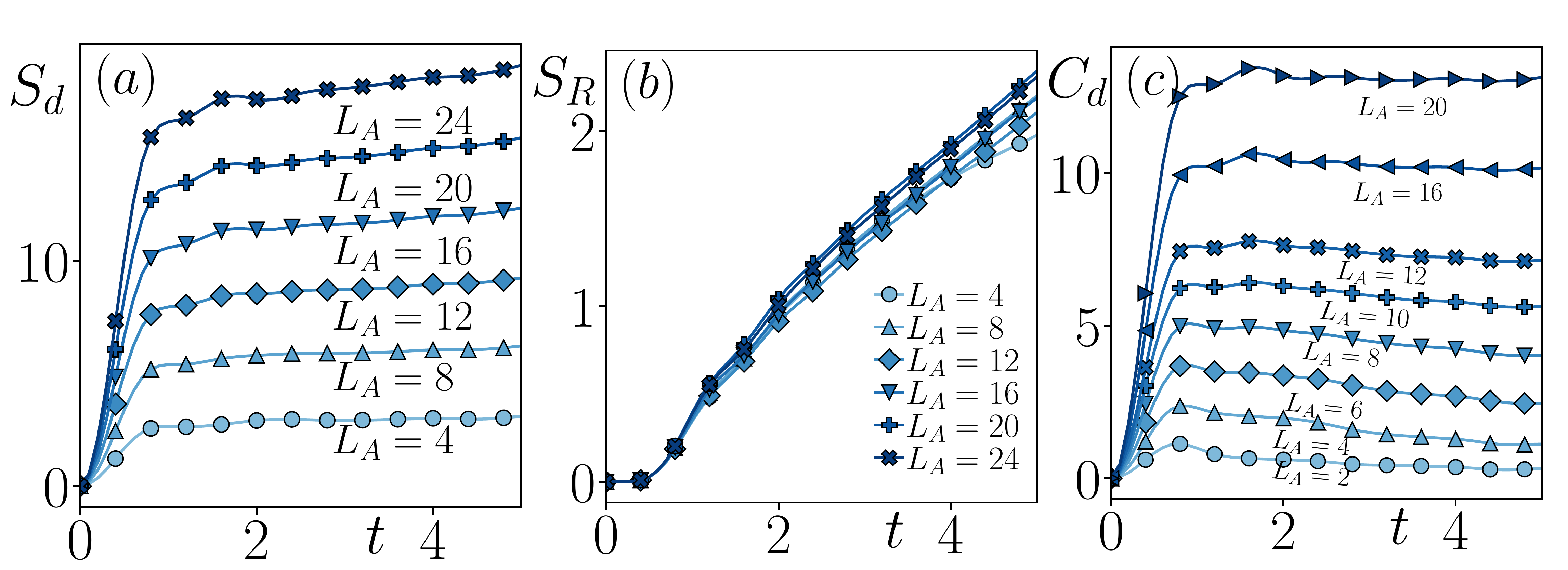}
\caption{{\bf Local coherence dynamics in the MFIM obtained using TN methods:}
Dynamics of local $S_d$, $S_R$, and $C_d$ for various subsystem sizes $L_A$ at $L=128$.
A well-defined peak time is observed for small $L_A$, which gradually smears out with increasing $L_A$, leading to a broad plateau in place of sharp coherence peaks. All results are averaged over $100$ product initial states; the bond dimension used for the TN simulations is $\chi=256$.}
\label{Ham_TDVP}
\end{figure}

\section{Coherence spreading in ergodic Hamiltonian dynamics}
\label{sec:mfim_results}
\subsection{Global dynamics of $S_d$}
\label{sec:mfim_global}
The global dynamics of $S_d$ in the mixed-field Ising chain were discussed in Ref.~\cite{tirrito2024anticoncentrationmagicspreadingergodic} in the context of anticoncentration. There it was shown that $S_d$ grows rapidly and, even at short times, approaches its long-time saturation value. Although this value differs from the Haar prediction, it still scales linearly with system size, $\propto L$, consistent with the ETH expectation for ergodic systems. Moreover, the deviation $\Delta S_d(t)$ was found to exhibit a power-law decay,
\begin{equation}
\Delta S_d(t)=A_{p}\,t^{-\beta_{p}},
\end{equation}
providing a clear indication that the global relaxation of coherence retains the hallmark of diffusive energy transport characteristic of generic local Hamiltonians~\cite{bertini2020scrambling}.
\subsection{Local dynamics of $S_d$}
\label{sec:mfim_local}
We now turn to the local dynamics of coherence under Hamiltonian evolution. As shown in Fig.~\ref{fig:Ham_ED}(a--c), the subsystem diagonal entropy $S_d(t)$ increases rapidly at early times, but subsequently exhibits a significantly slower approach to its saturation value. In contrast, the entanglement entropy $S_R(t)$ initially grows in a manner that is largely independent of the subsystem size $L_A$. However, as time progresses, the growth of $S_R(t)$ slows down further and does not reach its saturation value on the accessible timescales. As a result, the relative entropy of coherence $C_d(t)=S_d(t)-S_R(t)$ exhibits distinct peaks for smaller subsystem sizes $L_A$, which progressively broaden and evolve into extended plateaus as $L_A$ increases. These features are qualitatively distinct from those observed in symmetry-constrained random circuit dynamics, highlighting the fundamentally different mechanisms governing coherence spreading in Hamiltonian systems.

\begin{figure}[t!]
\includegraphics[width=0.7\columnwidth]{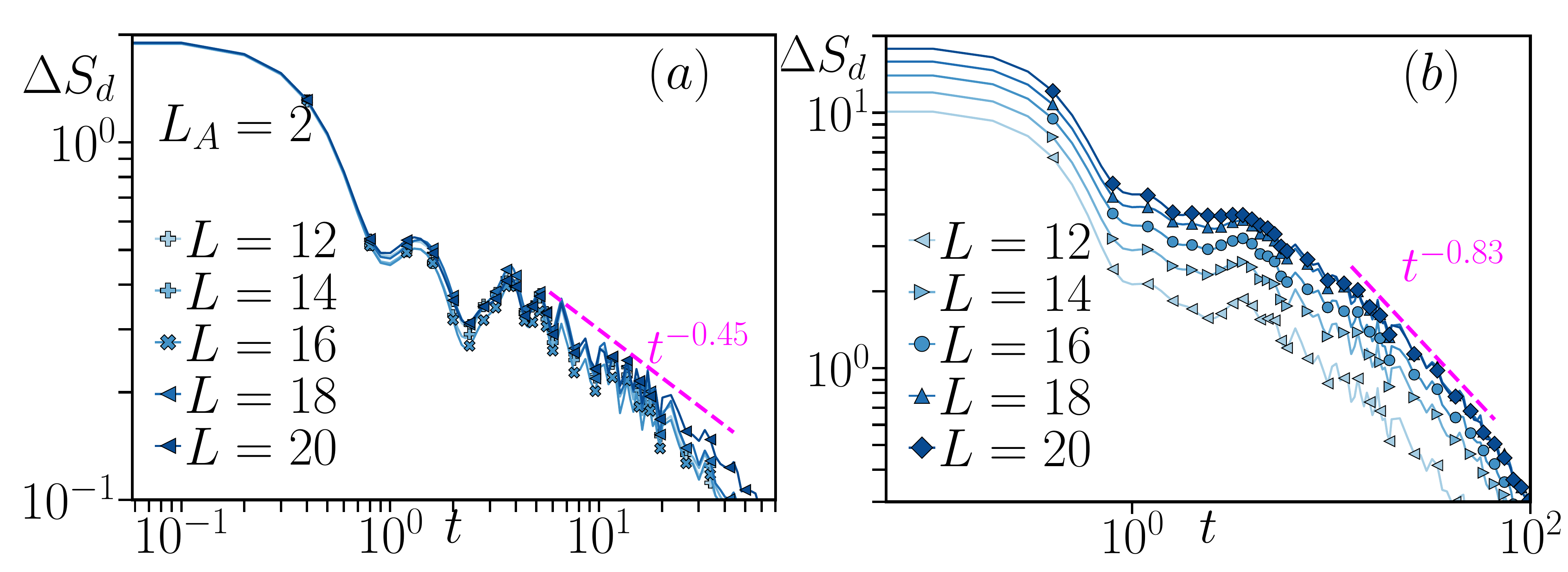}
\caption{{\bf Local and global $\Delta S_d$ decay in the MFIM:}
(a) The local $\Delta S_d$ exhibits power-law relaxation, $\sim t^{-\beta_{S_d}}$, with $\beta_{S_d} \simeq 0.45$ within the numerically accessible regime for $L_A=2$ and various system sizes $L$.
(b) For comparison, the global $\Delta S_d$ also shows power-law decay with exponent $\beta_{S_d} \simeq 0.83$ for different $L$. All the results are obtained using the Chebyshev method and averaged over $200$ $z$-basis product initial states.}
\label{Sd_decay_Ham_ED}
\end{figure}

These behaviors are consistently observed in both exact diagonalization (ED) and time-dependent variational principle (TDVP) simulations, as shown in Figs.~\ref{fig:Ham_ED} and~\ref{Ham_TDVP}, providing robust numerical evidence that the dynamics is governed by slow hydrodynamic modes rather than fast scrambling.
To gain a deeper understanding of this behavior, we analyze the time dependence of the deviation of $S_{d}$ from its long-time saturation value, where the latter is obtained by averaging over the time window $t\in [1000,5000]$. We observe that $\Delta S_{d}$ exhibits an algebraically slow relaxation $\sim A_{S_d}\,t^{-\beta_{S_d}}$ with $\beta_{S_d}\simeq 0.45$, as shown in Fig.~\ref{Sd_decay_Ham_ED}(a), reminiscent of the intermediate-time decay regime observed in random circuits with conservation laws. As a comparison, we also present the global $\Delta S_d$ decay, which exhibits an algebraic approach to saturation with $\beta_{S_d}\simeq 0.83$, as shown in Fig.~\ref{Sd_decay_Ham_ED}(b). Altogether, the local dynamics of coherence in the Hamiltonian case show markedly different behavior than symmetric circuits.

\section{Summary and outlook}
\label{sec:summary}
To summarize, we have characterized the spreading of quantum coherence, probed globally via the participation entropy $S_d$ and locally via the Rényi-$2$ relative entropy of coherence $C_d$, in three representative classes of many-body dynamics: $U(1)$-symmetric random circuits for spin-$\tfrac12$ and spin-$1$, charge- and dipole-conserving (fractonic) random circuits, and the mixed-field Ising chains as prototypes of ergodic Hamiltonian dynamics. Combining exact state-vector simulations, matrix-product-state evolution, and replica tensor-network methods, we found that conservation laws replace the logarithmic saturation characteristic of unconstrained circuits with a two-stage hydrodynamic decay of $\Delta S_d(t)$, consisting of an intermediate power law $\Delta S_d\sim t^{-\beta_p}$ followed by an exponential tail with timescale $\tau_p\propto L^{\alpha_p}$. As a consequence, the time $t_\epsilon$ needed for $S_d$ to saturate within a fixed tolerance grows as $t_\epsilon\propto L^{\alpha_\epsilon}$. Locally, the competition between the algebraic relaxation of $S_d$ and the ballistic growth of $S_R$ produces a rise--peak--fall profile of $C_d(t)$ in both $U(1)$- and dipole-conserving circuits, with a peak time $\tau_c^m\propto L_A^{\alpha_m}$ set by $\alpha_m=1/(\beta_{S_d}+1)$. The Hamiltonian case shares the same global phenomenology, but the sharp local peak observed at small $L_A$ broadens into an extended plateau at larger subsystems, signaling a distinct local relaxation mechanism. These numerical findings are supported by Haar-averaged expressions for the diagonal and subsystem purities and by a rare-region coupled-SSEP analysis that identifies the exponent $\beta_{S_d}\approx 1$ as the product of two independent diffusive modes.

Our results suggest several directions for future work. A natural next step is to develop a hydrodynamic theory of resource spreading that accounts both for the slow global relaxation and for the distinct local behaviors observed here; ongoing work on resource dynamics in many-body systems offers a concrete setting to pursue this program~\cite{rbt4-psfd, coffman2026groupfourierfilteringquantum, Deneris_2026, diaz2025unifiedapproachquantumresource, santra2025quantumresourcesnonabelianlattice, Varikuti2026impactofclifford, varikuti2025deepthermalizationmeasurementsquantum, varikuti2024quantuminformationscramblingchaos,y9r6-dx7p, aditya2025mpembaeffectsquantumcomplexity, aditya2025growthspreadingquantumresources}. A second direction is to test whether the same phenomenology extends to other nonclassical resources whose growth couples to symmetry and transport, such as magic~\cite{leone2022stabilizer, aaronson2004improved, gottesman1997stabilizer, gottesman1998theory, bravyi2005universal, zhu2016cliffordgroupfailsgracefully, Leone2021quantumchaosis, turkeshi2025magic, turkeshi2025pauli} and fermionic non-Gaussianity~\cite{Sierant26fermionic, Lumia24, Falco2026fermionic, collura2025quantummagicfermionicgaussian, sierant2026theorymatchgatecommutant, lastres2026geometryfreefermioncommutants, braccia2026commutantfermionicgaussianunitaries}. A third is to extend the analysis to richer symmetry settings, including non-Abelian charges, higher-moment conservation laws, and kinetically constrained or fragmented Hilbert spaces, where qualitatively new regimes may appear. Finally, initializing the dynamics from resourceful states evolved by the free operations of the corresponding theory, rather than from incoherent products, would disentangle resource generation from redistribution and transport.

\section*{Acknowledgements}
We thank Michael Knap and Marko \v{Z}nidari\v{c} for illuminating discussions. S.A acknowledges support from the Alexander Humboldt Foundation as a Humboldt postdoctoral fellow.
X.T. acknowledges support from the DFG under Germany’s Excellence Strategy, Cluster of Excellence \emph{Matter and Light for Quantum Computing} (ML4Q), EXC 2004/2, project no.~390534769; from the DFG Collaborative Research Center CRC 183, project no.~277101999, project B01; and from the DFG Emmy Noether Programme proposal ``\textit{Digital Quantum Matter Out-of-Equilibrium},'' project no.~560726973. P.S. acknowledges fellowship within the “Generación D” initiative, Red.es, Ministerio para la Transformación Digital y de la Función Pública, for talent attraction (C005/24-ED CV1), funded by the European Union NextGenerationEU funds, through PRTR.
E.T. was funded by the Swiss National Science Foundation
(SNSF) under Grant No. TMPFP2\_234754. E.T. acknowledges CINECA (Consorzio Interuniversitario per il Calcolo Automatico) award, under the ISCRA initiative and Leonardo early access program, for the availability of high-performance computing resources
and support.
We further thank the ITCC (IT Center University of Cologne) for providing computing resources on the DFG-funded HPC (High Performance Computing) system RAMSES (Research Accelerator for Modeling and Simulation with Enhanced Security) as well as support (DFG funding number: INST 216/512-1 FUGG).

\textbf{Code and Data Availability.} Code and data will be publicly shared at publication.  

\bibliography{biblio}

@inbook{potter2022entanglement,
   title={Entanglement Dynamics in Hybrid Quantum Circuits},
   ISBN={9783031039980},
   ISSN={2364-9062},
   url={http://dx.doi.org/10.1007/978-3-031-03998-0_9},
   DOI={10.1007/978-3-031-03998-0_9},
   booktitle={Entanglement in Spin Chains},
   publisher={Springer International Publishing},
   author={Potter, Andrew C. and Vasseur, Romain},
   year={2022},
   pages={211–249} }

@article{Magni2025quantumcomplexity,
  doi = {10.22331/q-2025-12-24-1956},
  url = {https://doi.org/10.22331/q-2025-12-24-1956},
  title = {Quantum {C}omplexity and {C}haos in {M}any-{Q}udit {D}oped {C}lifford {C}ircuits},
  author = {Magni, Beatrice and Turkeshi, Xhek},
  journal = {{Quantum}},
  issn = {2521-327X},
  publisher = {{Verein zur F{\"{o}}rderung des Open Access Publizierens in den Quantenwissenschaften}},
  volume = {9},
  pages = {1956},
  month = dec,
  year = {2025}
}

@article{p8dn-glcw,
  title = {Anticoncentration in Clifford Circuits and Beyond: From Random Tensor Networks to Pseudomagic States},
  author = {Magni, Beatrice and Christopoulos, Alexios and De Luca, Andrea and Turkeshi, Xhek},
  journal = {Phys. Rev. X},
  volume = {15},
  issue = {3},
  pages = {031071},
  numpages = {16},
  year = {2025},
  month = {Sep},
  publisher = {American Physical Society},
  doi = {10.1103/p8dn-glcw},
  url = {https://link.aps.org/doi/10.1103/p8dn-glcw}
}

@misc{tirrito2025universalspreadingnonstabilizernessquantum,
      title={Universal Spreading of Nonstabilizerness and Quantum Transport}, 
      author={Emanuele Tirrito and Poetri Sonya Tarabunga and Devendra Singh Bhakuni and Marcello Dalmonte and Piotr Sierant and Xhek Turkeshi},
      year={2025},
      eprint={2506.12133},
      archivePrefix={arXiv},
      primaryClass={quant-ph},
      url={https://arxiv.org/abs/2506.12133}, 
}

@article{rbt4-psfd,
  title = {Resource-Theoretical Unification of Mpemba Effects: Classical and Quantum},
  author = {Summer, Alessandro and Moroder, Mattia and Bettmann, Laetitia P. and Turkeshi, Xhek and Marvian, Iman and Goold, John},
  journal = {Phys. Rev. X},
  volume = {16},
  issue = {1},
  pages = {011065},
  numpages = {27},
  year = {2026},
  month = {Mar},
  publisher = {American Physical Society},
  doi = {10.1103/rbt4-psfd},
  url = {https://link.aps.org/doi/10.1103/rbt4-psfd}
}

@misc{iannotti2026nonstabilizernessu1symmetrychaotic,
      title={Non-stabilizerness and U(1) symmetry in chaotic many-body quantum systems}, 
      author={Daniele Iannotti and Angelo Russotto and Barbara Jasser and Jovan Odavić and Alioscia Hamma},
      year={2026},
      eprint={2603.28870},
      archivePrefix={arXiv},
      primaryClass={quant-ph},
      url={https://arxiv.org/abs/2603.28870}, 
}

@misc{sierant2026theorymatchgatecommutant,
      title={Theory of the Matchgate Commutant}, 
      author={Piotr Sierant and Xhek Turkeshi and Poetri Sonya Tarabunga},
      year={2026},
      eprint={2603.12392},
      archivePrefix={arXiv},
      primaryClass={quant-ph},
      url={https://arxiv.org/abs/2603.12392}, 
}

@misc{lastres2026geometryfreefermioncommutants,
      title={Geometry of Free Fermion Commutants}, 
      author={Marco Lastres and Sanjay Moudgalya},
      year={2026},
      eprint={2604.05031},
      archivePrefix={arXiv},
      primaryClass={quant-ph},
      url={https://arxiv.org/abs/2604.05031}, 
}

@misc{braccia2026commutantfermionicgaussianunitaries,
      title={The commutant of fermionic Gaussian unitaries}, 
      author={Paolo Braccia and N. L. Diaz and Martin Larocca and M. Cerezo and Diego García-Martín},
      year={2026},
      eprint={2603.19210},
      archivePrefix={arXiv},
      primaryClass={quant-ph},
      url={https://arxiv.org/abs/2603.19210}, 
}

@misc{coffman2026groupfourierfilteringquantum,
      title={Group Fourier filtering of quantum resources in quantum phase space}, 
      author={Luke Coffman and N. L. Diaz and Martin Larocca and Maria Schuld and M. Cerezo},
      year={2026},
      eprint={2601.14225},
      archivePrefix={arXiv},
      primaryClass={quant-ph},
      url={https://arxiv.org/abs/2601.14225}, 
}

@article{Deneris_2026,
   title={Analyzing the Free States of one Quantum Resource Theory as Resource States of Another},
   volume={9},
   ISSN={2511-9044},
   url={http://dx.doi.org/10.1002/qute.202500702},
   DOI={10.1002/qute.202500702},
   number={2},
   journal={Advanced Quantum Technologies},
   publisher={Wiley},
   author={Deneris, Andrew E. and Braccia, Paolo and Bermejo, Pablo and Diaz, N. L. and Mele, Antonio A. and Cerezo, M.},
   year={2026},
   month=feb }

@misc{diaz2025unifiedapproachquantumresource,
      title={A unified approach to quantum resource theories and a new class of free operations}, 
      author={N. L. Diaz and Antonio Anna Mele and Pablo Bermejo and Paolo Braccia and Andrew E. Deneris and Martin Larocca and M. Cerezo},
      year={2025},
      eprint={2507.10851},
      archivePrefix={arXiv},
      primaryClass={quant-ph},
      url={https://arxiv.org/abs/2507.10851}, 
}

@article{liu2024,
  title = {Symmetry Restoration and Quantum Mpemba Effect in Symmetric Random Circuits},
  author = {Liu, Shuo and Zhang, Hao-Kai and Yin, Shuai and Zhang, Shi-Xin},
  journal = {Phys. Rev. Lett.},
  volume = {133},
  issue = {14},
  pages = {140405},
  numpages = {7},
  year = {2024},
  month = {Oct},
  publisher = {American Physical Society},
  doi = {10.1103/PhysRevLett.133.140405},
  url = {https://link.aps.org/doi/10.1103/PhysRevLett.133.140405}
}

@misc{santra2025quantumresourcesnonabelianlattice,
      title={Quantum Resources in Non-Abelian Lattice Gauge Theories: Nonstabilizerness, Multipartite Entanglement, and Fermionic Non-Gaussianity}, 
      author={Gopal Chandra Santra and Julius Mildenberger and Edoardo Ballini and Alberto Bottarelli and Matteo M. Wauters and Philipp Hauke},
      year={2025},
      eprint={2510.07385},
      archivePrefix={arXiv},
      primaryClass={quant-ph},
      url={https://arxiv.org/abs/2510.07385}, 
}

@article{Varikuti2026impactofclifford,
  doi = {10.22331/q-2026-03-10-2017},
  url = {https://doi.org/10.22331/q-2026-03-10-2017},
  title = {Impact of {C}lifford operations on non-stabilizing power and quantum chaos},
  author = {Varikuti, Naga Dileep and Bandyopadhyay, Soumik and Hauke, Philipp},
  journal = {{Quantum}},
  issn = {2521-327X},
  publisher = {{Verein zur F{\"{o}}rderung des Open Access Publizierens in den Quantenwissenschaften}},
  volume = {10},
  pages = {2017},
  month = mar,
  year = {2026}
}

@misc{varikuti2025deepthermalizationmeasurementsquantum,
      title={Deep Thermalization and Measurements of Quantum Resources}, 
      author={Naga Dileep Varikuti and Soumik Bandyopadhyay and Philipp Hauke},
      year={2025},
      eprint={2512.09999},
      archivePrefix={arXiv},
      primaryClass={quant-ph},
      url={https://arxiv.org/abs/2512.09999}, 
}

@misc{varikuti2024quantuminformationscramblingchaos,
      title={Quantum Information Scrambling, Chaos, Sensitivity, and Emergent State Designs}, 
      author={Naga Dileep Varikuti},
      year={2024},
      eprint={2409.10182},
      archivePrefix={arXiv},
      primaryClass={quant-ph},
      url={https://arxiv.org/abs/2409.10182}, 
}

@misc{aditya2025mpembaeffectsquantumcomplexity,
      title={Mpemba Effects in Quantum Complexity}, 
      author={Sreemayee Aditya and Alessandro Summer and Piotr Sierant and Xhek Turkeshi},
      year={2025},
      eprint={2509.22176},
      archivePrefix={arXiv},
      primaryClass={quant-ph},
      url={https://arxiv.org/abs/2509.22176}, 
}

@article{y9r6-dx7p,
  title = {Stabilizer entropy in nonintegrable quantum evolutions},
  author = {Odavi\ifmmode \acute{c}\else \'{c}\fi{}, J. and Viscardi, M. and Hamma, A.},
  journal = {Phys. Rev. B},
  volume = {112},
  issue = {10},
  pages = {104301},
  numpages = {14},
  year = {2025},
  month = {Sep},
  publisher = {American Physical Society},
  doi = {10.1103/y9r6-dx7p},
  url = {https://link.aps.org/doi/10.1103/y9r6-dx7p}
}

@misc{aditya2025growthspreadingquantumresources,
      title={Growth and spreading of quantum resources under random circuit dynamics}, 
      author={Sreemayee Aditya and Xhek Turkeshi and Piotr Sierant},
      year={2025},
      eprint={2512.14827},
      archivePrefix={arXiv},
      primaryClass={quant-ph},
      url={https://arxiv.org/abs/2512.14827}, 
}

@misc{heinrich2025anticoncentrationalmostneed,
      title={Anti-concentration is (almost) all you need}, 
      author={Markus Heinrich and Jonas Haferkamp and Ingo Roth and Jonas Helsen},
      year={2025},
      eprint={2510.23719},
      archivePrefix={arXiv},
      primaryClass={quant-ph},
      url={https://arxiv.org/abs/2510.23719}, 
}

@misc{dowling2025freeindependenceunitarydesign,
      title={Free Independence and Unitary Design from Random Matrix Product Unitaries}, 
      author={Neil Dowling and Jacopo De Nardis and Markus Heinrich and Xhek Turkeshi and Silvia Pappalardi},
      year={2025},
      eprint={2508.00051},
      archivePrefix={arXiv},
      primaryClass={quant-ph},
      url={https://arxiv.org/abs/2508.00051}, 
}

@misc{Anonymous_2026,
      title={Universality in the Anticoncentration of Noisy Quantum Circuits at Finite Depths}, 
      author={Arman Sauliere and Guglielmo Lami and Corentin Boyer and Jacopo De Nardis and Andrea De Luca},
      year={2026},
      eprint={2508.14975},
      archivePrefix={arXiv},
      primaryClass={quant-ph},
      doi={10.1103/xl16-cdy9},
      url={https://arxiv.org/abs/2508.14975}, 
}

@misc{magni2025anticoncentrationstatedesigndoped,
      title={Anticoncentration and State Design of Doped Real Clifford Circuits and Tensor Networks}, 
      author={Beatrice Magni and Markus Heinrich and Lorenzo Leone and Xhek Turkeshi},
      year={2025},
      eprint={2512.15880},
      archivePrefix={arXiv},
      primaryClass={quant-ph},
      url={https://arxiv.org/abs/2512.15880}, 
}

@article{silva,
  title = {Colloquium: Nonequilibrium dynamics of closed interacting quantum systems},
  author = {Polkovnikov, Anatoli and Sengupta, Krishnendu and Silva, Alessandro and Vengalattore, Mukund},
  journal = {Rev. Mod. Phys.},
  volume = {83},
  issue = {3},
  pages = {863--883},
  numpages = {0},
  year = {2011},
  month = {Aug},
  publisher = {American Physical Society},
  doi = {10.1103/RevModPhys.83.863},
  url = {https://link.aps.org/doi/10.1103/RevModPhys.83.863}
}

@article{fisher2023random,
   title={Random Quantum Circuits},
   volume={14},
   ISSN={1947-5462},
   url={http://dx.doi.org/10.1146/annurev-conmatphys-031720-030658},
   DOI={10.1146/annurev-conmatphys-031720-030658},
   number={1},
   journal={Annu. Rev. Condens. Matter Phys.},
   publisher={Annual Reviews},
   author={Fisher, Matthew P.A. and Khemani, Vedika and Nahum, Adam and Vijay, Sagar},
   year={2023},
   month=mar, pages={335–379} }

@article{dalzell2022random,
  title = {Random Quantum Circuits Anticoncentrate in Log Depth},
  author = {Dalzell, Alexander M. and Hunter-Jones, Nicholas and Brand\~ao, Fernando G. S. L.},
  journal = {PRX Quantum},
  volume = {3},
  issue = {1},
  pages = {010333},
  numpages = {43},
  year = {2022},
  month = {Mar},
  publisher = {American Physical Society},
  doi = {10.1103/PRXQuantum.3.010333},
  url = {https://link.aps.org/doi/10.1103/PRXQuantum.3.010333}
}

@article{BKramer1993,
doi = {10.1088/0034-4885/56/12/001},
url = {https://doi.org/10.1088/0034-4885/56/12/001},
year = {1993},
month = {dec},
publisher = {},
volume = {56},
number = {12},
pages = {1469},
author = {B Kramer and A MacKinnon},
title = {Localization: theory and experiment},
journal = {Reports on Progress in Physics}
}

@article{hangleiter2023computational,
  title = {Computational advantage of quantum random sampling},
  author = {Hangleiter, Dominik and Eisert, Jens},
  journal = {Rev. Mod. Phys.},
  volume = {95},
  issue = {3},
  pages = {035001},
  numpages = {82},
  year = {2023},
  month = {Jul},
  publisher = {American Physical Society},
  doi = {10.1103/RevModPhys.95.035001},
  url = {https://link.aps.org/doi/10.1103/RevModPhys.95.035001}
}

@article{mace2019multifractal,
  title = {Multifractal Scalings Across the Many-Body Localization Transition},
  author = {Mac\'e, Nicolas and Alet, Fabien and Laflorencie, Nicolas},
  journal = {Phys. Rev. Lett.},
  volume = {123},
  issue = {18},
  pages = {180601},
  numpages = {6},
  year = {2019},
  month = {Oct},
  publisher = {American Physical Society},
  doi = {10.1103/PhysRevLett.123.180601},
  url = {https://link.aps.org/doi/10.1103/PhysRevLett.123.180601}
}

@article{sierant2022universal,
  title = {Universal Behavior beyond Multifractality of Wave Functions at Measurement-Induced Phase Transitions},
  author = {Sierant, Piotr and Turkeshi, Xhek},
  journal = {Phys. Rev. Lett.},
  volume = {128},
  issue = {13},
  pages = {130605},
  numpages = {7},
  year = {2022},
  month = {Apr},
  publisher = {American Physical Society},
  doi = {10.1103/PhysRevLett.128.130605},
  url = {https://link.aps.org/doi/10.1103/PhysRevLett.128.130605}
}

@article{boixo2018characterizing,
  title = {Characterizing quantum supremacy in near-term devices},
  volume = {14},
  ISSN = {1745-2481},
  url = {http://dx.doi.org/10.1038/s41567-018-0124-x},
  DOI = {10.1038/s41567-018-0124-x},
  number = {6},
  journal = {Nat. Phys.},
  publisher = {Springer Science and Business Media LLC},
  author = {Boixo,  Sergio and Isakov,  Sergei V. and Smelyanskiy,  Vadim N. and Babbush,  Ryan and Ding,  Nan and Jiang,  Zhang and Bremner,  Michael J. and Martinis,  John M. and Neven,  Hartmut},
  year = {2018},
  month = apr,
  pages = {595–600}
}

@article{arute2019quantum,
  title = {Quantum supremacy using a programmable superconducting processor},
  volume = {574},
  ISSN = {1476-4687},
  url = {http://dx.doi.org/10.1038/s41586-019-1666-5},
  DOI = {10.1038/s41586-019-1666-5},
  number = {7779},
  journal = {Nature},
  publisher = {Springer Science and Business Media LLC},
  author = {Arute,  Frank and Arya,  Kunal and others},
  year = {2019},
  month = oct,
  pages = {505–510}
}

@article{fava2025designs,
  title = {Designs via Free Probability},
  author = {Fava, Michele and Kurchan, Jorge and Pappalardi, Silvia},
  journal = {Phys. Rev. X},
  volume = {15},
  issue = {1},
  pages = {011031},
  numpages = {26},
  year = {2025},
  month = {Feb},
  publisher = {American Physical Society},
  doi = {10.1103/PhysRevX.15.011031},
  url = {https://link.aps.org/doi/10.1103/PhysRevX.15.011031}
}

@article{liu2024quantum,
  title = {Quantum algorithms for inverse participation ratio estimation in multiqubit and multiqudit systems},
  author = {Liu, Yingjian and Sierant, Piotr and Stornati, Paolo and Lewenstein, Maciej and P\l{}odzie\ifmmode \acute{n}\else \'{n}\fi{}, Marcin},
  journal = {Phys. Rev. A},
  volume = {111},
  issue = {5},
  pages = {052614},
  numpages = {13},
  year = {2025},
  month = {May},
  publisher = {American Physical Society},
  doi = {10.1103/PhysRevA.111.052614},
  url = {https://link.aps.org/doi/10.1103/PhysRevA.111.052614}
}

@article{turkeshi2024hilbert,
   title={Hilbert Space Delocalization under Random Unitary Circuits},
   volume={26},
   ISSN={1099-4300},
   url={http://dx.doi.org/10.3390/e26060471},
   DOI={10.3390/e26060471},
   number={6},
   journal={Entropy},
   publisher={MDPI AG},
   author={Turkeshi, Xhek and Sierant, Piotr},
   year={2024},
   month=may, pages={471} }

@article{luitz2014universal,
  title = {Universal Behavior beyond Multifractality in Quantum Many-Body Systems},
  author = {Luitz, David J. and Alet, Fabien and Laflorencie, Nicolas},
  journal = {Phys. Rev. Lett.},
  volume = {112},
  issue = {5},
  pages = {057203},
  numpages = {5},
  year = {2014},
  month = {Feb},
  publisher = {American Physical Society},
  doi = {10.1103/PhysRevLett.112.057203},
  url = {https://link.aps.org/doi/10.1103/PhysRevLett.112.057203}
}

@article{luitz2014participation,
  title = {Participation spectroscopy and entanglement Hamiltonian of quantum spin models},
  volume = {2014},
  ISSN = {1742-5468},
  url = {http://dx.doi.org/10.1088/1742-5468/2014/08/P08007},
  DOI = {10.1088/1742-5468/2014/08/p08007},
  number = {8},
  journal = {J. Stat. Mech. Theory Exp.},
  publisher = {IOP Publishing},
  author = {Luitz,  David J and Laflorencie,  Nicolas and Alet,  Fabien},
  year = {2014},
  month = aug,
  pages = {P08007}
}

@misc{hunterjones2019unitary,
      title={Unitary designs from statistical mechanics in random quantum circuits}, 
      author={Nicholas Hunter-Jones},
      year={2019},
      eprint={1905.12053},
      archivePrefix={arXiv},
      primaryClass={quant-ph},
      url={https://arxiv.org/abs/1905.12053}, 
}

@article{bertini2020scrambling,
  title = {Scrambling in random unitary circuits: Exact results},
  author = {Bertini, Bruno and Piroli, Lorenzo},
  journal = {Phys. Rev. B},
  volume = {102},
  issue = {6},
  pages = {064305},
  numpages = {25},
  year = {2020},
  month = {Aug},
  publisher = {American Physical Society},
  doi = {10.1103/PhysRevB.102.064305},
  url = {https://link.aps.org/doi/10.1103/PhysRevB.102.064305}
}

@Article{Chan2022,
author={Chan, Amos
and Shivam, Saumya
and Huse, David A.
and De Luca, Andrea},
title={Many-body quantum chaos and space-time translational invariance},
journal={Nature Commun.},
year={2022},
month={Dec},
day={05},
volume={13},
number={1},
pages={7484},
issn={2041-1723},
doi={10.1038/s41467-022-34318-1},
url={https://doi.org/10.1038/s41467-022-34318-1}
}

@article{christopoulos2024universal,
  title = {Universal distributions of overlaps from generic dynamics in quantum many-body systems},
  author = {Christopoulos, Alexios and Chan, Amos and De Luca, Andrea},
  journal = {Phys. Rev. Res.},
  volume = {7},
  issue = {4},
  pages = {043035},
  numpages = {12},
  year = {2025},
  month = {Oct},
  publisher = {American Physical Society},
  doi = {10.1103/kl64-xnsz},
  url = {https://link.aps.org/doi/10.1103/kl64-xnsz}
}

@article{claeys2025fockspace,
  title = {Fock-Space Delocalization and the Emergence of the Porter-Thomas Distribution from Dual-Unitary Dynamics},
  author = {Claeys, Pieter W. and De Tomasi, Giuseppe},
  journal = {Phys. Rev. Lett.},
  volume = {134},
  issue = {5},
  pages = {050405},
  numpages = {7},
  year = {2025},
  month = {Feb},
  publisher = {American Physical Society},
  doi = {10.1103/PhysRevLett.134.050405},
  url = {https://link.aps.org/doi/10.1103/PhysRevLett.134.050405}
}

@article{mark2024maximum,
  title = {Maximum Entropy Principle in Deep Thermalization and in Hilbert-Space Ergodicity},
  author = {Mark, Daniel K. and Surace, Federica and Elben, Andreas and Shaw, Adam L. and Choi, Joonhee and Refael, Gil and Endres, Manuel and Choi, Soonwon},
  journal = {Phys. Rev. X},
  volume = {14},
  issue = {4},
  pages = {041051},
  numpages = {49},
  year = {2024},
  month = {Nov},
  publisher = {American Physical Society},
  doi = {10.1103/PhysRevX.14.041051},
  url = {https://link.aps.org/doi/10.1103/PhysRevX.14.041051}
}

@article{mark2023benchmarking,
  title = {Benchmarking Quantum Simulators Using Ergodic Quantum Dynamics},
  author = {Mark, Daniel K. and Choi, Joonhee and Shaw, Adam L. and Endres, Manuel and Choi, Soonwon},
  journal = {Phys. Rev. Lett.},
  volume = {131},
  issue = {11},
  pages = {110601},
  numpages = {7},
  year = {2023},
  month = {Sep},
  publisher = {American Physical Society},
  doi = {10.1103/PhysRevLett.131.110601},
  url = {https://link.aps.org/doi/10.1103/PhysRevLett.131.110601}
}

@article{schollwock2011the,
title = {The density-matrix renormalization group in the age of matrix product states},
journal = {Ann. Phys.},
volume = {326},
number = {1},
pages = {96-192},
year = {2011},
doi = {https://doi.org/10.1016/j.aop.2010.09.012},
url = {https://www.sciencedirect.com/science/article/pii/S0003491610001752},
author = {Ulrich Schollwöck},
}

@Article{silvi2019the,
	title={{The Tensor Networks Anthology: Simulation techniques for many-body quantum lattice systems}},
	author={Pietro Silvi and Ferdinand Tschirsich and Matthias Gerster and Johannes Jünemann and Daniel Jaschke and Matteo Rizzi and Simone Montangero},
	journal={SciPost Phys. Lect. Notes},
	pages={8},
	year={2019},
	publisher={SciPost},
	doi={10.21468/SciPostPhysLectNotes.8},
	url={https://scipost.org/10.21468/SciPostPhysLectNotes.8},
}

@article{Haah2025short,
  doi = {10.22331/q-2025-12-11-1940},
  url = {https://doi.org/10.22331/q-2025-12-11-1940},
  title = {Short remarks on shallow unitary circuits},
  author = {Haah, Jeongwan},
  journal = {{Quantum}},
  issn = {2521-327X},
  publisher = {{Verein zur F{\"{o}}rderung des Open Access Publizierens in den Quantenwissenschaften}},
  volume = {9},
  pages = {1940},
  month = dec,
  year = {2025}
}

@article{Yang20tdvp,
  title = {Time-dependent variational principle with ancillary Krylov subspace},
  author = {Yang, Mingru and White, Steven R.},
  journal = {Phys. Rev. B},
  volume = {102},
  issue = {9},
  pages = {094315},
  numpages = {6},
  year = {2020},
  month = {Sep},
  publisher = {American Physical Society},
  doi = {10.1103/PhysRevB.102.094315},
  url = {https://link.aps.org/doi/10.1103/PhysRevB.102.094315}
}

@article{orus2014a,
title = {A practical introduction to tensor networks: Matrix product states and projected entangled pair states},
journal = {Ann. Phys.},
volume = {349},
pages = {117-158},
year = {2014},
issn = {0003-4916},
doi = {https://doi.org/10.1016/j.aop.2014.06.013},
url = {https://www.sciencedirect.com/science/article/pii/S0003491614001596},
author = {Román Orús},}

@article{turkeshi2025pauli,
  title = {Pauli spectrum and nonstabilizerness of typical quantum many-body states},
  author = {Turkeshi, Xhek and Dymarsky, Anatoly and Sierant, Piotr},
  journal = {Phys. Rev. B},
  volume = {111},
  issue = {5},
  pages = {054301},
  numpages = {12},
  year = {2025},
  month = {Feb},
  publisher = {American Physical Society},
  doi = {10.1103/PhysRevB.111.054301},
  url = {https://link.aps.org/doi/10.1103/PhysRevB.111.054301}
}

@article{nahum2017quantum,
  title = {Quantum Entanglement Growth under Random Unitary Dynamics},
  author = {Nahum, Adam and Ruhman, Jonathan and Vijay, Sagar and Haah, Jeongwan},
  journal = {Phys. Rev. X},
  volume = {7},
  issue = {3},
  pages = {031016},
  numpages = {30},
  year = {2017},
  month = {Jul},
  publisher = {American Physical Society},
  doi = {10.1103/PhysRevX.7.031016},
  url = {https://link.aps.org/doi/10.1103/PhysRevX.7.031016}
}

@article{Ezer84,
    author = {Tal‐Ezer, H. and Kosloff, R.},
    title = {An accurate and efficient scheme for propagating the time dependent Schrödinger equation},
    journal = {The Journal of Chemical Physics},
    volume = {81},
    number = {9},
    pages = {3967-3971},
    year = {1984},
    month = {11},
    issn = {0021-9606},
    doi = {10.1063/1.448136}
}

@article{Sierant22chal,
  title = {Challenges to observation of many-body localization},
  author = {Sierant, Piotr and Zakrzewski, Jakub},
  journal = {Phys. Rev. B},
  volume = {105},
  issue = {22},
  pages = {224203},
  numpages = {18},
  year = {2022},
  month = {Jun},
  publisher = {American Physical Society},
  doi = {10.1103/PhysRevB.105.224203},
  url = {https://link.aps.org/doi/10.1103/PhysRevB.105.224203}
}

@article{morvan2024phase,
   title={Phase transitions in random circuit sampling},
   volume={634},
   ISSN={1476-4687},
   url={http://dx.doi.org/10.1038/s41586-024-07998-6},
   DOI={10.1038/s41586-024-07998-6},
   number={8033},
   journal={Nature},
   publisher={Springer Science and Business Media LLC},
   author={Morvan, A. and Villalonga, B. and others},
   year={2024},
   month=oct, pages={328–333} }

@misc{ware2023sharp,
      title={A sharp phase transition in linear cross-entropy benchmarking}, 
      author={Brayden Ware and Abhinav Deshpande and Dominik Hangleiter and Pradeep Niroula and Bill Fefferman and Alexey V. Gorshkov and Michael J. Gullans},
      year={2023},
      eprint={2305.04954},
      archivePrefix={arXiv},
      primaryClass={quant-ph},
      url={https://arxiv.org/abs/2305.04954}, 
}

@article{dalzell2024random,
  title = {Random Quantum Circuits Transform Local Noise into Global White Noise},
  volume = {405},
  ISSN = {1432-0916},
  url = {http://dx.doi.org/10.1007/s00220-024-04958-z},
  DOI = {10.1007/s00220-024-04958-z},
  pages={78},
  number = {3},
  journal = {Commun. Math. Phys.},
  publisher = {Springer Science and Business Media LLC},
  author = {Dalzell,  Alexander M. and Hunter-Jones,  Nicholas and Brandao,  Fernando G. S. L.},
  year = {2024},
  month = mar 
}

@article{Sierant26fermionic,
  title = {Fermionic Magic Resources of Quantum Many-Body Systems},
  author = {Sierant, Piotr and Stornati, Paolo and Turkeshi, Xhek},
  journal = {PRX Quantum},
  volume = {7},
  issue = {1},
  pages = {010302},
  numpages = {41},
  year = {2026},
  month = {Jan},
  publisher = {American Physical Society},
  doi = {10.1103/3yx4-1j27},
  url = {https://link.aps.org/doi/10.1103/3yx4-1j27}
}

@misc{Falco2026fermionic,
      title={Fermionic magic resources in disordered quantum spin chains}, 
      author={Pedro R. Nicácio Falcão and Jakub Zakrzewski and Piotr Sierant},
      year={2026},
      eprint={2602.00245},
      archivePrefix={arXiv},
      primaryClass={quant-ph},
      url={https://arxiv.org/abs/2602.00245}, 
}

@article{liu2022manybody,
  title = {Many-Body Quantum Magic},
  author = {Liu, Zi-Wen and Winter, Andreas},
  journal = {PRX Quantum},
  volume = {3},
  issue = {2},
  pages = {020333},
  numpages = {18},
  year = {2022},
  month = {May},
  publisher = {American Physical Society},
  doi = {10.1103/PRXQuantum.3.020333},
  url = {https://link.aps.org/doi/10.1103/PRXQuantum.3.020333}
}

@article{chitambar2019quantum,
  title = {Quantum resource theories},
  author = {Chitambar, Eric and Gour, Gilad},
  journal = {Rev. Mod. Phys.},
  volume = {91},
  issue = {2},
  pages = {025001},
  numpages = {48},
  year = {2019},
  month = {Apr},
  publisher = {American Physical Society},
  doi = {10.1103/RevModPhys.91.025001},
  url = {https://link.aps.org/doi/10.1103/RevModPhys.91.025001}
}

@article{bravyi2005universal,
  title = {Universal quantum computation with ideal Clifford gates and noisy ancillas},
  author = {Bravyi, Sergey and Kitaev, Alexei},
  journal = {Phys. Rev. A},
  volume = {71},
  issue = {2},
  pages = {022316},
  numpages = {14},
  year = {2005},
  month = {Feb},
  publisher = {American Physical Society},
  doi = {10.1103/PhysRevA.71.022316},
  url = {https://link.aps.org/doi/10.1103/PhysRevA.71.022316}
}

@article{gottesman1998theory,
  title = {Theory of fault-tolerant quantum computation},
  author = {Gottesman, Daniel},
  journal = {Phys. Rev. A},
  volume = {57},
  issue = {1},
  pages = {127--137},
  numpages = {0},
  year = {1998},
  month = {Jan},
  publisher = {American Physical Society},
  doi = {10.1103/PhysRevA.57.127},
  url = {https://link.aps.org/doi/10.1103/PhysRevA.57.127}
}

@article{aaronson2004improved,
  title = {Improved simulation of stabilizer circuits},
  author = {Aaronson, Scott and Gottesman, Daniel},
  journal = {Phys. Rev. A},
  volume = {70},
  issue = {5},
  pages = {052328},
  numpages = {14},
  year = {2004},
  month = {Nov},
  publisher = {American Physical Society},
  doi = {10.1103/PhysRevA.70.052328},
  url = {https://link.aps.org/doi/10.1103/PhysRevA.70.052328}
}

@article{zhou2019emergent,
  title = {Emergent statistical mechanics of entanglement in random unitary circuits},
  author = {Zhou, Tianci and Nahum, Adam},
  journal = {Phys. Rev. B},
  volume = {99},
  issue = {17},
  pages = {174205},
  numpages = {28},
  year = {2019},
  month = {May},
  publisher = {American Physical Society},
  doi = {10.1103/PhysRevB.99.174205},
  url = {https://link.aps.org/doi/10.1103/PhysRevB.99.174205}
}

@article{zhou2020entanglement,
  title = {Entanglement Membrane in Chaotic Many-Body Systems},
  author = {Zhou, Tianci and Nahum, Adam},
  journal = {Phys. Rev. X},
  volume = {10},
  issue = {3},
  pages = {031066},
  numpages = {37},
  year = {2020},
  month = {Sep},
  publisher = {American Physical Society},
  doi = {10.1103/PhysRevX.10.031066},
  url = {https://link.aps.org/doi/10.1103/PhysRevX.10.031066}
}

@article{leone2022stabilizer,
  title = {Stabilizer R\'enyi Entropy},
  author = {Leone, Lorenzo and Oliviero, Salvatore F. E. and Hamma, Alioscia},
  journal = {Phys. Rev. Lett.},
  volume = {128},
  issue = {5},
  pages = {050402},
  numpages = {5},
  year = {2022},
  month = {Feb},
  publisher = {American Physical Society},
  doi = {10.1103/PhysRevLett.128.050402},
  url = {https://link.aps.org/doi/10.1103/PhysRevLett.128.050402}
}

@article{sierant2023ent,
  title = {Entanglement Growth and Minimal Membranes in ($d+1$) Random Unitary Circuits},
  author = {Sierant, Piotr and Schir\`o, Marco and Lewenstein, Maciej and Turkeshi, Xhek},
  journal = {Phys. Rev. Lett.},
  volume = {131},
  issue = {23},
  pages = {230403},
  numpages = {7},
  year = {2023},
  month = {Dec},
  publisher = {American Physical Society},
  doi = {10.1103/PhysRevLett.131.230403},
  url = {https://link.aps.org/doi/10.1103/PhysRevLett.131.230403}
}

@article{sauliere2025universalityanticoncentrationchaoticquantum,
  title = {Universality in the anticoncentration of chaotic quantum circuits},
  author = {Sauliere, Arman and Magni, Beatrice and Lami, Guglielmo and Turkeshi, Xhek and De Nardis, Jacopo},
  journal = {Phys. Rev. B},
  volume = {112},
  issue = {13},
  pages = {134312},
  numpages = {14},
  year = {2025},
  month = {Oct},
  publisher = {American Physical Society},
  doi = {10.1103/lkwg-4dbt},
  url = {https://link.aps.org/doi/10.1103/lkwg-4dbt}
}

@article{chan2018,
  title = {Solution of a Minimal Model for Many-Body Quantum Chaos},
  author = {Chan, Amos and De Luca, Andrea and Chalker, J. T.},
  journal = {Phys. Rev. X},
  volume = {8},
  issue = {4},
  pages = {041019},
  numpages = {17},
  year = {2018},
  month = {Nov},
  publisher = {American Physical Society},
  doi = {10.1103/PhysRevX.8.041019},
  url = {https://link.aps.org/doi/10.1103/PhysRevX.8.041019}
}

@misc{grevink2025glueshortdepthdesignsunitary,
      title={Will it glue? On short-depth designs beyond the unitary group}, 
      author={Lorenzo Grevink and Jonas Haferkamp and Markus Heinrich and Jonas Helsen and Marcel Hinsche and Thomas Schuster and Zoltán Zimborás},
      year={2025},
      eprint={2506.23925},
      archivePrefix={arXiv},
      primaryClass={quant-ph},
      url={https://arxiv.org/abs/2506.23925}, 
}

@article{turkeshi2025quantum,
  title = {Quantum Mpemba Effect in Random Circuits},
  author = {Turkeshi, Xhek and Calabrese, Pasquale and De Luca, Andrea},
  journal = {Phys. Rev. Lett.},
  volume = {135},
  issue = {4},
  pages = {040403},
  numpages = {10},
  year = {2025},
  month = {Jul},
  publisher = {American Physical Society},
  doi = {10.1103/5d6p-8d1b},
  url = {https://link.aps.org/doi/10.1103/5d6p-8d1b}
}

@article{ippoliti2023dynamical,
  title = {Dynamical Purification and the Emergence of Quantum State Designs from the Projected Ensemble},
  author = {Ippoliti, Matteo and Ho, Wen Wei},
  journal = {PRX Quantum},
  volume = {4},
  issue = {3},
  pages = {030322},
  numpages = {28},
  year = {2023},
  month = {Aug},
  publisher = {American Physical Society},
  doi = {10.1103/PRXQuantum.4.030322},
  url = {https://link.aps.org/doi/10.1103/PRXQuantum.4.030322}
}

@article{backer2019multifractal,
  title = {Multifractal dimensions for random matrices, chaotic quantum maps, and many-body systems},
  author = {B\"acker, Arnd and Haque, Masudul and Khaymovich, Ivan M.},
  journal = {Phys. Rev. E},
  volume = {100},
  issue = {3},
  pages = {032117},
  numpages = {14},
  year = {2019},
  month = {Sep},
  publisher = {American Physical Society},
  doi = {10.1103/PhysRevE.100.032117},
  url = {https://link.aps.org/doi/10.1103/PhysRevE.100.032117}
}

@article{Haferkamp2022randomquantum,
  doi = {10.22331/q-2022-09-08-795},
  url = {https://doi.org/10.22331/q-2022-09-08-795},
  title = {Random quantum circuits are approximate unitary {$t$}-designs in depth {$O\left(nt^{5+o(1)}\right)$}},
  author = {Haferkamp, Jonas},
  journal = {{Quantum}},
  issn = {2521-327X},
  publisher = {{Verein zur F{\"{o}}rderung des Open Access Publizierens in den Quantenwissenschaften}},
  volume = {6},
  pages = {795},
  month = sep,
  year = {2022}
}

@article{collins2006integration,
  title = {Integration with Respect to the Haar Measure on Unitary,  Orthogonal and Symplectic Group},
  volume = {264},
  ISSN = {1432-0916},
  url = {http://dx.doi.org/10.1007/s00220-006-1554-3},
  DOI = {10.1007/s00220-006-1554-3},
  number = {3},
  journal = {Commun. Math. Phys.},
  publisher = {Springer Science and Business Media LLC},
  author = {Collins,  Benoît and Śniady,  Piotr},
  year = {2006},
  month = mar,
  pages = {773–795}
}

@article{haferkamp2022,
   title={Efficient Unitary Designs with a System-Size Independent Number of Non-Clifford Gates},
   volume={397},
   ISSN={1432-0916},
   url={http://dx.doi.org/10.1007/s00220-022-04507-6},
   DOI={10.1007/s00220-022-04507-6},
   number={3},
   journal={Commun. Math. Phys.},
   publisher={Springer Science and Business Media LLC},
   author={Haferkamp, J. and Montealegre-Mora, F. and Heinrich, M. and Eisert, J. and Gross, D. and Roth, I.},
   year={2022},
   month=nov, pages={995–1041} }

@misc{zhu2016cliffordgroupfailsgracefully,
      title={The Clifford group fails gracefully to be a unitary 4-design}, 
      author={Huangjun Zhu and Richard Kueng and Markus Grassl and David Gross},
      year={2016},
      eprint={1609.08172},
      archivePrefix={arXiv},
      primaryClass={quant-ph},
      url={https://arxiv.org/abs/1609.08172}, 
}

@article{Huang2020predicting,
   title={Predicting many properties of a quantum system from very few measurements},
   volume={16},
   ISSN={1745-2481},
   url={http://dx.doi.org/10.1038/s41567-020-0932-7},
   DOI={10.1038/s41567-020-0932-7},
   number={10},
   journal={Nat. Phys.},
   publisher={Springer Science and Business Media LLC},
   author={Huang, Hsin-Yuan and Kueng, Richard and Preskill, John},
   year={2020},
   month=jun, pages={1050–1057} }

@article{Leone2021quantumchaosis,
  doi = {10.22331/q-2021-05-04-453},
  url = {https://doi.org/10.22331/q-2021-05-04-453},
  title = {Quantum {C}haos is {Q}uantum},
  author = {Leone, Lorenzo and Oliviero, Salvatore F. E. and Zhou, You and Hamma, Alioscia},
  journal = {{Quantum}},
  issn = {2521-327X},
  publisher = {{Verein zur F{\"{o}}rderung des Open Access Publizierens in den Quantenwissenschaften}},
  volume = {5},
  pages = {453},
  month = may,
  year = {2021}
}

@article{tirrito2024anticoncentrationmagicspreadingergodic,
  title   = {Anticoncentration and nonstabilizerness spreading under ergodic quantum dynamics},
  author  = {Tirrito, Emanuele and Turkeshi, Xhek and Sierant, Piotr},
  journal = {Phys. Rev. Lett.},
  volume  = {135},
  number  = {22},
  pages   = {220401},
  year    = {2025},
  doi     = {10.1103/1jzy-sk9r},
}

@article{Hangleiter2018anticoncentration,
  doi = {10.22331/q-2018-05-22-65},
  url = {https://doi.org/10.22331/q-2018-05-22-65},
  title = {Anticoncentration theorems for schemes showing a quantum speedup},
  author = {Hangleiter, Dominik and Bermejo-Vega, Juan and Schwarz, Martin and Eisert, Jens},
  journal = {{Quantum}},
  issn = {2521-327X},
  publisher = {{Verein zur F{\"{o}}rderung des Open Access Publizierens in den Quantenwissenschaften}},
  volume = {2},
  pages = {65},
  month = may,
  year = {2018}
}

@article{Bouland2018on,
  title = {On the complexity and verification of quantum random circuit sampling},
  volume = {15},
  ISSN = {1745-2481},
  url = {http://dx.doi.org/10.1038/s41567-018-0318-2},
  DOI = {10.1038/s41567-018-0318-2},
  number = {2},
  journal = {Nat. Phys.},
  publisher = {Springer Science and Business Media LLC},
  author = {Bouland,  Adam and Fefferman,  Bill and Nirkhe,  Chinmay and Vazirani,  Umesh},
  year = {2018},
  month = oct,
  pages = {159–163}
}

@article{nahum2018operator,
  title = {Operator Spreading in Random Unitary Circuits},
  author = {Nahum, Adam and Vijay, Sagar and Haah, Jeongwan},
  journal = {Phys. Rev. X},
  volume = {8},
  issue = {2},
  pages = {021014},
  numpages = {30},
  year = {2018},
  month = {Apr},
  publisher = {American Physical Society},
  doi = {10.1103/PhysRevX.8.021014},
  url = {https://link.aps.org/doi/10.1103/PhysRevX.8.021014}
}

@misc{fefferman2024anticoncentrationunitaryhaarmeasure,
      title={Anti-Concentration for the Unitary Haar Measure and Applications to Random Quantum Circuits}, 
      author={Bill Fefferman and Soumik Ghosh and Wei Zhan},
      year={2024},
      eprint={2407.19561},
      archivePrefix={arXiv},
      primaryClass={quant-ph},
      url={https://arxiv.org/abs/2407.19561}, 
}

@article{Keyserlingk2018operator,
  title = {Operator Hydrodynamics, OTOCs, and Entanglement Growth in Systems without Conservation Laws},
  author = {von Keyserlingk, C. W. and Rakovszky, Tibor and Pollmann, Frank and Sondhi, S. L.},
  journal = {Phys. Rev. X},
  volume = {8},
  issue = {2},
  pages = {021013},
  numpages = {19},
  year = {2018},
  month = {Apr},
  publisher = {American Physical Society},
  doi = {10.1103/PhysRevX.8.021013},
  url = {https://link.aps.org/doi/10.1103/PhysRevX.8.021013}
}

@article{fefferman2024effect,
  title = {Effect of Nonunital Noise on Random-Circuit Sampling},
  author = {Fefferman, Bill and Ghosh, Soumik and Gullans, Michael and Kuroiwa, Kohdai and Sharma, Kunal},
  journal = {PRX Quantum},
  volume = {5},
  issue = {3},
  pages = {030317},
  numpages = {41},
  year = {2024},
  month = {Jul},
  publisher = {American Physical Society},
  doi = {10.1103/PRXQuantum.5.030317},
  url = {https://link.aps.org/doi/10.1103/PRXQuantum.5.030317}
}

@article{Braccia2024computing,
   title={Computing exact moments of local random quantum circuits via tensor networks},
   volume={6},
   ISSN={2524-4914},
   url={http://dx.doi.org/10.1007/s42484-024-00187-8},
   DOI={10.1007/s42484-024-00187-8},
   number={2},
   pages = {54},
   journal={Quantum Mach. Intell.},
   publisher={Springer Science and Business Media LLC},
   author={Braccia, Paolo and Bermejo, Pablo and Cincio, Lukasz and Cerezo, M.},
   year={2024},
   month=sep }

@article{Lumia24,
  title = {Measurement-induced transitions beyond Gaussianity: A single particle description},
  author = {Lumia, Luca and Tirrito, Emanuele and Fazio, Rosario and Collura, Mario},
  journal = {Phys. Rev. Res.},
  volume = {6},
  issue = {2},
  pages = {023176},
  numpages = {10},
  year = {2024},
  month = {May},
  publisher = {American Physical Society},
  doi = {10.1103/PhysRevResearch.6.023176},
  url = {https://link.aps.org/doi/10.1103/PhysRevResearch.6.023176}
}

@article{collura2025quantummagicfermionicgaussian,
  doi = {10.22331/q-2026-03-23-2036},
  url = {https://doi.org/10.22331/q-2026-03-23-2036},
  title = {The non-stabilizerness of fermionic {G}aussian states},
  author = {Collura, Mario and Nardis, Jacopo De and Alba, Vincenzo and Lami, Guglielmo},
  journal = {{Quantum}},
  issn = {2521-327X},
  publisher = {{Verein zur F{\"{o}}rderung des Open Access Publizierens in den Quantenwissenschaften}},
  volume = {10},
  pages = {2036},
  month = mar,
  year = {2026}
}

@misc{gottesman1997stabilizer,
      title={Stabilizer Codes and Quantum Error Correction}, 
      author={Daniel Gottesman},
      year={1997},
      eprint={quant-ph/9705052},
      archivePrefix={arXiv},
      primaryClass={quant-ph},
      url={https://arxiv.org/abs/quant-ph/9705052}, 
}

@article{schuster2025randomunitariesextremelylow,
  title = {Random unitaries in extremely low depth},
  volume = {389},
  ISSN = {1095-9203},
  url = {http://dx.doi.org/10.1126/science.adv8590},
  DOI = {10.1126/science.adv8590},
  number = {6755},
  journal = {Science},
  publisher = {American Association for the Advancement of Science (AAAS)},
  author = {Schuster,  Thomas and Haferkamp,  Jonas and Huang,  Hsin-Yuan},
  year = {2025},
  month = July,
  pages = {92–96}
}

@Article{itensor,
	title={{The ITensor Software Library for Tensor Network Calculations}},
	author={Matthew Fishman and Steven R. White and E. Miles Stoudenmire},
	journal={SciPost Phys. Codebases},
	pages={4},
	year={2022},
	publisher={SciPost},
	doi={10.21468/SciPostPhysCodeb.4},
	url={https://scipost.org/10.21468/SciPostPhysCodeb.4},
}

@article{lami2025anticoncentration,
  title = {Anticoncentration and State Design of Random Tensor Networks},
  author = {Lami, Guglielmo and De Nardis, Jacopo and Turkeshi, Xhek},
  journal = {Phys. Rev. Lett.},
  volume = {134},
  issue = {1},
  pages = {010401},
  numpages = {8},
  year = {2025},
  month = {Jan},
  publisher = {American Physical Society},
  doi = {10.1103/PhysRevLett.134.010401},
  url = {https://link.aps.org/doi/10.1103/PhysRevLett.134.010401}
}

@article{Hangleiter2024bell,
  title = {Bell Sampling from Quantum Circuits},
  author = {Hangleiter, Dominik and Gullans, Michael J.},
  journal = {Phys. Rev. Lett.},
  volume = {133},
  issue = {2},
  pages = {020601},
  numpages = {7},
  year = {2024},
  month = {Jul},
  publisher = {American Physical Society},
  doi = {10.1103/PhysRevLett.133.020601},
  url = {https://link.aps.org/doi/10.1103/PhysRevLett.133.020601}
}

@article{baumgratz2014quantifying,
  title = {Quantifying Coherence},
  author = {Baumgratz, T. and Cramer, M. and Plenio, M. B.},
  journal = {Phys. Rev. Lett.},
  volume = {113},
  issue = {14},
  pages = {140401},
  year = {2014},
  month = {Sep},
  publisher = {American Physical Society},
  doi = {10.1103/PhysRevLett.113.140401}
}

@article{streltsov2017colloquium,
  title = {Colloquium: Quantum coherence as a resource},
  author = {Streltsov, Alexander and Adesso, Gerardo and Plenio, Martin B.},
  journal = {Rev. Mod. Phys.},
  volume = {89},
  issue = {4},
  pages = {041003},
  year = {2017},
  month = {Oct},
  publisher = {American Physical Society},
  doi = {10.1103/RevModPhys.89.041003}
}

@article{rakovszky2018diffusive,
  title = {Diffusive Hydrodynamics of Out-of-Time-Ordered Correlators with Charge Conservation},
  author = {Rakovszky, Tibor and Pollmann, Frank and von Keyserlingk, C. W.},
  journal = {Phys. Rev. X},
  volume = {8},
  issue = {3},
  pages = {031058},
  year = {2018},
  month = {Sep},
  publisher = {American Physical Society},
  doi = {10.1103/PhysRevX.8.031058}
}

@article{khemani2018operator,
  title = {Operator Spreading and the Emergence of Dissipative Hydrodynamics under Unitary Evolution with Conservation Laws},
  author = {Khemani, Vedika and Vishwanath, Ashvin and Huse, David A.},
  journal = {Phys. Rev. X},
  volume = {8},
  issue = {3},
  pages = {031057},
  year = {2018},
  month = {Sep},
  publisher = {American Physical Society},
  doi = {10.1103/PhysRevX.8.031057}
}

@article{rakovszky2019sub,
  title = {Sub-ballistic Growth of {R}\'enyi Entropies due to Diffusion},
  author = {Rakovszky, Tibor and Pollmann, Frank and von Keyserlingk, C. W.},
  journal = {Phys. Rev. Lett.},
  volume = {122},
  issue = {25},
  pages = {250602},
  year = {2019},
  month = {Jun},
  publisher = {American Physical Society},
  doi = {10.1103/PhysRevLett.122.250602}
}

@article{huang2020dynamics,
  title = {Dynamics of {R}\'enyi entanglement entropy in diffusive qudit systems},
  author = {Huang, Yichen},
  journal = {IOP SciNotes},
  volume = {1},
  pages = {035205},
  year = {2020},
  doi = {10.1088/2633-1357/abd1e2}
}

@article{znidaric2020entanglement,
  title = {Entanglement growth in diffusive systems},
  author = {Žnidari\v{c}, Marko},
  journal = {Commun. Phys.},
  volume = {3},
  pages = {100},
  year = {2020},
  doi = {10.1038/s42005-020-0366-7}
}

@article{pai2019localization,
  title = {Localization in Fractonic Random Circuits},
  author = {Pai, Shriya and Pretko, Michael and Nandkishore, Rahul M.},
  journal = {Phys. Rev. X},
  volume = {9},
  issue = {2},
  pages = {021003},
  year = {2019},
  month = {Apr},
  publisher = {American Physical Society},
  doi = {10.1103/PhysRevX.9.021003}
}

@article{sala2020ergodicity,
  title = {Ergodicity Breaking Arising from Hilbert Space Fragmentation in Dipole-Conserving Hamiltonians},
  author = {Sala, Pablo and Rakovszky, Tibor and Verresen, Ruben and Knap, Michael and Pollmann, Frank},
  journal = {Phys. Rev. X},
  volume = {10},
  issue = {1},
  pages = {011047},
  year = {2020},
  month = {Feb},
  publisher = {American Physical Society},
  doi = {10.1103/PhysRevX.10.011047}
}

@article{khemani2020localization,
  title = {Localization from Hilbert space shattering: From theory to physical realizations},
  author = {Khemani, Vedika and Hermele, Michael and Nandkishore, Rahul},
  journal = {Phys. Rev. B},
  volume = {101},
  issue = {17},
  pages = {174204},
  year = {2020},
  month = {May},
  publisher = {American Physical Society},
  doi = {10.1103/PhysRevB.101.174204}
}

@article{moudgalya2022quantum,
  title = {Quantum many-body scars and {H}ilbert space fragmentation: a review of exact results},
  author = {Moudgalya, Sanjay and Bernevig, B. Andrei and Regnault, Nicolas},
  journal = {Rep. Prog. Phys.},
  volume = {85},
  pages = {086501},
  year = {2022},
  doi = {10.1088/1361-6633/ac73a0}
}

@article{feldmeier2020anomalous,
  title = {Anomalous Diffusion in Dipole- and Higher-Moment-Conserving Systems},
  author = {Feldmeier, Johannes and Sala, Pablo and De Tomasi, Giuseppe and Pollmann, Frank and Knap, Michael},
  journal = {Phys. Rev. Lett.},
  volume = {125},
  issue = {24},
  pages = {245303},
  year = {2020},
  month = {Dec},
  publisher = {American Physical Society},
  doi = {10.1103/PhysRevLett.125.245303}
}

@article{iaconis2021multipole,
  title = {Multipole Conservation Laws and Subdiffusion in Any Dimension},
  author = {Iaconis, Jason and Lucas, Andrew and Nandkishore, Rahul},
  journal = {Phys. Rev. E},
  volume = {103},
  pages = {022142},
  year = {2021},
  doi = {10.1103/PhysRevE.103.022142}
}

@article{morningstar2020kinetically,
  title = {Kinetically constrained freezing transition in a dipole-conserving system},
  author = {Morningstar, Alan and Khemani, Vedika and Huse, David A.},
  journal = {Phys. Rev. B},
  volume = {101},
  issue = {21},
  pages = {214205},
  year = {2020},
  month = {Jun},
  publisher = {American Physical Society},
  doi = {10.1103/PhysRevB.101.214205}
}

@article{pretko2017subdimensional,
  title = {Subdimensional particle structure of higher rank {U(1)} spin liquids},
  author = {Pretko, Michael},
  journal = {Phys. Rev. B},
  volume = {95},
  issue = {11},
  pages = {115139},
  year = {2017},
  month = {Mar},
  publisher = {American Physical Society},
  doi = {10.1103/PhysRevB.95.115139}
}

@article{nandkishore2019fractons,
  title = {Fractons},
  author = {Nandkishore, Rahul M. and Hermele, Michael},
  journal = {Annu. Rev. Condens. Matter Phys.},
  volume = {10},
  pages = {295},
  year = {2019},
  doi = {10.1146/annurev-conmatphys-031218-013604}
}

@article{kim2013ballistic,
  title = {Ballistic Spreading of Entanglement in a Diffusive Nonintegrable System},
  author = {Kim, Hyungwon and Huse, David A.},
  journal = {Phys. Rev. Lett.},
  volume = {111},
  issue = {12},
  pages = {127205},
  year = {2013},
  month = {Sep},
  publisher = {American Physical Society},
  doi = {10.1103/PhysRevLett.111.127205}
}

@article{kim2014testing,
  title = {Testing Whether All Eigenstates Obey the Eigenstate Thermalization Hypothesis},
  author = {Kim, Hyungwon and Ikeda, Tatsuhiko N. and Huse, David A.},
  journal = {Phys. Rev. E},
  volume = {90},
  pages = {052105},
  year = {2014},
  doi = {10.1103/PhysRevE.90.052105}
}

@article{dalessio2016quantum,
  title = {From quantum chaos and eigenstate thermalization to statistical mechanics and thermodynamics},
  author = {D'Alessio, Luca and Kafri, Yariv and Polkovnikov, Anatoli and Rigol, Marcos},
  journal = {Adv. Phys.},
  volume = {65},
  pages = {239},
  year = {2016},
  doi = {10.1080/00018732.2016.1198134}
}

@article{haegeman2011time,
  title = {Time-Dependent Variational Principle for Quantum Lattices},
  author = {Haegeman, Jutho and Cirac, J. Ignacio and Osborne, Tobias J. and Pi\v{z}orn, Iztok and Verschelde, Henri and Verstraete, Frank},
  journal = {Phys. Rev. Lett.},
  volume = {107},
  issue = {7},
  pages = {070601},
  year = {2011},
  month = {Aug},
  publisher = {American Physical Society},
  doi = {10.1103/PhysRevLett.107.070601}
}

@article{haegeman2016unifying,
  title = {Unifying time evolution and optimization with matrix product states},
  author = {Haegeman, Jutho and Lubich, Christian and Oseledets, Ivan and Vandereycken, Bart and Verstraete, Frank},
  journal = {Phys. Rev. B},
  volume = {94},
  pages = {165116},
  year = {2016},
  doi = {10.1103/PhysRevB.94.165116}
}

@article{Jonay24slow,
  title = {Slow thermalization and subdiffusion in $U(1)$ conserving Floquet random circuits},
  author = {Jonay, Cheryne and Rodriguez-Nieva, Joaquin F. and Khemani, Vedika},
  journal = {Phys. Rev. B},
  volume = {109},
  issue = {2},
  pages = {024311},
  numpages = {11},
  year = {2024},
  month = {Jan},
  publisher = {American Physical Society},
  doi = {10.1103/PhysRevB.109.024311},
  url = {https://link.aps.org/doi/10.1103/PhysRevB.109.024311}
}

@article{prosen2007chaos,
  title = {Chaos and complexity of quantum motion},
  author = {Prosen, Tomaž},
  journal = {J. Phys. A: Math. Theor.},
  volume = {40},
  pages = {7881},
  year = {2007},
  doi = {10.1088/1751-8113/40/28/S02}
}

@article{bertini2019exact,
  title = {Exact Spectral Form Factor in a Minimal Model of Many-Body Quantum Chaos},
  author = {Bertini, Bruno and Kos, Pavel and Prosen, Toma\v{z}},
  journal = {Phys. Rev. Lett.},
  volume = {121},
  issue = {26},
  pages = {264101},
  year = {2018},
  month = {Dec},
  publisher = {American Physical Society},
  doi = {10.1103/PhysRevLett.121.264101}
}

@article{deutsch1991quantum,
  title = {Quantum statistical mechanics in a closed system},
  author = {Deutsch, J. M.},
  journal = {Phys. Rev. A},
  volume = {43},
  pages = {2046},
  year = {1991},
  doi = {10.1103/PhysRevA.43.2046}
}

@article{srednicki1994chaos,
  title = {Chaos and quantum thermalization},
  author = {Srednicki, Mark},
  journal = {Phys. Rev. E},
  volume = {50},
  pages = {888},
  year = {1994},
  doi = {10.1103/PhysRevE.50.888}
}

@article{rigol2008thermalization,
  title = {Thermalization and its mechanism for generic isolated quantum systems},
  author = {Rigol, Marcos and Dunjko, Vanja and Olshanii, Maxim},
  journal = {Nature},
  volume = {452},
  pages = {854},
  year = {2008},
  doi = {10.1038/nature06838}
}

@misc{heinrich2026criticalbehaviorsmagicparticipation,
      title={Critical behaviors of magic and participation entropy at measurement induced phase transitions}, 
      author={Eliot Heinrich and Hanchen Liu and Tianci Zhou and Xiao Chen},
      year={2026},
      eprint={2603.12626},
      archivePrefix={arXiv},
      primaryClass={quant-ph},
      url={https://arxiv.org/abs/2603.12626}, 
}

@Article{10.21468/SciPostPhys.15.6.250,
	title={{Coherence requirements for quantum communication from hybrid circuit dynamics}},
	author={Shane P. Kelly and Ulrich Poschinger and Ferdinand Schmidt-Kaler and Matthew P. A. Fisher and Jamir Marino},
	journal={SciPost Phys.},
	volume={15},
	pages={250},
	year={2023},
	publisher={SciPost},
	doi={10.21468/SciPostPhys.15.6.250},
	url={https://scipost.org/10.21468/SciPostPhys.15.6.250},
}

@article{turkeshi2025magic,
  title = {Magic Spreading in Random Quantum Circuits},
  author = {Turkeshi, Xhek and Tirrito, Emanuele and Sierant, Piotr},
  journal = {Nat. Commun.},
  volume = {16},
  pages = {2575},
  year = {2025},
  doi = {10.1038/s41467-025-57704-x}
}

@article{Singh2021subdiffusion,
  title   = {Subdiffusion and Many-Body Quantum Chaos with Kinetic Constraints},
  author  = {Singh, Hansveer and Ware, Brayden A. and Vasseur, Romain and Friedman, Aaron J.},
  journal = {Phys. Rev. Lett.},
  volume  = {127},
  number  = {23},
  pages   = {230602},
  year    = {2021},
  month   = dec,
  publisher = {American Physical Society},
  doi     = {10.1103/PhysRevLett.127.230602},
}

@article{Aditya2024subspace,
  title = {Subspace-restricted thermalization in a correlated-hopping model with strong Hilbert space fragmentation characterized by irreducible strings},
  author = {Aditya, Sreemayee and Dhar, Deepak and Sen, Diptiman},
  journal = {Phys. Rev. B},
  volume = {110},
  issue = {4},
  pages = {045418},
  numpages = {19},
  year = {2024},
  month = {Jul},
  publisher = {American Physical Society},
  doi = {10.1103/PhysRevB.110.045418},
  url = {https://link.aps.org/doi/10.1103/PhysRevB.110.045418}
}

@article{Ganguli2025East,
  title = {Aspects of Hilbert space fragmentation in the quantum East model: Fragmentation, subspace-restricted quantum scars, and effects of density-density interactions},
  author = {Ganguli, Maitri and Aditya, Sreemayee and Sen, Diptiman},
  journal = {Phys. Rev. B},
  volume = {111},
  issue = {4},
  pages = {045411},
  numpages = {25},
  year = {2025},
  month = {Jan},
  publisher = {American Physical Society},
  doi = {10.1103/PhysRevB.111.045411},
  url = {https://link.aps.org/doi/10.1103/PhysRevB.111.045411}
}

@article{Aditya2025East2,
  title = {Diagnostics of Hilbert space fragmentation, freezing transition, and its effects in the family of quantum East models involving varying range of constraints},
  author = {Aditya, Sreemayee},
  journal = {Phys. Rev. B},
  volume = {112},
  issue = {19},
  pages = {195413},
  numpages = {31},
  year = {2025},
  month = {Nov},
  publisher = {American Physical Society},
  doi = {10.1103/7j6x-74f1},
  url = {https://link.aps.org/doi/10.1103/7j6x-74f1}
}

@article{Burchards2022coupledhydro,
  title = {Coupled hydrodynamics in dipole-conserving quantum systems},
  author = {Burchards, Ansgar G. and Feldmeier, Johannes and Schuckert, Alexander and Knap, Michael},
  journal = {Phys. Rev. B},
  volume = {105},
  issue = {20},
  pages = {205127},
  numpages = {14},
  year = {2022},
  month = {May},
  publisher = {American Physical Society},
  doi = {10.1103/PhysRevB.105.205127},
  url = {https://link.aps.org/doi/10.1103/PhysRevB.105.205127}
}

@article{Feldmeier2021criticalslow,
  title = {Critically Slow Operator Dynamics in Constrained Many-Body Systems},
  author = {Feldmeier, Johannes and Knap, Michael},
  journal = {Phys. Rev. Lett.},
  volume = {127},
  issue = {23},
  pages = {235301},
  numpages = {6},
  year = {2021},
  month = {Dec},
  publisher = {American Physical Society},
  doi = {10.1103/PhysRevLett.127.235301},
  url = {https://link.aps.org/doi/10.1103/PhysRevLett.127.235301}
}

@article{saxena2020coherence,
  title = {Dynamical resource theory of quantum coherence},
  author = {Saxena, Gaurav and Chitambar, Eric and Gour, Gilad},
  journal = {Phys. Rev. Res.},
  volume = {2},
  issue = {2},
  pages = {023298},
  numpages = {27},
  year = {2020},
  month = {Jun},
  publisher = {American Physical Society},
  doi = {10.1103/PhysRevResearch.2.023298},
  url = {https://link.aps.org/doi/10.1103/PhysRevResearch.2.023298}
}

\appendix
\section{Haar-averaged subsystem purities}
\label{app:haar}

\subsection{General formulas}
\label{app:general_diag_renyi}
\label{app:haar_fragment_purity}

The saturation values used throughout the main text follow from two general formulas, one for the subsystem diagonal purity and one for the full subsystem purity, of a Haar-random pure state restricted to a symmetry sector (or, equivalently, to a Krylov fragment). Throughout, by \emph{Haar-averaged} we mean the ensemble average $\overline{\,\cdot\,}$ taken with respect to the Haar measure on the unitary group acting on the relevant invariant subspace. We derive both formulas here in a single unified framework.

\paragraph{Setup.} Fix an invariant subspace $\mathcal{H}_\mathfrak{q}$ of total dimension $D_\mathfrak{q}$: a symmetry sector, a Krylov fragment, or any other dynamically accessible Hilbert space. Under a bipartition $A\cup B$, it decomposes as
\begin{equation}
\mathcal{H}_\mathfrak{q} = \bigoplus_\lambda \mathcal{H}_A^{(\lambda)}\otimes\mathcal{H}_B^{(\lambda)},
\end{equation}
where $\lambda$ labels the effective sub-sector crossing the cut, with subsystem dimensions $d_A(\lambda)=\dim\mathcal{H}_A^{(\lambda)}$ and $d_B(\lambda)=\dim\mathcal{H}_B^{(\lambda)}$, so that $D_\mathfrak{q}=\sum_\lambda d_A(\lambda)\,d_B(\lambda)$. A Haar-random state on $\mathcal{H}_\mathfrak{q}$ takes the form
\begin{equation}
\ket{\Psi} = \sum_\lambda \sum_{a=1}^{d_A(\lambda)}\sum_{b=1}^{d_B(\lambda)} c_{ab}^{(\lambda)}\,\ket{a,\lambda}_A\otimes\ket{b,\lambda}_B,
\end{equation}
with Haar-distributed coefficients $c_{ab}^{(\lambda)}$. The reduced density matrix on $A$ reads
\begin{equation}
\rho_A = \mathrm{Tr}_B\ket{\Psi}\!\bra{\Psi} = \sum_\lambda \sum_{a,a'=1}^{d_A(\lambda)}\sum_{b=1}^{d_B(\lambda)} c_{ab}^{(\lambda)}\,c_{a'b}^{(\lambda)\,*}\,\ket{a,\lambda}\!\bra{a',\lambda}.
\end{equation}
The two purities we target are then
\begin{align}
\mathrm{Tr}\,\rho_A^2
&= \sum_{\lambda,\lambda'}\sum_{a,a'}\sum_{b,b'} c_{ab}^{(\lambda)}\,c_{a'b}^{(\lambda)\,*}\,c_{a'b'}^{(\lambda')}\,c_{ab'}^{(\lambda')\,*},\label{eq:purity_expanded}\\
\mathrm{Tr}\bigl[(\rho_{A,\mathrm{diag}})^2\bigr]
&= \sum_\lambda\sum_{a=1}^{d_A(\lambda)} p_{a,\lambda}^{\,2},\qquad p_{a,\lambda}=\sum_{b=1}^{d_B(\lambda)}|c_{ab}^{(\lambda)}|^2.\label{eq:app_a_purity_def}
\end{align}

\paragraph{Haar fourth-moment identity.} Both averages are controlled by the standard quartic moment on $\mathcal{H}_\mathfrak{q}$,
\begin{equation}
\overline{c_\mu c_\nu^* c_{\mu'} c_{\nu'}^*}
= \frac{\delta_{\mu\nu}\delta_{\mu'\nu'}+\delta_{\mu\nu'}\delta_{\nu\mu'}}{D_\mathfrak{q}(D_\mathfrak{q}+1)},
\label{eq:haar_quartic}
\end{equation}
where $\mu=(\lambda,a,b)$ is a composite index. Its diagonal specialization is
\begin{equation}
\overline{|c_\mu|^2\,|c_\nu|^2} = \frac{1+\delta_{\mu\nu}}{D_\mathfrak{q}(D_\mathfrak{q}+1)}.
\label{eq:app_a_fourth_moment}
\end{equation}

\paragraph{Full subsystem purity.} Applying Eq.~\eqref{eq:haar_quartic} to Eq.~\eqref{eq:purity_expanded} and summing the two resulting Wick contractions yields the general Haar-averaged formula
\begin{equation}
\overline{\mathrm{Tr}\,\rho_A^2}
= \frac{\sum_\lambda d_A(\lambda)^2\,d_B(\lambda) + \sum_\lambda d_A(\lambda)\,d_B(\lambda)^2}{D_\mathfrak{q}(D_\mathfrak{q}+1)}.
\label{eq:general_purity_fragment}
\end{equation}

\paragraph{Diagonal subsystem purity.} Applying Eq.~\eqref{eq:app_a_fourth_moment} to Eq.~\eqref{eq:app_a_purity_def} and summing the $b,b'$ indices within each $(\lambda,a)$ block gives
\begin{equation}
\overline{p_{a,\lambda}^{\,2}} = \sum_{b,b'=1}^{d_B(\lambda)} \frac{1+\delta_{bb'}}{D_\mathfrak{q}(D_\mathfrak{q}+1)} = \frac{d_B(\lambda)^2 + d_B(\lambda)}{D_\mathfrak{q}(D_\mathfrak{q}+1)}.
\end{equation}
Summing over $a$ and $\lambda$, and using $\sum_\lambda d_A(\lambda)\,d_B(\lambda)=D_\mathfrak{q}$ to collect the linear term, yields the compact result
\begin{equation}
\overline{\mathrm{Tr}\bigl[(\rho_{A,\mathrm{diag}})^2\bigr]} = \frac{\sum_\lambda d_A(\lambda)\,d_B(\lambda)^2 + D_\mathfrak{q}}{D_\mathfrak{q}(D_\mathfrak{q}+1)}.
\label{eq:general_diag_purity_sector}
\end{equation}
The corresponding Haar-averaged subsystem diagonal entropy follows directly,
\begin{equation}
S_{d,\mathrm{Haar}}(L,L_A) = -\log_2\!\left[\frac{\sum_\lambda d_A(\lambda)\,d_B(\lambda)^2 + D_\mathfrak{q}}{D_\mathfrak{q}(D_\mathfrak{q}+1)}\right].
\label{eq:general_diag_renyi_sector}
\end{equation}

Equations~\eqref{eq:general_purity_fragment} and~\eqref{eq:general_diag_renyi_sector} reduce, via the appropriate choice of sub-sector labels and multiplicities, to the saturation values quoted in the main text: the full $(Q=0)$ sector (Appendix~\ref{app:u1_q0}), the full $(Q,P)=(0,0)$ sector (Appendix~\ref{app:qp_full_sector}), and the dipole-swap fragment (Appendix~\ref{app:fragment_resolved}). In the fragment case, $\mathcal{H}_\mathfrak{q}$ and $D_\mathfrak{q}$ should be read as the fragment Hilbert space $\mathcal{H}_F$ and its dimension $D_F$.

\subsection{$U(1)$ sector at $Q=0$}
\label{app:u1_q0}

We now specialize the general formulas of Appendices~\ref{app:general_diag_renyi} to a global $U(1)$ symmetry with total charge fixed to $Q=0$. The sector label $\lambda$ appearing in the general formulas corresponds, after a bipartition $L=L_A+L_B$, to the conserved charge in subsystem $A$, and the multiplicities $d_A(\lambda),d_B(\lambda)$ reduce to binomial or trinomial-like counts of configurations compatible with that charge. The spin-$\tfrac{1}{2}$ and spin-$1$ cases differ only in these counts.

\subsubsection{Spin-$\tfrac{1}{2}$}

For a spin-$\tfrac{1}{2}$ chain of length $L$, the $Q=0$ sector corresponds to half filling and has total dimension
\begin{equation}
D_{Q=0}=\binom{L}{L/2}.
\end{equation}
Under a bipartition of sizes $L_A$ and $L_B=L-L_A$, a configuration in the $Q=0$ sector is uniquely specified by the number of up spins $n_A$ in subsystem $A$, with the complementary subsystem carrying $L/2-n_A$ up spins. The range of allowed values is
\begin{equation}
n_A \in \bigl[\max(0,\,L/2-L_B),\,\min(L_A,\,L/2)\bigr],
\label{eq:nA_range}
\end{equation}
and the corresponding subsystem multiplicities are
\begin{equation}
d_A(n_A)=\binom{L_A}{n_A},\qquad d_B(L/2-n_A)=\binom{L_B}{L/2-n_A}.
\label{eq:u1_spin_half_mults}
\end{equation}
With the convention that binomial coefficients vanish when their arguments lie outside the range above, the sums below may be taken over all integers $n_A$ without loss of generality.

Substituting these multiplicities into Eq.~\eqref{eq:general_diag_renyi_sector} gives the Haar-averaged-average diagonal entropy,
\begin{equation}
S_{d,\mathrm{Haar}}
=-\log_2\!\left[
\frac{\displaystyle\sum_{n_A}\binom{L_A}{n_A}\binom{L_B}{L/2-n_A}^{2}+\binom{L}{L/2}}
{\displaystyle\binom{L}{L/2}\!\left[\binom{L}{L/2}+1\right]}
\right].
\label{eq:Sd_spinhalf}
\end{equation}
Likewise, Eq.~\eqref{eq:general_purity_fragment} yields the Haar-averaged subsystem purity,
\begin{equation}
\overline{\Tr\rho_A^2}
=\frac{\displaystyle\sum_{n_A}\binom{L_A}{n_A}^{2}\binom{L_B}{L/2-n_A}+\sum_{n_A}\binom{L_A}{n_A}\binom{L_B}{L/2-n_A}^{2}}
{\displaystyle\binom{L}{L/2}\!\left[\binom{L}{L/2}+1\right]},
\label{eq:purity_spinhalf}
\end{equation}
and the corresponding Haar-averaged saturation entropy is $S_R^{\mathrm{sat}}(L_A)=-\log_2\overline{\Tr\rho_A^2}$.

\subsubsection{Spin-$1$}

For a spin-$1$ chain, each site carries a local charge $Z_i\in\{-1,0,+1\}$, and the global constraint is $\sum_i Z_i=0$. Each configuration in this sector is specified by the number of $+1$ sites, which equals the number of $-1$ sites (call their common value $k$), with the remaining $L-2k$ sites in the $Z_i=0$ state. The sector dimension is therefore
\begin{equation}
D_{Q=0}^{(\spinlabel=1)}
=\sum_{k=0}^{\lfloor L/2\rfloor}\frac{L!}{k!\,k!\,(L-2k)!},
\label{eq:DS1_total}
\end{equation}
where the multinomial factor counts placements of $k$ particles of charge $+1$, $k$ of charge $-1$, and $L-2k$ of charge $0$.
The bipartition is now labelled by the subsystem charge $q_A=\sum_{i\in A}Z_i \in\{-L_A,\ldots,L_A\}$. A configuration of subsystem $A$ with charge $q_A$ is specified by the number $k$ of minority-species sites in $A$: when $q_A\geq 0$ this is the number of $-1$ sites and there are $k+q_A$ sites of charge $+1$; when $q_A<0$ the roles are reversed. In either case the count of $0$-sites is $L_A-2k-|q_A|$, and the multiplicity is
\begin{equation}
d_A(q_A)
=\sum_{k=0}^{\lfloor(L_A-|q_A|)/2\rfloor}\frac{L_A!}{k!\,(k+|q_A|)!\,(L_A-2k-|q_A|)!}.
\label{eq:dA_spin1}
\end{equation}
The complementary subsystem $B$ carries charge $-q_A$ and has multiplicity $d_B(-q_A)$ given by the same expression with $L_A\to L_B$. The total dimension can equivalently be written as
\begin{equation}
D_{Q=0}^{(\spinlabel=1)}=\sum_{q_A}d_A(q_A)\,d_B(-q_A),
\label{eq:DS1_split}
\end{equation}
with the sum running over $q_A\in[-\min(L_A,L_B),\min(L_A,L_B)]$.

Specializing Eq.~\eqref{eq:general_diag_renyi_sector} with these multiplicities gives the Haar-averaged diagonal entropy,
\begin{equation}
S_{d,\mathrm{Haar}}
=-\log_2\!\left[
\frac{\displaystyle\sum_{q_A}d_A(q_A)\,d_B(-q_A)^2+D_{Q=0}^{(\spinlabel=1)}}
{\displaystyle D_{Q=0}^{(\spinlabel=1)}\!\left(D_{Q=0}^{(\spinlabel=1)}+1\right)}
\right],
\label{eq:Sd_spin1}
\end{equation}
and Eq.~\eqref{eq:general_purity_fragment} gives the Haar-averaged subsystem purity,
\begin{equation}
\overline{\Tr\rho_A^2}
=\frac{\displaystyle\sum_{q_A}d_A(q_A)^2\,d_B(-q_A)+\sum_{q_A}d_A(q_A)\,d_B(-q_A)^2}
{\displaystyle D_{Q=0}^{(\spinlabel=1)}\!\left(D_{Q=0}^{(\spinlabel=1)}+1\right)},
\label{eq:purity_spin1}
\end{equation}
with saturation entropy $S_R^{\mathrm{sat}}(L_A)=-\log_2\overline{\Tr\rho_A^2}$. Equations~\eqref{eq:Sd_spinhalf}--\eqref{eq:purity_spinhalf} and~\eqref{eq:Sd_spin1}--\eqref{eq:purity_spin1} provide the U(1) saturation values used in the main text.

\subsection{$U(1)_Q\times U(1)_P$ sector at $(\mathfrak{q},\mathfrak{p})=(0,0)$}
\label{app:qp_full_sector}

The saturation values discussed in Sec.~5 for the charge- and dipole-conserving circuit are benchmarked against those of a Haar-random state in the \emph{full} $(Q,P)=(0,0)$ sector, that is, without restricting the dynamics to a particular Hilbert-space fragment. Evaluating these benchmarks requires counting configurations compatible with both conservation laws simultaneously, a task naturally handled by generating functions. The construction we develop in this appendix treats spin-$\tfrac{1}{2}$ and spin-$1$ in parallel: only the local factor encoding the on-site states changes between the two cases. Throughout the derivation we keep the total charge and dipole as generic eigenvalues $\mathfrak{q}$ and $\mathfrak{p}$ of the conserved operators $Q$ and $P$, and specialize to $(\mathfrak{q},\mathfrak{p})=(0,0)$ only in the final formulas.

We begin with the sector decomposition. A bipartition $L=L_A+L_B$ splits the $(\mathfrak{q},\mathfrak{p})$ Hilbert space into blocks labelled by the subsystem charge and dipole,
\begin{equation}
\mathcal{H}_{\mathfrak{q},\mathfrak{p}}=\bigoplus_{q_A,p_A}\mathcal{H}_A^{(q_A,p_A)}\otimes\mathcal{H}_B^{(\mathfrak{q}-q_A,\mathfrak{p}-p_A)},
\end{equation}
with subsystem dimensions $d_A(q_A,p_A)=\dim\mathcal{H}_A^{(q_A,p_A)}$ and $d_B(\mathfrak{q}-q_A,\mathfrak{p}-p_A)$, and total dimension $D_{\mathfrak{q},\mathfrak{p}}=\sum_{q_A,p_A}d_A(q_A,p_A)\,d_B(\mathfrak{q}-q_A,\mathfrak{p}-p_A)$. The multiplicities $d_A(q_A,p_A)$ are the sole non-trivial input to Eqs.~\eqref{eq:general_diag_renyi_sector} and~\eqref{eq:general_purity_fragment}; once we have them, the saturation values follow.

The multiplicities themselves are easiest to extract by packaging the information about each site into a formal polynomial. Each site at coordinate $x_i\in\{1,\dots,L\}$ carries a local charge $Z_i$ and contributes $Z_i$ to $q_A$ and $x_iZ_i$ to $p_A$ whenever $i\in A$. Associating a formal variable $z$ to charge and $y$ to dipole, we encode this as a site-local factor $\sum_{Z_i}z^{Z_i}y^{x_iZ_i}$, whose monomials are in one-to-one correspondence with the allowed on-site states. Multiplying these local factors over all sites of a subsystem generates every many-body configuration as a distinct monomial in $z$ and $y$, with the exponents recording the total charge and dipole; the multiplicity $d_A(q_A,p_A)$ is then the coefficient of the appropriate monomial.

The spin-$\tfrac{1}{2}$ case is slightly more involved than spin-$1$ because the on-site charges are half-integer, $Z_i\in\{-\tfrac{1}{2},+\tfrac{1}{2}\}$, so the local factor initially carries half-integer powers,
\begin{equation}
z^{-1/2}y^{-x_i/2}+z^{+1/2}y^{+x_i/2}=z^{-1/2}y^{-x_i/2}\bigl(1+z\,y^{x_i}\bigr).
\label{eq:spinhalf_local}
\end{equation}
Factoring out the half-integer prefactor restores integer powers inside the product, at the cost of a bookkeeping shift. Introducing the up-spin occupation $n_i=Z_i+\tfrac{1}{2}\in\{0,1\}$ and writing $n_A=\sum_{i\in A}n_i$, $\pi_A=\sum_{i\in A}x_in_i$, $\Sigma_A=\sum_{i\in A}x_i$, the subsystem charge and dipole in the physical variables are
\begin{equation}
q_A=n_A-\tfrac{L_A}{2},\qquad p_A=\pi_A-\tfrac{\Sigma_A}{2},
\label{eq:qA_pA_occ}
\end{equation}
and the coefficient extraction becomes
\begin{equation}
d_A(q_A,p_A)=\bigl[z^{q_A+L_A/2}\,y^{p_A+\Sigma_A/2}\bigr]\prod_{i\in A}\bigl(1+z\,y^{x_i}\bigr),
\label{eq:gen_fn_spinhalf}
\end{equation}
with the analogous expression for $d_B(q_B,p_B)$. Although $q_A$ and $p_A$ are individually half-integer, the shifts $q_A+L_A/2$ and $p_A+\Sigma_A/2$ are integers for every allowed configuration, so the coefficient extraction is unambiguous. The total dimension follows by the same recipe on the full chain,
\begin{equation}
D_{\mathfrak{q},\mathfrak{p}}=[z^{\mathfrak{q}+L/2}\,y^{\mathfrak{p}+\Sigma/2}]\prod_{i=1}^{L}\bigl(1+z\,y^{x_i}\bigr),\qquad\Sigma=\sum_{i=1}^{L}x_i.
\label{eq:DQP_spinhalf}
\end{equation}

For spin-$1$ the charges $Z_i\in\{-1,0,+1\}$ are already integer-valued, so no shift is needed. The local factor now carries three terms, one for each on-site state, and the multiplicities are obtained directly,
\begin{equation}
d_A(q_A,p_A)=[z^{q_A}y^{p_A}]\prod_{i\in A}\!\bigl(1+z\,y^{x_i}+z^{-1}y^{-x_i}\bigr),
\label{eq:gen_fn_spin1}
\end{equation}
with the analogous $d_B(q_B,p_B)$ and a total dimension
\begin{equation}
D_{\mathfrak{q},\mathfrak{p}}^{(\spinlabel=1)}=[z^{\mathfrak{q}} y^{\mathfrak{p}}]\prod_{i=1}^{L}\!\bigl(1+z\,y^{x_i}+z^{-1}y^{-x_i}\bigr)
\label{eq:DQP_spin1}
\end{equation}
generated by the same product on the full chain. Specializing to $(\mathfrak{q},\mathfrak{p})=(0,0)$ in either case returns the sector dimension $D_{0,0}$ that enters Eq.~\eqref{eq:general_diag_renyi_sector} and Eq.~\eqref{eq:general_purity_fragment}.

With the multiplicities in hand, the saturation values are immediate. The Haar-averaged diagonal entropy of the subsystem is
\begin{equation}
S_{d,\mathrm{Haar}}=-\log_2\!\left[\frac{\displaystyle\sum_{q_A,p_A}d_A(q_A,p_A)\,d_B(\mathfrak{q}-q_A,\mathfrak{p}-p_A)^{2}+D_{\mathfrak{q},\mathfrak{p}}}{\displaystyle D_{\mathfrak{q},\mathfrak{p}}\bigl(D_{\mathfrak{q},\mathfrak{p}}+1\bigr)}\right],
\label{eq:sd_qp_full}
\end{equation}
and the Haar-averaged full purity is
\begin{equation}
\overline{\Tr\rho_A^2}=\frac{\displaystyle\sum_{q_A,p_A}d_A(q_A,p_A)^{2}\,d_B(\mathfrak{q}-q_A,\mathfrak{p}-p_A)+\sum_{q_A,p_A}d_A(q_A,p_A)\,d_B(\mathfrak{q}-q_A,\mathfrak{p}-p_A)^{2}}{\displaystyle D_{\mathfrak{q},\mathfrak{p}}\bigl(D_{\mathfrak{q},\mathfrak{p}}+1\bigr)},
\label{eq:purity_qp_full}
\end{equation}
with $S_R^{\mathrm{sat}}(L_A)=-\log_2\overline{\Tr\rho_A^2}$. Specializing to $(\mathfrak{q},\mathfrak{p})=(0,0)$ yields the benchmarks quoted in the main text. In the main text, the measured saturation of $S_d$ in the dipole-swap fragment falls well below Eq.~\eqref{eq:sd_qp_full}, confirming that the accessible Hilbert space is substantially smaller than the full $(Q,P)=(0,0)$ sector, a direct dynamical fingerprint of Hilbert-space fragmentation.

\subsection{Dimer-swap fragment}
\label{app:fragment_resolved}

The saturation values for the charge- and dipole-conserving circuit quoted in Sec.~5.2 are obtained by specializing the general formulas of Appendices~\ref{app:general_diag_renyi}  to the connected Krylov fragment selected by the initial state. Working within this fragment introduces one genuinely new feature compared with the earlier appendices: whether the bipartition cuts between two dimers or through a single dimer affects the structure of the sector counting. We develop both cases here in a unified way.

Within the fragment generated by the dipole-preserving swap $|+1,-1,-1,+1\rangle\leftrightarrow|-1,+1,+1,-1\rangle$, the physical chain is naturally grouped into dimers, with $N_d=L/2$ dimers in total. Following the dimer mapping of Sec.~5.1, we label each dimer by whether it is of type $A\equiv|+1,-1\rangle$ or $B\equiv|-1,+1\rangle$. The only allowed move is the exchange $AB\leftrightarrow BA$, so the dynamics is restricted to strings of $A$ and $B$ dimers on an effective chain of length $N_d$. The root state $|+1,-1,-1,+1,\cdots,+1,-1,-1,+1\rangle$ places $K=L/4$ dimers of type $A$ (and equally many of type $B$); any other fragment element is a permutation of this assignment. The fragment dimension is therefore $D_F=\binom{N_d}{K}$.

Under a bipartition $L=L_A+L_B$, configurations in the fragment are classified by the number $r$ of $A$-dimers that fall in subsystem $A$, with the complementary subsystem carrying $K-r$ of them. The allowed values are
\begin{equation}
r\in\bigl[\max(0,\,K-(N_d-n)),\;\min(n,\,K)\bigr],\qquad n=\lfloor L_A/2\rfloor,
\label{eq:r_range}
\end{equation}
with the convention $\binom{a}{b}=0$ for $b<0$ or $b>a$ so that sums over $r$ may be taken over all integers without loss of generality.

When $L_A=2n$ is even, the bipartition falls between two dimers, and every dimer belongs entirely to either $A$ or $B$. The subsystem and complementary multiplicities are straightforward,
\begin{equation}
d_A(r)=\binom{n}{r},\qquad d_B(r)=\binom{N_d-n}{K-r},
\label{eq:dimer_mults_even}
\end{equation}
and the specialization of Appendices~\ref{app:general_diag_renyi}  is immediate. The Haar-averaged diagonal purity is
\begin{equation}
\overline{P_{A,\mathrm{diag}}}=\frac{\displaystyle\sum_{r}\binom{n}{r}\binom{N_d-n}{K-r}^{2}+D_F}{D_F\bigl(D_F+1\bigr)},
\label{eq:diag_even}
\end{equation}
and the Haar-averaged full purity is
\begin{equation}
\overline{\Tr\rho_A^2}=\frac{\displaystyle\sum_{r}\binom{n}{r}^{2}\binom{N_d-n}{K-r}+\sum_{r}\binom{n}{r}\binom{N_d-n}{K-r}^{2}}{D_F\bigl(D_F+1\bigr)}.
\label{eq:full_even}
\end{equation}

The odd case $L_A=2n+1$ is more subtle, because the bipartition passes through a dimer rather than between two dimers. Label the dimer that straddles the cut as the \emph{boundary dimer}: it occupies dimer-slot $n+1$ (counting from the left), with its top site in $A$ and its bottom site in $B$. Because the boundary dimer is split, it is not one of the bulk dimers counted by $r$; its state must instead be summed over, with two possibilities.

If the boundary dimer is a $B$-dimer ($|-1,+1\rangle$), so that its top site in $A$ is $-1$, it does not contribute to the $K$ total $A$-dimers, and all $K-r$ remaining $A$-dimers must be distributed among the $M=N_d-n-1$ bulk dimers of $B$. The number of such configurations is
\begin{equation}
m_1(r)=\binom{M}{K-r},
\label{eq:m1_def}
\end{equation}
where the subscript $1$ tracks the occupation $n_i=Z_i+\tfrac{1}{2}=1$ of the site of the boundary dimer that lies in $A$; i.e., $m_1$ counts configurations for which the $A$-side of the boundary dimer is the spin-down site of a $|-1,+1\rangle$ pair. Conversely, if the boundary dimer is an $A$-dimer ($|+1,-1\rangle$), it contributes to the $K$ total, leaving only $K-r-1$ $A$-dimers to be distributed among the bulk dimers of $B$:
\begin{equation}
m_0(r)=\binom{M}{K-r-1}.
\label{eq:m0_def}
\end{equation}
The subsystem-$A$ multiplicity $d_A(r)=\binom{n}{r}$, counting placements of the $r$ bulk $A$-dimers among the $n$ bulk dimer slots of $A$, is unchanged from the even case.

How these boundary multiplicities enter the purity depends on whether the observable couples identical configurations or independent ones in the two replicas. The diagonal purity $\overline{P_{A,\mathrm{diag}}}$ compares each configuration with itself in both copies, so the state of the boundary dimer must match across the two replicas. Each boundary channel therefore contributes its multiplicity squared, and one obtains
\begin{equation}
\overline{P_{A,\mathrm{diag}}}=\frac{\displaystyle\sum_{r}\binom{n}{r}\bigl[m_1(r)^{2}+m_0(r)^{2}\bigr]+D_F}{D_F\bigl(D_F+1\bigr)}.
\label{eq:diag_odd}
\end{equation}

The full subsystem purity involves both a diagonal and a crossed Wick contraction [cf.\ Eq.~\eqref{eq:general_purity_fragment}]. The diagonal contraction again locks the boundary dimer state across the two replicas and produces the sum of squares $m_1^2+m_0^2$; the crossed contraction, by contrast, permits the boundary dimer to be in different states in the two copies, and sees only the \emph{total} complementary multiplicity $d_B(r)=m_1(r)+m_0(r)$. The two contributions together yield
\begin{equation}
\overline{\Tr\rho_A^2}=\frac{\displaystyle\sum_{r}\binom{n}{r}^{2}\bigl[m_1(r)+m_0(r)\bigr]+\sum_{r}\binom{n}{r}\bigl[m_1(r)^{2}+m_0(r)^{2}\bigr]}{D_F\bigl(D_F+1\bigr)},
\label{eq:full_odd}
\end{equation}
and the corresponding saturation entropies are $S_{d,\mathrm{Haar}}(L_A)=-\log_2\overline{P_{A,\mathrm{diag}}}$ and $S_R^{\mathrm{sat}}(L_A)=-\log_2\overline{\Tr\rho_A^2}$.

Equations~\eqref{eq:diag_even}--\eqref{eq:full_odd} give the exact saturation values within the dipole-swap fragment. They are the analytic benchmark against which the numerical $\Delta S_d(t)$ curves in the main text are measured: the agreement of the simulation with these values confirms that the dynamics explores the full Krylov fragment generated by the root configuration.

\section{Replica tensor network}
\label{app:rtn}

We now outline the computational framework used for the \(n=2\) replica calculation~\cite{zhou2019emergent, zhou2020entanglement, Braccia2024computing, turkeshi2025magic}. We consider a chain of qudits with local Hilbert space \(\mathcal H_q\simeq\mathbb C^q\), focusing on the cases \(q=2\) and \(q=3\), corresponding respectively to spin-\(\tfrac12\) and spin-1 degrees of freedom. A convenient local basis is denoted by \(\{|x\rangle\}\equiv\{|\sigma\rangle\}\), where \(\sigma=\pm\tfrac12\) for \(q=2\), and \(\sigma\in\{-1,0,+1\}\) for \(q=3\).

Our analysis is formulated in the doubled Hilbert-space representation, in which operators are treated as vectors. Explicitly, an operator
\[
A=\sum_{x,y}A_{x,y}\,|x\rangle\langle y|
\]
acting on the many-body Hilbert space \(\mathcal H=\mathcal H_q^{\otimes N}\) is mapped to
\[
|A\rangle\!\rangle=\sum_{x,y}A_{x,y}\,|x,y\rangle\!\rangle.
\]
In this language, the Hilbert--Schmidt inner product becomes \(\mathrm{Tr}(B^\dagger A)=\langle\!\langle B|A\rangle\!\rangle\), while unitary conjugation is represented by the action of \(U\otimes U^*\). This makes the replica construction particularly convenient, since both the time evolution and the swap observables can be written as ordinary matrix elements in the enlarged space.

Starting from an initial pure state \(|\Psi_0\rangle\), the system evolves under the circuit \(U_t\) to \(\rho=U_t|\Psi_0\rangle\langle\Psi_0|U_t^\dagger\). The purity of the reduced density matrix on a subsystem \(A\) is computed using the swap operator \(\mathcal F_A=\prod_{i\in A}\mathcal F_i\), where each local swap exchanges the two replicas. In the vectorized representation, the subsystem purity takes the form
\begin{equation}
P_A=
\langle\!\langle \mathcal F_A |
(U_t\otimes U_t^*)^{\otimes 2}
|\rho_0^{\otimes 2}\rangle\!\rangle.
\end{equation}
Similarly, the purity of the diagonal ensemble is obtained by replacing the swap boundary condition with a diagonal projector,
\begin{equation}
P_{A,\mathrm{diag}}=
\langle\!\langle \Lambda_A |
(U_t\otimes U_t^*)^{\otimes 2}
|\rho_0^{\otimes 2}\rangle\!\rangle,
\end{equation}
where \(|\Lambda_A\rangle\!\rangle=\prod_{i\in A}|+\rangle\!\rangle_i\), with \(|+\rangle\!\rangle=\sum_j |j\rangle\langle j|\). This boundary vector projects onto diagonal matrix elements and therefore gives direct access to the diagonal purity.

After averaging over circuit realizations, both the subsystem purity \(\mathcal P_A=\mathbb E(P_A)\) and the diagonal purity \(\mathcal P_{A,\mathrm{diag}}=\mathbb E(P_{A,\mathrm{diag}})\) reduce to the evaluation of the second moment of the local Haar-random gates,
\[
\mathcal T_{i,i+1}=\mathbb E\!\left[(U_{i,i+1}\otimes U_{i,i+1}^*)^{\otimes 2}\right].
\]
This object plays the role of a local transfer operator and forms the basic ingredient of the replica tensor-network construction.

In the \(U(1)\)-symmetric case, the two-site Hilbert space decomposes into charge sectors labelled by \(Z\). The Haar average therefore splits according to how the replicas are paired across these sectors. One finds
\begin{eqnarray}
\mathcal{T}_{i,i+1}
&=&
\sum_{Z_1\neq Z_2}
\mathbb{E}_{\mathrm{Haar}}\!\left(
U_{Z_1}\otimes U_{Z_1}^{*}\otimes U_{Z_2}\otimes U_{Z_2}^{*}
\right)
\nonumber\\
&&+
\sum_{Z_1\neq Z_2}
\mathbb{E}_{\mathrm{Haar}}\!\left(
U_{Z_1}\otimes U_{Z_2}^{*}\otimes U_{Z_2}\otimes U_{Z_1}^{*}
\right)
\nonumber\\
&&+
\sum_{Z}
\mathbb{E}_{\mathrm{Haar}}\!\left(
U_{Z}\otimes U_{Z}^{*}\otimes U_{Z}\otimes U_{Z}^{*}
\right).
\label{eq:av_formula}
\end{eqnarray}
The first two sums correspond to pairings between distinct symmetry sectors, while the last term describes the contribution from identical sectors and therefore requires the full fourth Haar moment.

For \(Z_1\neq Z_2\), the first contraction projects onto the invariant state
\begin{equation}
\mathbb{E}_{\mathrm{Haar}}\!\left(
U_{Z_1}\otimes U_{Z_1}^{*}\otimes U_{Z_2}\otimes U_{Z_2}^{*}
\right)
=
\frac{1}{d_{Z_1} d_{Z_2}}\,|\mathcal{I}^{+}_{Z_1 Z_2}\rangle\!\rangle\langle\!\langle \mathcal{I}^{+}_{Z_1 Z_2}|,
\end{equation}
with
$|\mathcal{I}^{+}_{Z_1 Z_2}\rangle\!\rangle
=
\sum_{\alpha \in \mathcal{H}_{Z_1}}
\sum_{\beta \in \mathcal{H}_{Z_2}}
|\alpha \alpha \beta \beta\rangle\!\rangle$.
The crossed pairing similarly gives
\begin{equation}
\mathbb{E}_{\mathrm{Haar}}\!\left(
U_{Z_1}\otimes U_{Z_2}^{*}\otimes U_{Z_2}\otimes U_{Z_1}^{*}
\right)
=
\frac{1}{d_{Z_1} d_{Z_2}}\,|\mathcal{I}^{-}_{Z_1 Z_2}\rangle\!\rangle\langle\!\langle \mathcal{I}^{-}_{Z_1 Z_2}|,
\end{equation}
with
$|\mathcal{I}^{-}_{Z_1 Z_2}\rangle\!\rangle
=
\sum_{\alpha \in \mathcal{H}_{Z_1}}
\sum_{\beta \in \mathcal{H}_{Z_2}}
|\alpha \beta \beta \alpha\rangle\!\rangle$.
These two vectors encode the two inequivalent ways of pairing the replicas across different symmetry sectors.

For the diagonal contribution \(Z_1=Z_2\equiv Z\), one must use the fourth Haar moment within a single sector. This gives
\begin{align}
\mathbb{E}_{\mathrm{Haar}}\!\left(
U_{Z} \otimes U_{Z}^{*} \otimes U_{Z} \otimes U_{Z}^{*}
\right)
&=
\frac{1}{d_{Z}^{2}-1}
\Bigg[
|\mathcal{I}_{ZZ}^{+}\rangle\!\rangle\langle\!\langle \mathcal{I}_{ZZ}^{+}|
+
|\mathcal{I}_{ZZ}^{-}\rangle\!\rangle\langle\!\langle \mathcal{I}_{ZZ}^{-}|
\nonumber\\
&\qquad
-\frac{1}{d_{Z}}
\left(
|\mathcal{I}_{ZZ}^{+}\rangle\!\rangle\langle\!\langle \mathcal{I}_{ZZ}^{-}|
+
|\mathcal{I}_{ZZ}^{-}\rangle\!\rangle\langle\!\langle \mathcal{I}_{ZZ}^{+}|
\right)
\Bigg].
\end{align}
This is the familiar Weingarten-type structure: the two invariant pairings appear again, now with coefficients fixed by the dimension of the symmetry sector. Eq.~\eqref{eq:av_formula} can be further rewritten in a more compact form by recasting $
|\mathcal{J}_{Z_1 Z_2}\rangle\!\rangle
\equiv
|\mathcal{I}^{-}_{Z_1 Z_2}\rangle\!\rangle
-\frac{\delta_{Z_1 Z_2}}{d_{Z_1}}
|\mathcal{I}^{+}_{Z_1 Z_2}\rangle\!\rangle$ and renaming $|\mathcal{I}^{+}_{Z_1 Z_2}\rangle\!\rangle\rightarrow|\mathcal{I}_{Z_1 Z_2}\rangle\!\rangle$
as

\begin{equation}
\mathbb{E}_{\text{Haar}}\left(U\otimes U^{*}\otimes U\otimes U^{*}\right)=\sum_{Z_{1}Z_{2}}\frac{1}{d_{Z_1}d_{Z_2}}|\mathcal{I}_{Z_1 Z_2}\rangle\!\rangle\langle\!\langle \mathcal{I}_{Z_1 Z_2}|+\sum_{Z_{1}Z_{2}}\frac{1}{d_{Z_1}d_{Z_2}-\delta_{Z_1 Z_2}}|\mathcal{J}_{Z_1 Z_2}\rangle\!\rangle\langle\!\langle \mathcal{J}_{Z_1 Z_2}|.
\label{eq:transbigZ}
\end{equation}

For practical use in the tensor-network construction, it is useful to rewrite these invariant states in terms of local single-site degrees of freedom living on four replicas. Using the symmetry constraint, one can then recast them as
\begin{equation}
|\mathcal I^{+}_{Z_1 Z_2}\rangle\!\rangle
=
\sum_{\alpha,\beta,\gamma,\delta}
|\alpha\alpha\beta\beta\rangle\!\rangle_{1}\,
|\gamma\gamma\delta\delta\rangle\!\rangle_{2}\,
\delta_{\alpha+\gamma,Z_1}\delta_{\beta+\delta,Z_2},
\end{equation}
and
\begin{equation}
|\mathcal I^{-}_{Z_1 Z_2}\rangle\!\rangle
=
\sum_{\alpha,\beta,\gamma,\delta}
|\alpha\beta\beta\alpha\rangle\!\rangle_{1}\,
|\gamma\delta\delta\gamma\rangle\!\rangle_{2}\,
\delta_{\alpha+\gamma,Z_1}\delta_{\beta+\delta,Z_2}.
\end{equation}

For the \(q=2\) case, the single-site basis in the four-replica space consists of six local states,
\begin{equation}
\begin{aligned}
|\textbf{0}\rangle\!\rangle &= |{+1,+1,+1,+1}\rangle\!\rangle,\qquad
|+_1\rangle\!\rangle = |{+1,-1,+1,-1}\rangle\!\rangle,\qquad
|+_2\rangle\!\rangle= |{-_1,+1,-1,+1}\rangle\!\rangle, \\
|-_1\rangle\!\rangle&= |{+1,-1,-1,+1}\rangle\!\rangle,\qquad
|-_2\rangle\!\rangle= |{-1,+1,+1,-1}\rangle\!\rangle,\qquad
|\textbf{1}\rangle\!\rangle= |{-1,-1,-1,-1}\rangle\!\rangle.
\end{aligned}
\end{equation}

For the \(q=3\) case, the corresponding single-site basis consists of fifteen local states,
\begin{equation}
\begin{aligned}
|\textbf{1}\rangle\!\rangle &= |{+1,+1,+1,+1}\rangle\!\rangle,\qquad
|\textbf{0}\rangle\!\rangle= |{0,0,0,0}\rangle\!\rangle,\qquad
|{\textbf{-1}}\rangle\!\rangle=|{-1,-1,-1,-1}\rangle\!\rangle, \\
|+_1\rangle\!\rangle &=|{+1,0,+1,0}\rangle\!\rangle,\qquad
|+_2\rangle\!\rangle =|{0,+1,0,+1}\rangle\!\rangle,\qquad
|+_3\rangle\!\rangle = |{+1,-1,+1,-1}\rangle\!\rangle, \\
|+_4\rangle\!\rangle &= |{-1,+1,-1,+1}\rangle\!\rangle,\qquad
|+_5\rangle\!\rangle = |{0,-1,0,-1}\rangle\!\rangle,\qquad
|+_6\rangle\!\rangle = |{-1,0,-1,0}\rangle\!\rangle, \\
|-_1\rangle\!\rangle &= |{+1,0,0,+1}\rangle\!\rangle,\qquad
|-_2\rangle\!\rangle = |{0,+1,+1,0}\rangle\!\rangle,\qquad
|-_3\rangle\!\rangle = |{+1,-1,-1,+1}\rangle\!\rangle, \\
|-_4\rangle\!\rangle&= |{-1,+1,+1,-1}\rangle\!\rangle,\qquad
|-_5\rangle\!\rangle = |{0,-1,-1,0}\rangle\!\rangle,\qquad
|-_6\rangle\!\rangle= |{-1,0,0,-1}\rangle\!\rangle.
\end{aligned}
\end{equation}

These states form the local building blocks of the replica transfer matrix. Since the averaged gate acts on two neighboring physical sites, it is natural to use the tensor-product basis built from the single-site states above. Thus, if \(I,J,K,L\) denote local basis states on the two sites \(i\) and \(i+1\), we write
\[
|IJ\rangle\!\rangle \equiv |I\rangle\!\rangle_i\otimes |J\rangle\!\rangle_{i+1},
\qquad
|KL\rangle\!\rangle \equiv |K\rangle\!\rangle_i\otimes |L\rangle\!\rangle_{i+1}.
\]
In this basis, the local transfer operator takes the form
\begin{equation}
\mathcal T_{i,i+1}
=
\sum_{I,J,K,L}
T_{IJ;KL}\,|IJ\rangle\!\rangle\langle\!\langle KL|,
\end{equation}
where
\begin{equation}
T_{IJ;KL}
=
\langle\!\langle IJ|\mathcal T_{i,i+1}|KL\rangle\!\rangle.
\end{equation}

Using the compact decomposition in Eq.~\eqref{eq:transbigZ}, these matrix elements are
\begin{equation}
T_{IJ;KL}
=
\sum_{Z_1,Z_2}
\frac{
\langle\!\langle IJ|\mathcal I_{Z_1Z_2}\rangle\!\rangle
\langle\!\langle \mathcal I_{Z_1Z_2}|KL\rangle\!\rangle
}{
d_{Z_1}d_{Z_2}
}
+
\sum_{Z_1,Z_2}
\frac{
\langle\!\langle IJ|\mathcal J_{Z_1Z_2}\rangle\!\rangle
\langle\!\langle \mathcal J_{Z_1Z_2}|KL\rangle\!\rangle
}{
d_{Z_1}d_{Z_2}-\delta_{Z_1Z_2}
}.
\label{eq:T_IJKL}
\end{equation}

Equivalently, introducing the coefficients
$\Gamma^{(I)}_{Z_1Z_2}(I,J)=\langle\!\langle IJ|\mathcal I_{Z_1Z_2}\rangle\!\rangle$ and $
\Gamma^{(J)}_{Z_1Z_2}(I,J)=\langle\!\langle IJ|\mathcal J_{Z_1Z_2}\rangle\!\rangle$,
one may write
\begin{equation}
T_{IJ;KL}
=
\sum_{Z_1,Z_2}
\frac{
\Gamma^{(I)}_{Z_1Z_2}(I,J)\Gamma^{(I)}_{Z_1Z_2}(K,L)
}{
d_{Z_1}d_{Z_2}
}
+
\sum_{Z_1,Z_2}
\frac{
\Gamma^{(J)}_{Z_1Z_2}(I,J)\Gamma^{(J)}_{Z_1Z_2}(K,L)
}{
d_{Z_1}d_{Z_2}-\delta_{Z_1Z_2}
}.
\end{equation}

Using this local basis, the transfer matrix is a \(36\times 36\) matrix for \(q=2\) and a \(225\times 225\) matrix for \(q=3\). In this representation, the initial states also take a simple form in terms of the one-site replica basis. For the N\'eel state in the \(q=2\) case, one has \( |\rho_{0}^{\otimes 2}\rangle\!\rangle = \bigotimes_{i=1}^{L/2} |0\rangle\!\rangle_{2i-1}|1\rangle\!\rangle_{2i} \), while for the period-3 N\'eel state in the \(q=3\) case, \( |\rho_{0}^{\otimes 2}\rangle\!\rangle = \bigotimes_{i=1}^{L/3} |{-1}\rangle\!\rangle_{3i-2}|0\rangle\!\rangle_{3i-1}|1\rangle\!\rangle_{3i} \).

Similarly, the final state relevant for the purity is $|\mathcal{F}_{A}\rangle\!\rangle = \bigotimes_{i\in A}\left(\sum_{k=1}^{q^2-q}|-_k\rangle\!\rangle \right)_{i}$, whereas for the diagonal purity it is \( |\Gamma_A\rangle\!\rangle = \bigotimes_{i\in A}\left(\sum_{k\in \mathcal{D}}|k\rangle\!\rangle\right)_i \). Here, \(\mathcal{D}\) denotes the diagonal one-site replica basis: for \(q=2\), \(\mathcal{D}=\{|0\rangle\!\rangle,|1\rangle\!\rangle\}\), while for \(q=3\), \(\mathcal{D}=\{|{-1}\rangle\!\rangle,|0\rangle\!\rangle,|1\rangle\!\rangle\}\).

Putting everything together, after applying the replica trick and Haar averaging, each layer of the original circuit maps to \( \mathcal{T}=\left(\prod_{i=1}^{L/2}\mathcal{T}_{2i-1}\right)\left(\prod_{i=1}^{L/2}\mathcal{T}_{2i}\right) \). The purity and diagonal purity are then given by \( \mathcal{P}_A=\langle\!\langle\mathcal{F}_A|\mathcal{T}^t|\rho_0^{\otimes 2}\rangle\!\rangle \) and \( \mathcal{P}_{A,\mathrm{diag}}=\langle\!\langle\Gamma_A|\mathcal{T}^t|\rho_0^{\otimes 2}\rangle\!\rangle \), respectively. This formulation allows \(\mathcal{T}\) to be recast as a matrix product operator, while the initial and final states become matrix product states, so that both purities can be computed efficiently using tensor-network contractions.

\section{Rare-region effects and the role of the initial state}
\label{app:z_basis_average}

In this appendix we explain why the Haar-averaged entanglement entropy $S_R$ grows ballistically for $Z$-basis product states but exhibits a sub-ballistic $\sqrt{t}$ correction for tilted states, and why the diagonal entropy $S_d$ retains a power-law tail $\Delta S_d\sim t^{-1}$ for $Z$-basis initial states. Section~\ref{app:rare_intuition} gives the physical picture in terms of rare frozen regions, following Refs.~\cite{rakovszky2018diffusive, rakovszky2019sub}. Sections~\ref{app:eff_model_setup}--\ref{app:eff_model_SR} develop a quantitative argument using the effective stochastic-model framework of Ref.~\cite{turkeshi2025quantum}. Sections~\ref{app:eff_model_gammaA}--\ref{app:eff_model_mode_sum} extend that framework to the diagonal purity; this extension goes beyond the scope of Ref.~\cite{turkeshi2025quantum} and is one of the contributions of this work.

\subsection{Physical picture: rare frozen regions}
\label{app:rare_intuition}

Consider a one-dimensional spin-$\spinlabel$ circuit with conserved total magnetization $Q=\sum_{i=1}^{L}Z_i$ and dynamics satisfying $[U(t),Q]=0$. Let the entanglement cut lie across the bond $(x,x+1)$. Because the dynamics preserves the total $Z$, entanglement across the cut can grow only if the local configuration near the cut belongs to a sector in which magnetization can be exchanged between the two sides. If instead the neighborhood of the cut is locally \emph{frozen} (all sites in an interval of length $\ell$ around the cut occupying the same extremal $Z$ state), then no local gate inside that interval transports charge across the cut, and the cut remains inactive until a magnetization fluctuation reaches it from outside.

Since the conserved magnetization spreads diffusively, the time required to activate an inactive interval of length $\ell$ scales as
\begin{equation}
t_\ell \sim \ell^2/D,
\label{eq:rare_activation}
\end{equation}
with $D$ the diffusion constant. Large frozen regions therefore act as bottlenecks for entanglement growth~\cite{rakovszky2018diffusive, rakovszky2019sub}.

The crucial distinction between initial states is whether the wavefunction contains many $Z$-basis configurations coherently. For tilted product states such as $\ket{+}^{\otimes L}$ or $\bigl(e^{-iY\theta/2}\ket{0}\bigr)^{\otimes L}$, the wavefunction is a coherent superposition of exponentially many $Z$-basis configurations, including rare ones in which the cut lies inside a large frozen interval. Higher R\'enyi entropies are particularly sensitive to such weakly entangled components, and these rare configurations can dominate the average dynamics, producing sub-ballistic growth~\cite{rakovszky2018diffusive, rakovszky2019sub},
\begin{equation}
S_{\alpha>1}(t)\sim \sqrt{t}.
\label{eq:sqrtt_tilted}
\end{equation}

By contrast, a single $Z$-basis product state contains only one classical configuration. The rare-region mechanism is therefore not intrinsically built into the wavefunction: for a given realization, the cut either lies inside a large frozen interval or it does not. For a typical $Z$-basis product state at density $\rho$, the probability that the cut lies inside a fully polarized interval of length $\ell$ is
\begin{equation}
P(\ell)\sim \max(\rho,1-\rho)^{\ell},
\label{eq:zbasis_rare_probability}
\end{equation}
i.e., exponentially suppressed. At half filling ($\rho=1/2$) the suppression rate is $c=\ln 2$. Consequently the typical distance from the cut to the nearest active region is $\mathcal{O}(1)$, so the cut activates on a microscopic timescale $t_{\mathrm{act}}\sim\mathcal{O}(1)$, and the subsequent entanglement growth is governed by ordinary local scrambling,
\begin{equation}
S_{\alpha>1}(t)\propto t.
\label{eq:ballistic_zbasis}
\end{equation}

The same logic applies after averaging over an ensemble of $Z$-basis product states: configurations with anomalously large frozen intervals exist in the ensemble but are exponentially suppressed by Eq.~\eqref{eq:zbasis_rare_probability}, so the average is dominated by configurations with microscopic activation times, again yielding ballistic growth. Diffusive magnetization transport still governs the late-time relaxation to equilibrium, but it does not control the intermediate-time growth of higher R\'enyi entropies for $Z$-basis product-state ensembles.

In Secs.~\ref{app:eff_model_setup}--\ref{app:eff_model_SR} we make this distinction quantitative using the effective model of Ref.~\cite{turkeshi2025quantum}.

\subsection{Effective-model setup}
\label{app:eff_model_setup}

We adopt the effective-model construction introduced by Turkeshi, Calabrese, and De Luca~\cite{turkeshi2025quantum} for $U(1)$-symmetric random circuits. We summarize only the elements we use; the full derivation and the supplementary identities quoted below are in Ref.~\cite{turkeshi2025quantum} and its supplemental material.

We consider the $U(1)$-symmetric circuit on $N$ sites with local Hilbert space $\mathcal{H}_{2,c}=\mathbb{C}^2\otimes\mathbb{C}^c$. The $\mathbb{C}^2$ factor carries the conserved spin, with charge density $Z_j=\sigma^z_j\otimes I_c$; the $\mathbb{C}^c$ factor is a non-conserved \emph{color} sector. The circuit-averaged Haar-averaged purity of a subsystem $A$ of size $L_A$ is
\begin{equation}
P_A(t)=\langle\!\langle \mathcal{F}_A|\,T^t\,|\rho_0^{\otimes 2}\rangle\!\rangle,
\label{eq:PA_def}
\end{equation}
with $T=\mathbb{E}_{\mathrm{Haar}}[(U_{i,i+1}\otimes U^*_{i,i+1})^{\otimes 2}]$ and $|\rho_0^{\otimes 2}\rangle\!\rangle=\bigotimes_j|\Theta^{(j)}\rangle\!\rangle$.

The single-site four-replica space $(\mathcal{H}_{2,c})^{\otimes 4}$ is spanned, for $c\geq 2$, by the $8$-dimensional basis $|\mu,r,b\rangle\!\rangle$ with $\mu=\pm$, $r,b=\pm\tfrac{1}{2}$:
\begin{align}
|+,r,b\rangle\!\rangle &= \sum_{\substack{x_1=(\sigma_1,\alpha_1)\\ x_2=(\sigma_2,\alpha_2)}} |x_1,x_1,x_2,x_2\rangle\!\rangle\,\delta_{s_1,r}\,\delta_{s_2,b}, \label{eq:basis_plus}\\
|-,r,b\rangle\!\rangle &= \sum_{\substack{x_1=(\sigma_1,\alpha_1)\\ x_2=(\sigma_2,\alpha_2)}} |x_1,x_2,x_2,x_1\rangle\!\rangle\,\delta_{s_1,r}\,\delta_{s_2,b}, \label{eq:basis_minus}
\end{align}
where the four entries in each ket are (ket 1, bra 1, ket 2, bra 2), $s_i$ is the spin part of $x_i$, and color indices are summed freely. The indices $r$ and $b$ are the physical spins on replicas 1 and 2, and $\mu=+$ (resp. $-$) corresponds to the identity (resp. swap) pairing of the color sector. The norms are $\langle\!\langle\mu,r,b|\mu',r',b'\rangle\!\rangle=c^2\,\delta_{\mu\mu'}\delta_{rr'}\delta_{bb'}$ for $c\geq 2$.

Following Refs.~\cite{rakovszky2018diffusive, turkeshi2025quantum}, one introduces the \emph{dressed swap operators}
\begin{equation}
\mathcal{F}_{r,b}(x)=\prod_{j\leq L_A}(Z_j\otimes I)^{n_j^{(r)}}(I\otimes Z_j)^{n_j^{(b)}}\,\mathcal{F}(x),
\label{eq:dressed_swap}
\end{equation}
where $x$ is the interface position, $n_j^{(r)},n_j^{(b)}\in\{0,1\}$ are red and blue particle occupations, and $\mathcal{F}(x)=\prod_{j\leq x}\mathcal{F}_j$ is the bare swap string. A red (blue) particle at site $j$ corresponds to a $Z_j$ insertion on replica 1 (replica 2). The bare purity in Eq.~\eqref{eq:PA_def} is $\mathcal{F}_{0,0}(L_A)$, i.e., the zero-particle sector.

For a product initial state $|\Psi_0\rangle=\bigotimes_y|\psi_y\rangle$, the doubled initial state factorises over sites, and the overlap with the dressed swap at site $y$ reduces, after using the swap identity $\mathcal{F}(\rho\otimes\sigma)=\sigma\otimes\rho$ and the cyclicity of the trace, to
\begin{equation}
\langle\!\langle \mathcal{F}_{r,b}|\rho_y^{\otimes 2}\rangle\!\rangle=\langle Z_y\rangle^{n_y^{(r)}+n_y^{(b)}},
\label{eq:site_overlap}
\end{equation}
with $\langle Z_y\rangle=\Tr[\rho_y Z_y]$. Hence the full boundary weight over all sites is
\begin{equation}
W_{\mathrm{init}}[H]=\prod_y\langle Z_y\rangle^{n_y^{(r)}+n_y^{(b)}}.
\label{eq:W_init}
\end{equation}

In the $c\to\infty$ limit, Ref.~\cite{turkeshi2025quantum} shows that the transfer matrix generates a Markov process on the classical variables $\{n_j^{(r)},n_j^{(b)},x\}$ governed by three stochastic matrices,
\begin{align}
M_{j,j+1} &= \tfrac{1}{4}\bigl(1+S^{(r)}_{j,j+1}\bigr)\bigl(1+S^{(b)}_{j,j+1}\bigr), \label{eq:M}\\
R_{j,j+1} &= \tfrac{1}{2}\bigl(M_{j,j+1}+R^{(r)}_{j,j+1}R^{(b)}_{j,j+1}\bigr), \label{eq:R}\\
L_{j,j+1} &= \tfrac{1}{2}\bigl(M_{j,j+1}+L^{(r)}_{j,j+1}L^{(b)}_{j,j+1}\bigr), \label{eq:L}
\end{align}
with $S^{(\alpha)}$ the color-$\alpha$ exchange and $L^{(\alpha)},R^{(\alpha)}$ the interface matrices defined in the supplemental material of Ref.~\cite{turkeshi2025quantum}. Physically, $M$ governs the bulk: two independent symmetric simple exclusion processes (SSEP), with each particle hopping left or right with equal probability subject to the same-color hard-core constraint, and no particles created or destroyed. $L$ and $R$ act at the interface: the interface moves by one site, and simultaneously a red--blue pair is created or annihilated; thus, pair creation/annihilation is locked to interface motion. The Haar-averaged purity is a weighted sum over histories,
\begin{equation}
P_A(t)=\Bigl(\frac{2}{c}\Bigr)^{t}\sum_{H}W_{\mathrm{bulk}}[H]\,W_{\mathrm{init}}[H],
\label{eq:PA_history}
\end{equation}
with the prefactor $(2/c)^t$ from the color sector.

We stress two points. (i) The framework computes the \emph{purity}; it does not modify the underlying Haar-averaged bulk scrambling. The initial-state dependence enters entirely through $W_{\mathrm{init}}$. (ii) The quantity $N_r+N_b$ counted along a history is the number of charge insertions required by the dressed swap, not a physical particle number; suppressing histories with many particles therefore confines the interface, rather than restricting the accessible bulk dynamics.

\subsection{Entanglement entropy: tilted vs $Z$-basis states}
\label{app:eff_model_SR}

\paragraph{Tilted states.} For the tilted ferromagnetic state $|\Psi_0\rangle=(e^{-iY\theta/2}|0\rangle\otimes|0\rangle)^{\otimes N}$ with uniform $\langle Z_y\rangle=\cos\theta$, the boundary weight~\eqref{eq:W_init} becomes
\begin{equation}
W_{\mathrm{init}}[H]=(\cos\theta)^{N_r+N_b},\qquad N_{r,b}=\sum_j n_j^{(r,b)}.
\end{equation}
The macroscopic fluctuation theory analysis of Ref.~\cite{turkeshi2025quantum} yields
\begin{equation}
\lim_{t\to\infty}\frac{1}{\sqrt{t}}\ln P_A(t)=4\mu\bigl(\ln\cos\theta;\tfrac{1}{2},0\bigr),
\label{eq:mu_rate}
\end{equation}
with $\mu$ the rate function computed in the supplemental material of Ref.~\cite{turkeshi2025quantum}. For small $\theta$ this rate function is negative, with leading behavior of order $-\theta^2/\sqrt{\pi}$; see Ref.~\cite{turkeshi2025quantum} for the explicit perturbative expansion. Consequently
\begin{equation}
\ln P_A(t)\simeq \log(c/2)\,t-\gamma(\theta)\sqrt{Dt},\qquad \gamma(\theta)>0,
\label{eq:SR_tilted}
\end{equation}
so that $S_R=-\log_2 P_A$ picks up a positive $\sqrt{Dt}$ correction on top of the ballistic growth. Physically, every red or blue particle created by $L,R$ at the interface picks up a factor $|\cos\theta|<1$, suppressing histories with many particles and confining the interface to a strip of width $\sim\sqrt{Dt}$ around the cut. This quantitatively realises the rare-region mechanism of Sec.~\ref{app:rare_intuition}: the rare frozen configurations that delay activation of the cut are encoded in the effective model as the $\cos\theta$ reweighting of histories.

\paragraph{$Z$-basis product states.} For a general $Z$-basis state $|\Psi_0\rangle=\bigotimes_y|s_y\rangle$ with $s_y=\,+1$ or $-1$, the local magnetisation is $\langle Z_y\rangle=\pm 1$, so
\begin{equation}
|W_{\mathrm{init}}[H]|=1
\label{eq:W_init_zbasis}
\end{equation}
for every history $H$. This holds at the level of the discrete Markov process, without any coarse-graining or translation-invariance assumption. This further implies that no history is suppressed, and the interface is not confined to a $\sqrt{Dt}$ strip around the cut. The ballistic growth $S_R\propto t$ is the standard random-circuit result~\cite{nahum2018operator}; the role of the effective model here is to show that, unlike the tilted case, \emph{no subleading $\sqrt{t}$ correction is generated} for $Z$-basis initial states, because no history is reweighted.

\subsection{Extension to the diagonal purity: the boundary $\Gamma_A$}
\label{app:eff_model_gammaA}

We now turn to the diagonal purity $P_{A,\mathrm{diag}}=\Tr[(\rho_{A,\mathrm{diag}})^2]$, with $\rho_{A,\mathrm{diag}}=\sum_s\langle s|\rho_A|s\rangle\,|s\rangle\langle s|$. The framework of Ref.~\cite{turkeshi2025quantum} was developed for the full purity $P_A$; in what follows we extend it to $P_{A,\mathrm{diag}}$. In the replica formalism (cf.\ Appendix~\ref{app:rtn}),
\begin{equation}
P_{A,\mathrm{diag}}(t)=\langle\!\langle\Gamma_A|\,T^t\,|\rho_0^{\otimes 2}\rangle\!\rangle,
\label{eq:PAD}
\end{equation}
with $\Gamma_A=\bigotimes_{i\in A}\Gamma^{(i)}\otimes\bigotimes_{i\in\bar A}I^{(i)}$, where $I^{(i)}$ is the identity boundary on $\bar A$ and the diagonal projector at a single site acts on the four-replica space as
\begin{equation}
\Gamma^{(i)}=\sum_\sigma|\sigma,\sigma,\sigma,\sigma\rangle\langle\sigma,\sigma,\sigma,\sigma|,
\label{eq:Gamma_i}
\end{equation}
with $\sigma=(\sigma_s,\alpha)$ running over spin and color. Computing the overlap of $\Gamma^{(i)}$ with the basis states~\eqref{eq:basis_plus}--\eqref{eq:basis_minus} using the diagonal constraint yields:
\begin{equation}
\langle\!\langle+,r,b|\Gamma^{(i)}\rangle\!\rangle=c\,\delta_{r,b},
\qquad
\langle\!\langle-,r,b|\Gamma^{(i)}\rangle\!\rangle=c\,\delta_{r,b}.
\label{eq:Gamma_overlaps}
\end{equation}
Inverting the Gram matrix on the $b=r$ subspace yields
\begin{equation}
|\Gamma^{(i)}\rangle\!\rangle=\frac{1}{c+1}\sum_{r=\pm 1/2}\bigl(|+,r,r\rangle\!\rangle+|-,r,r\rangle\!\rangle\bigr).
\label{eq:Gamma_expansion}
\end{equation}

Two features drive what follows: first, both $\mu=+$ and $\mu=-$ enter with equal amplitude; there is no boundary suppression of either sector at $t=0$. Second, the constraint $b=r$ means both replicas carry the same spin at every site of $A$, so $\sum_r|\mu,r,r\rangle\!\rangle\propto(1+Z_j^{(r)}Z_j^{(b)})|\mu,-\tfrac{1}{2},-\tfrac{1}{2}\rangle\!\rangle$, i.e., $\Gamma_A$ injects a red--blue pair at every site of $A$.

\subsection{Diagonal relaxation: the $\mu=-$ sector and the mode sum}
\label{app:eff_model_mode_sum}

The time evolution exponentially suppresses the $\mu=-$ sector relative to $\mu=+$. In $\mu=+$, the color indices pair $(\alpha,\alpha,\beta,\beta)$; the Haar average acts as identity on the color sector with per-step eigenvalue $\lambda_+=1$. In $\mu=-$, the color indices cross, $(\alpha,\beta,\beta,\alpha)$; the Haar average produces a Weingarten-type suppression with leading large-$c$ behaviour
\begin{equation}
\lambda_-\sim\frac{1}{c}\qquad(c\to\infty).
\label{eq:lambda_minus}
\end{equation}
The $\mathcal{O}(1)$ numerical coefficient is given by the Weingarten contraction detailed in the supplemental material of Ref.~\cite{turkeshi2025quantum} and does not affect the scaling argument that follows. Since $\Gamma_A$ populates both sectors equally [Eq.~\eqref{eq:Gamma_expansion}], the diagonal purity splits as
\begin{equation}
P_{A,\mathrm{diag}}(t)=\underbrace{P_{A,\mathrm{diag},+}(t)}_{\text{eigenvalue }1}+\underbrace{P_{A,\mathrm{diag},-}(t)}_{\text{exponentially suppressed in }t},
\label{eq:PAD_split}
\end{equation}
and at late times $P_{A,\mathrm{diag}}(t)\to P_{A,\mathrm{diag},+}(t)$.

In the surviving $\mu=+$ sector, the permutation index is $\mu=+$ on both $A$ and $\bar A$, so no domain wall is present and the interface operators $L,R$ are never triggered. The bulk dynamics is controlled entirely by $M$ in Eq.~\eqref{eq:M}: pure symmetric exclusion, with no particle creation or annihilation. The red--blue pairs injected by $\Gamma_A$ at every site of $A$ at the boundary time $t$ propagate backward under $M$, their total number conserved. At the boundary, the particle density is a step function, $\rho_0$ for $y\in A$ and $0$ for $y\in\bar A$, which under $M$ relaxes diffusively, with edges smoothing over a width $\sim\sqrt{Dt}$. For $Z$-basis initial states, the injected particles arrive at the initial-state boundary with unit weight [Eq.~\eqref{eq:W_init_zbasis}], so no configuration is suppressed, and $\Delta P_{A,\mathrm{diag}}(t)$ is controlled entirely by the incomplete diffusive relaxation.

\paragraph{Mode-sum argument.} The deviation admits a mode decomposition over the diffusive eigenmodes of the SSEP, labelled by momentum $k$ with eigenvalue $e^{-Dk^2t}$. The key input is that $\Gamma_A$ injects $Z_j^{(r)}Z_j^{(b)}$ at each site, a \emph{product} of two independent charge operators, one per replica. Since red and blue particles evolve independently under $M$ (which factorises as $M=\tfrac{1}{4}(1+S^{(r)})(1+S^{(b)})$), the leading contribution to $\Delta P_{A,\mathrm{diag}}$ is a product of two independent mode integrals,
\begin{equation}
\Delta P_{A,\mathrm{diag}}(t)\sim
\underbrace{\int\frac{dk}{2\pi}\,a_r(k)\,e^{-Dk^2t}}_{\text{red leg}}
\times
\underbrace{\int\frac{dk'}{2\pi}\,a_b(k')\,e^{-D{k'}^2t}}_{\text{blue leg}}.
\label{eq:mode_sum}
\end{equation}
If the spectral weights $a_{r,b}(k)$ are smooth and nonzero at $k=0$, as expected for a local charge injection against a translation-invariant SSEP background, each integral gives $\sim(Dt)^{-1/2}$, so
\begin{equation}
\Delta P_{A,\mathrm{diag}}(t)\sim\frac{1}{Dt},\qquad
\Delta S_d(t)\sim t^{-\beta_{S_d}}\text{ with }\beta_{S_d}\approx 1.
\label{eq:DeltaSd_t1}
\end{equation}
This is consistent with the numerically observed exponent $\beta_{S_d}\simeq 1.10$ reported in the main text for the spin-$\tfrac{1}{2}$ $U(1)$ circuit.

\section{Tensor-network computation of the relative entropy of coherence}
\label{app:tn_coherence}

In this appendix, we describe the algorithm used to compute the relative entropy of coherence \(C_d(\rho_A)=S_d(\rho_A)-S_R(\rho_A)\) directly from a matrix product state (MPS) representation of \(|\Psi\rangle\), as employed for the Hamiltonian dynamics simulations. The entanglement entropy \(S_R(\rho_A)=-\log_2\sum_\alpha\lambda_\alpha^4\) is obtained from the Schmidt values \(\lambda_\alpha\) at the \(A|\bar{A}\) bipartition via a standard singular value decomposition. The nontrivial ingredient is the diagonal entropy \(S_d(\rho_A)\), whose evaluation requires the diagonal elements of the reduced density matrix in the computational basis,
\begin{equation}
p_{\boldsymbol{\sigma}_A}
=
\langle \boldsymbol{\sigma}_A | \rho_A | \boldsymbol{\sigma}_A \rangle
=
\sum_{\boldsymbol{\sigma}_{\bar{A}}} \left| \langle \boldsymbol{\sigma}_A, \boldsymbol{\sigma}_{\bar{A}} | \Psi \rangle \right|^2,
\end{equation}
where \(\boldsymbol{\sigma}_A = (\sigma_1, \ldots, \sigma_{L_A})\) labels a computational-basis configuration of subsystem~\(A\). The diagonal purity is
\begin{equation}
\mathrm{Tr}\!\left[(\rho_{A,\mathrm{diag}})^2\right] = \sum_{\boldsymbol{\sigma}_A} p_{\boldsymbol{\sigma}_A}^2.
\label{eq:diag_purity_mps}
\end{equation}

Na\"ively, evaluating Eq.~\eqref{eq:diag_purity_mps} requires summing over all \(q^{L_A}\) configurations. We circumvent this by constructing an MPS representation of the probability vector \(\{p_{\boldsymbol{\sigma}_A}\}\) and computing the diagonal purity as its squared norm.

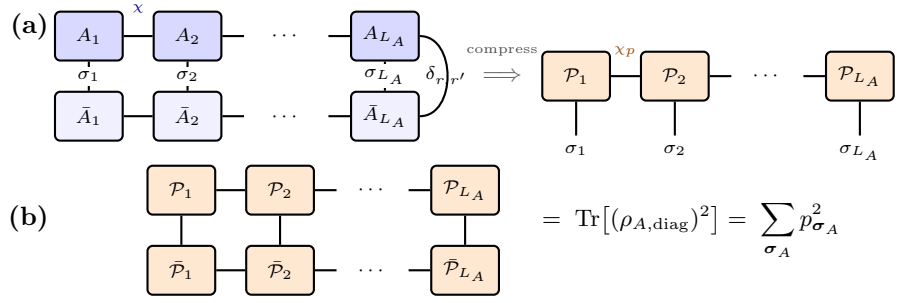
\begin{figure}[t!]
\centering
\begin{tikzpicture}[
 scale=0.82,
 ket/.style={draw, thick, fill=blue!15, rounded corners=2pt,
 minimum width=9mm, minimum height=6.5mm, inner sep=0.5pt},
 bra/.style={draw, thick, fill=blue!6, rounded corners=2pt,
 minimum width=9mm, minimum height=6.5mm, inner sep=0.5pt},
 prob/.style={draw, thick, fill=orange!18, rounded corners=2pt,
 minimum width=9mm, minimum height=6.5mm, inner sep=0.5pt},
 lbl/.style={font=\scriptsize},
 every path/.style={thick},
 >=latex
]

\node[font=\normalsize\bfseries, anchor=east] at (-0.5, 0.8) {(a)};

\node[ket, lbl] (k1) at (0, 0.65) {$A_1$};
\node[ket, lbl] (k2) at (1.6, 0.65) {$A_2$};
\node[ket, lbl] (kN) at (4.8, 0.65) {$A_{L_A}$};

\node[bra, lbl] (b1) at (0, -0.65) {$\bar{A}_1$};
\node[bra, lbl] (b2) at (1.6, -0.65) {$\bar{A}_2$};
\node[bra, lbl] (bN) at (4.8, -0.65) {$\bar{A}_{L_A}$};

\draw (k1.east) -- (k2.west);
\draw (k2.east) -- ++(0.35,0);
\node[lbl] at (3.2, 0.65) {$\cdots$};
\draw (kN.west) -- ++(-0.35,0);

\draw (b1.east) -- (b2.west);
\draw (b2.east) -- ++(0.35,0);
\node[lbl] at (3.2, -0.65) {$\cdots$};
\draw (bN.west) -- ++(-0.35,0);

\draw (k1.south) -- (b1.north);
\draw (k2.south) -- (b2.north);
\draw (kN.south) -- (bN.north);
\node[fill=white, inner sep=1.5pt, lbl] at (0, 0) {$\sigma_1$};
\node[fill=white, inner sep=1.5pt, lbl] at (1.6, 0) {$\sigma_2$};
\node[fill=white, inner sep=1.5pt, lbl] at (4.8, 0) {$\sigma_{L_A}$};

\draw (kN.east) .. controls ++(0.6,0) and ++(0.6,0) .. (bN.east);
\node[lbl] at (5.8, 0) {$\delta_{r,r'}$};

\node[font=\tiny, blue!60!black] at (0.8, 1.1) {$\chi$};

\node[font=\small, gray!80!black] at (6.7, 0) {$\Longrightarrow$};
\node[font=\tiny, gray!60!black] at (6.7, 0.4) {compress};

\node[prob, lbl] (p1) at (7.9, 0) {$\mathcal{P}_1$};
\node[prob, lbl] (p2) at (9.5, 0) {$\mathcal{P}_2$};
\node[prob, lbl] (pN) at (12.5, 0) {$\mathcal{P}_{L_A}$};

\draw (p1.east) -- (p2.west);
\draw (p2.east) -- ++(0.35,0);
\node[lbl] at (11.0, 0) {$\cdots$};
\draw (pN.west) -- ++(-0.35,0);

\draw (p1.south) -- ++(0,-0.55) node[below, lbl] {$\sigma_1$};
\draw (p2.south) -- ++(0,-0.55) node[below, lbl] {$\sigma_2$};
\draw (pN.south) -- ++(0,-0.55) node[below, lbl] {$\sigma_{L_A}$};

\node[font=\tiny, orange!60!black] at (8.7, 0.4) {$\chi_p$};

\node[font=\normalsize\bfseries, anchor=east] at (-0.5, -2.3) {(b)};

\node[prob, lbl] (q1) at (1.5, -1.85) {$\mathcal{P}_1$};
\node[prob, lbl] (q2) at (3.1, -1.85) {$\mathcal{P}_2$};
\node[prob, lbl] (qN) at (6.1, -1.85) {$\mathcal{P}_{L_A}$};

\node[prob, lbl] (r1) at (1.5, -3.15) {$\bar{\mathcal{P}}_1$};
\node[prob, lbl] (r2) at (3.1, -3.15) {$\bar{\mathcal{P}}_2$};
\node[prob, lbl] (rN) at (6.1, -3.15) {$\bar{\mathcal{P}}_{L_A}$};

\draw (q1.east) -- (q2.west);
\draw (q2.east) -- ++(0.35,0);
\node[lbl] at (4.6, -1.85) {$\cdots$};
\draw (qN.west) -- ++(-0.35,0);

\draw (r1.east) -- (r2.west);
\draw (r2.east) -- ++(0.35,0);
\node[lbl] at (4.6, -3.15) {$\cdots$};
\draw (rN.west) -- ++(-0.35,0);

\draw (q1.south) -- (r1.north);
\draw (q2.south) -- (r2.north);
\draw (qN.south) -- (rN.north);

\node[font=\small, anchor=west] at (7.2, -2.5)
 {$=\;\mathrm{Tr}\!\left[(\rho_{A,\mathrm{diag}})^2\right]
 =\;\displaystyle\sum_{\boldsymbol{\sigma}_A} p_{\boldsymbol{\sigma}_A}^2$};

\end{tikzpicture}
\caption{
\textbf{Tensor-network construction of the diagonal purity from an MPS.}
(a)~The probability vector \(p_{\boldsymbol{\sigma}_A}\) is obtained by pairing the MPS tensors \(A_j\) (ket, dark blue) with their conjugates \(\bar{A}_j\) (bra, light blue) at each site of subsystem~\(A\). Both layers are projected onto the same computational-basis state~\(\sigma_j\), which becomes the physical index of the resulting \emph{probability MPS}. The right boundary condition \(\delta_{r,r'}\) implements the partial trace over the complement~\(\bar{A}\) (enabled by right-canonical gauge for sites \(j>L_A\)). Truncated SVD compression at each bond reduces the doubled bond dimension~\(\chi^2\) to a manageable value~\(\chi_p\).
(b)~The diagonal purity \(\mathrm{Tr}[(\rho_{A,\mathrm{diag}})^2]\) is computed as the squared norm \(\langle\mathcal{P}|\mathcal{P}\rangle\) of the probability MPS, with physical indices contracted between the ket and bra copies.}
\label{fig:prob_mps}
\end{figure}

Let \(A^{[\sigma_j]}_{l_j r_j}\) denote the MPS matrix at site~\(j\) obtained by projecting the local tensor onto \(|\sigma_j\rangle\), with bond indices \(l_j\) and \(r_j\) of dimension~\(\chi_j\). We gauge the MPS so that sites \(j > L_A\) are right-canonical, whereby the partial trace over the complement reduces to a contraction of the bond indices at site~\(L_A\),
\begin{equation}
\sum_{\boldsymbol{\sigma}_{\bar{A}}} \prod_{j=L_A+1}^{L} A^{[\sigma_j]} \otimes \bar{A}^{[\sigma_j]} = I.
\end{equation}
The probability then takes the form
\begin{equation}
p_{\boldsymbol{\sigma}_A}
=
\sum_{r}
\left( \prod_{j=1}^{L_A} A^{[\sigma_j]} \right)_{1,r}
\overline{\left( \prod_{j=1}^{L_A} A^{[\sigma_j]} \right)_{1,r}}.
\end{equation}

This expression has a natural matrix product structure in the \emph{doubled} bond space. At each site~\(j \in A\), define the local tensor
\begin{equation}
\mathcal{P}^{[\sigma_j]}_{(l,l'),(r,r')}
=
A^{[\sigma_j]}_{l,r} \, \bar{A}^{[\sigma_j]}_{l',r'},
\end{equation}
with combined indices \((l,l')\) and \((r,r')\), each of dimension up to~\(\chi^2\). The probability is
\begin{equation}
p_{\boldsymbol{\sigma}_A}
=
\mathrm{Tr}_{r,r'}\!\left[
\prod_{j=1}^{L_A} \mathcal{P}^{[\sigma_j]}
\right],
\end{equation}
where \(\mathrm{Tr}_{r,r'}\) denotes the contraction \(\delta_{r_{L_A},r'_{L_A}}\) at the right boundary. Viewing \(\sigma_j\) as a physical index, this defines an MPS whose amplitudes are the probabilities \(p_{\boldsymbol{\sigma}_A}\), and the diagonal purity is the squared norm of this probability MPS,
\begin{equation}
\mathrm{Tr}\!\left[(\rho_{A,\mathrm{diag}})^2\right]
=
\langle \mathcal{P} | \mathcal{P} \rangle
=
\sum_{\boldsymbol{\sigma}_A} p_{\boldsymbol{\sigma}_A}^2.
\end{equation}
The construction is illustrated schematically in Fig.~\ref{fig:prob_mps}.

To keep the bond dimension manageable, the probability MPS is compressed via truncated SVD at each step, retaining at most~\(\chi_p\) singular values. The algorithm sweeps from left to right through sites \(j=1,\ldots,L_A\). At each site and for each basis state~\(\sigma_j\), the accumulated left environment~\(L_{j-1}\) (of dimensions \(\chi_p \times \chi \times \chi\)) is contracted first with \(A^{[\sigma_j]}\) and then with \(\bar{A}^{[\sigma_j]}\). This two-step contraction avoids forming the full \(\chi^2 \times \chi^2\) doubled transfer matrix, reducing the per-site cost from \(O(d\,\chi_p\,\chi^4)\) to \(O(d\,\chi_p\,\chi^3)\). The resulting block is decomposed via truncated SVD into a local MPS tensor and an updated environment; at the boundary site~\(j = L_A\), the bond indices are contracted (\(\delta_{r,r'}\)) to perform the partial trace. The diagonal entropy is then
\begin{equation}
S_d(\rho_A) = -\log_2 \langle \mathcal{P} | \mathcal{P} \rangle,
\end{equation}
and the total cost of the algorithm scales as \(O(L_A \, d \, \chi_p \, \chi^3)\).

\end{document}